\documentclass[11pt]{article}

\usepackage[utf8]{inputenc}  
\usepackage[T1]{fontenc}
\usepackage{setspace}

\usepackage{lmodern}
\usepackage{microtype}
\usepackage{amsmath,amssymb}%,amsfonts,amssymb,amsthm
\usepackage{mathrsfs}
\usepackage[table]{xcolor} 
\usepackage{graphicx}
\usepackage{braket}
\usepackage{mleftright} %\mleftright
\usepackage{array}
\usepackage{booktabs}
\usepackage{slashed}
\usepackage{color}   

\usepackage[english]{babel} 
\usepackage{caption} 
\usepackage{subcaption} %for subfigures
\usepackage{bm} %for making symbols bold by \bm command   
\usepackage{multirow} %for drawing table
\usepackage{calc} 
\usepackage{makecell} %for adjusting table /hline thickness
\usepackage{physics}
\usepackage{url}
\usepackage{hyperref}
\usepackage{cleveref}
\usepackage{ragged2e}     % command \RaggedRight
\usepackage{verbatim} 

\definecolor{blizzardblue}{rgb}{0.93, 0.93, 0.93} % light gray
\newcolumntype{?}{!{\vrule width 0.8pt}} %makes vertical lines in tables with thickness 0.8pt
\mathchardef\mhyphen="2D %small hyphen in math mode

\newcommand{\RN}[1]{ %command for writing roman numbers 
  \textup{\uppercase\expandafter{\romannumeral#1}}%  
\renewcommand\thesubfigure{(\alph{subfigure})} %for referencing to subfigure as 2(a) instead of 2a
\captionsetup[sub]{labelformat=simple} 
}
\newcommand{\ra}{ %command for writing roman numbers 
\rightarrow
}

\begin{document}

\begin{center}

\vspace*{15mm}
\vspace{1cm}
{\Large \bf 
\begin{onehalfspace}
Search for lepton-flavor-violating ALPs at a future muon collider and utilization of polarization-induced effects
\end{onehalfspace}
}

\vspace{1cm}

\small{\bf Gholamhossein Haghighat$^{\dagger}$\footnote{h.haghighat@ipm.ir}, Mojtaba Mohammadi Najafabadi$^{\dagger}$\footnote{mojtaba@ipm.ir}}

 \vspace*{0.5cm}
 
{\small\sl 
$^{\dagger}$ School of Particles and Accelerators, Institute for Research in Fundamental Sciences (IPM), \\P.O. Box 19395-5531, Tehran, Iran}

\vspace*{.2cm}
\end{center}

\vspace*{10mm}

%***********************************************************
\begin{abstract}\label{abstract}
Axion-Like Particles (ALPs) are pseudo Nambu-Goldstone bosons associated with spontaneously broken global $U(1)$ symmetries. Such particles can have lepton-flavor-violating (LFV) couplings to the SM charged leptons. LFV ALPs provide the possibility to address some of the SM long-lasting problems. We investigate the sensitivity of a future muon collider suggested by the Muon Accelerator Program (MAP) to the production of LFV ALPs in the ALP mass range $m_a\leq1$ MeV. ALPs are assumed to be produced through the LFV decay $\tau\ra\ell a$ ($\ell=e,\mu$) of one of the tau leptons produced in the muon-anti muon annihilation. Performing a realistic detector simulation and deploying a multivariate technique, we constrain the LFV couplings $c_{\tau e}$ and $c_{\tau \mu}$ for both the cases of unpolarized and polarized muon beams. Three different chiral structures are considered for the LFV ALP coupling and the muon collider is assumed to operate at the center-of-mass energies of 126, 350 and 1500 GeV. We present a procedure to search for LFV ALPs at colliders which takes advantage of tau polarization-induced effects. Polarized tau leptons which produce such effects can be produced when the initial muon beams are polarized. Utilizing the properties of polarized tau decays, the main SM background which overwhelms the ALP production in the case of unpolarized muon beams can be significantly suppressed. We present 95$\%$ CL expected limits on the LFV couplings and show that the present analysis can improve current experimental limits on the ALP LFV couplings by roughly one order of magnitude. 
\end{abstract}

\newpage

%***********************************************************
\section{Introduction}
\label{sec:introduction}
The observed neutrino oscillations \cite{Lipari:677618} suggest the possibility of charged lepton-flavor-violating (LFV) decays. Such decays are, however, strongly suppressed by the neutrino mass-squared differences and their predicted branching fractions \cite{Petcov:1976ff,Hernandez-Tome:2018fbq} are many orders of magnitude smaller than the present experimental limits \cite{Bellgardt:1987du,TheMEG:2016wtm}. Absence of a significant source of lepton-flavor violation in the Standard Model (SM) provides a strong motivation for searching for LFV processes as any observable LFV signal must come from physics beyond SM (BSM).

Axion-Like Particles (ALPs) are pseudo Nambu-Goldstone bosons that appear as a result of the spontaneous breaking of global $U(1)$ symmetries \cite{Peccei:1977hh,DiLuzio:2020wdo,Brivio:2017ije}. ALPs, their couplings to the SM particles (including LFV couplings) and their applications have been investigated in many studies to date. These studies suggest that ALPs are capable of addressing some of the long-lasting SM problems. These particles provide a solution to the strong CP problem if they couple to gluons \cite{Dine:2000cj,Hook:2018dlk}. They can be good candidates for non-thermal Dark Matter (DM) \cite{Preskill:1982cy,Abbott:1982af,Dine:1982ah} and can also be used  to explain the observed matter-antimatter asymmetry \cite{Jeong:2018jqe,Co:2019wyp}. Furthermore, ALPs provide a possibility to explain the anomalous magnetic dipole moment of muon \cite{Bauer:2019gfk,Cornella:2019uxs}. A significant region of the ALPs parameter space has already been constrained or will be accessible by future searches \cite{Bauer:2019gfk,Choi:2020rgn,Athron:2020maw,Han:2020dwo,Mimasu:2014nea,Bauer:2017ris,Aloni:2019ruo,Haghighat:2020nuh,Ebadi:2019gij,Inan:2020aal,ARGUSlimit,Baldini:2020okg,Aad:2019ugc,Calibbi:2020jvd,Iguro:2020rby,Endo:2020mev}. 

ALPs with flavor-violating couplings to the SM charged leptons can be produced through the LFV decays of the $\tau$ and $\mu$ leptons. If kinematically allowed and if relevant ALP couplings are non-zero, the produced ALPs can decay into the SM particles. The LFV ALPs parameter space in terms of the ALP mass and its LFV couplings to the SM charged leptons have been probed to some extent in recent decades \cite{Bauer:2019gfk,ARGUSlimit,Baldini:2020okg,Aad:2019ugc,Calibbi:2020jvd,Iguro:2020rby,Endo:2020mev}. In particular, the ALP-tau-electron and ALP-tau-muon couplings have been constrained for a wide range of ALP masses including the relatively light ALPs with masses $m_a\leq 1$ MeV \cite{Bauer:2019gfk,ARGUSlimit}. The light LFV ALP has attracted great interest particularly as its presence can be motivated by addressing the strong CP problem, Dark Matter, the SM flavor puzzle and neutrino masses in a broad class of models. The LFV QCD axion, the leptonic familon, the LFV axiflavon and the majoron are several examples of such well-motivated LFV ALP models which show how pseudo Nambu-Goldstone bosons associated with a global symmetry can naturally have flavor-violating couplings to leptons and how incorporating a global symmetry in the SM Lagrangian can address some of the long-lasting SM problems. The LFV QCD axion in DFSZ models \cite{Zhitnitsky:1980tq,Dine:1981rt} elegantly solves the strong CP problem and also makes a good DM candidate  \cite{Preskill:1982cy,Abbott:1982af,Dine:1982ah,diCortona:2015ldu}. The leptonic familon is the pseudo Nambu-Goldstone boson of a spontaneously broken $U(1)$ lepton flavor symmetry which provides an explanation for the hierarchies among the charged leptons using the Froggatt-Nielsen mechanism \cite{Froggatt:1978nt}. Unlike the QCD axion mass which stringently depends on the characteristic scale of global symmetry breaking, the mass of the leptonic familon is taken to be a free parameter and, consequently, the leptonic familon doesn't solve the strong CP problem. The strong CP problem and the SM flavor puzzle can be simultaneously addressed in a similar model with the LFV Axiflavon \cite{Calibbi:2016hwq,Ema:2016ops,Linster:2018avp}. In the LFV Axiflavon model, the PQ symmetry is identified with the $U(1)$ subgroup of the $U(2)$ flavor symmetry which explains the fermion mass hierarchies and mixings. Another example of a well-motivated LFV ALP model is a seesaw majoron model in which the majoron is the pseudo Nambu-Goldstone boson from the spontaneous breakdown of the lepton number \cite{Chikashige:1980ui, Schechter:1981cv}. In such a model, large LFV couplings can arise for the majoron and the neutrino masses are suppressed by a generalized lepton number. The light LFV ALP in any of the above-mentioned models can also make a good DM candidate. The mass range for which the ALP is viable in these models is model-dependent and varies with the values chosen for the free parameters of each model. However, based on the detailed study in Ref. \cite{Calibbi:2020jvd}, it is seen that for any mass value from sub-eV to MeV scale, at least one of the above-mentioned models provides a viable LFV ALP capable of addressing one or more SM problems. Motivated by such a rich theoretical foundation, we search for LFV ALPs in the mass range $m_a\leq 1$ MeV in this study.

Constraints on the ALP-tau-electron and ALP-tau-muon couplings have been derived from collider searches for the LFV tau decay $\tau\ra e/\mu+a$, where $a$ denotes the ALP \cite{ARGUSlimit}. Light ALPs with $m_a\leq 1$ MeV ($\approx 2m_e$) can only decay into a photon pair. The ALP coupling to photons is, however, strongly constrained \cite{Bauer:2017ris} and therefore, the majority of ALPs produced at a collider escape the detector before they decay. Searches for such ALPs produced in the LFV decays of the tau lepton at colliders are, therefore, based on the $e/\mu + \slashed{E}$ signature. The big challenge in these searches is to overcome the huge background from the SM tau decay $\tau\ra e/\mu+\nu \bar{\nu}$. In general, the differential decay rate of the tau lepton into an electron (or a muon) and an ALP depends on the polarization direction of the decaying tau lepton. This is also the case for the SM decay $\tau\ra e/\mu+\nu \bar{\nu}$. Comparing the properties of the polarized LFV and SM tau decays provides a way to discriminate the LFV signal from the SM background based on polarization-induced effects. Previous collider searches for LFV ALPs have not utilized such effects and are generally based on unpolarized decays. In this work, in addition to studying the unpolarized tau decays, we also consider the polarization-induced effects in polarized decays of the tau lepton and show that utilizing such effects can improve the sensitivity of the search significantly. We search for ALPs produced through the muon-anti muon annihilation into a tau pair followed by the LFV decay $\tau\ra e/\mu+a$. We study both the $e^\pm+ a$ and $\mu^\pm+ a$ final states of the tau decay in a model-independent fashion and compute the expected upper limits on the ALP-tau-muon and ALP-tau-electron couplings for ALP masses $m_a\leq 1$ MeV. Taking the effects from the tau polarization into account, the limits become sensitive to the chiral structure assumed for the LFV ALP coupling. We assume three different chiral structures, i.e. the ALP coupling to right-handed leptons (V+A), the ALP coupling to left-handed leptons (V-A) and the ALP with a non-chiral coupling to the SM leptons (V/A), and study them independently. Finally, comparing the obtained expected limits on the LFV couplings with present experimental limits, we will show that the analysis presented in this study is capable of improving the present experimental limits by roughly one order of magnitude. 

We assume a future muon collider suggested by the Muon Accelerator Program (MAP) \cite{muoncollider2,muoncollider1}. Muon colliders are not only capable of testing new physics but also offer considerable advantages over other proposed alternatives \cite{Ali:2021xlw,Yin:2020afe}. Searching for BSM signals at muon colliders is mostly motivated by the fact that they provide a relatively clean environment with less background compared with the hadron colliders, and that muon colliders are much more efficient than electron-positron colliders as muon beams produce very small synchrotron radiation compared with electron beams. Furthermore, unlike the hadron colliders, there is no ambiguity about energies of the colliding particles at a muon collider. 

This paper is structured as follows: In Section \ref{sec:lagrangian}, we provide the theoretical framework for LFV ALPs. In Section \ref{sec:productionprocess}, we discuss the ALP production process and its main SM background assuming a muon collider. In particular, in Section \ref{sec:mumutautau}, we discuss the main properties of the muon-anti muon annihilation into a tau pair process for the two cases of unpolarized and polarized muon beams, and in Sections \ref{sec:BSMtaudecay} and \ref{sec:SMtaudecay}, we focus on the LFV and SM tau decays $\tau\ra e/\mu+a$ and $\tau\ra e/\mu+\nu \bar{\nu}$. In Section \ref{sec:muoncollider}, we provide the basic information regarding the assumed future muon collider. In Section \ref{sec:eventgeneration}, SM background processes relevant to the ALP production process and the event generation method are discussed. Sections \ref{sec:eventselection} and \ref{sec:analysis} are respectively devoted to the employed event selection and event analysis methods, and Section \ref{sec:prospects} provides the prospects for upper limits on the ALP LFV couplings.

%***********************************************************
\section{Lepton-flavor-violating ALPs}  
\label{sec:lagrangian}
The spontaneously broken global $U\mathrm{(1)}_{\mathrm{PQ}}$ symmetry and the resulting QCD Axion were originally suggested by Peccei and Quinn and provide a way to address some of the SM problems \cite{Peccei:1977hh,DiLuzio:2020wdo}. In general, any model with a global $U\mathrm{(1)}$ symmetry which is spontaneously broken possesses a pseudo Nambu-Goldstone boson, called the ALP \cite{Brivio:2017ije}. ALPs exist in many extensions of the SM. They are scalar, odd under the CP transformation, and derivatively coupled to the SM fermions. Unlike the Axion mass, the mass of the ALP is not related to its couplings and hence the ALP is theoretically allowed to have a wide range of masses. The interaction of an ALP with photons and charged SM leptons is described by the effective Lagrangian \cite{Bauer:2019gfk,Calibbi:2020jvd}
\begin{align}
& \mathcal{L}_{\rm eff} = 
c_{\gamma\gamma} \frac{a}{f_a} F_{\mu\nu} \tilde{F}^{\mu\nu}
\, + \,
\sum_{i}\frac{\partial_\mu a}{2 f_a} \, \bar \ell_i c^A_{\ell_i \ell_i} \gamma^\mu\gamma_5 \ell_i  
\, + \,
\sum_{i\ne j}\frac{\partial_\mu a}{2 f_a} \, \bar \ell_i \gamma^\mu (c^V_{\ell_i \ell_j} + c^A_{\ell_i \ell_j} \gamma_5 ) \ell_j  \, , 
\label{lagrangian}
\end{align}
where $f_a$ is the scale associated with the breakdown of the global $U\mathrm{(1)}$ symmetry, $F_{\mu\nu}$ is the electromagnetic field strength tensor, $\tilde{F}^{\mu\nu}$ is the dual field strength tensor defined as $\tilde{F}^{\mu\nu} \equiv \frac{1}{2} \epsilon^{\mu \nu \rho \sigma} F_{\rho \sigma}$, with $\epsilon^{\mu \nu \rho \sigma} $ being the Levi-Civita symbol, $c_{\gamma\gamma}$ is a real constant (ALP-photon coupling), $c^{A}_{\ell_i \ell_i}$ is a real diagonal matrix, $c^{V,A}_{\ell_i \ell_j}$ are hermitian matrices in flavor space and $i,j$ run over the integers $1,2,3$ with $\ell_i=e,\mu,\tau$. In this study, we assume that the LFV couplings $c^{V,A}_{\ell_i \ell_j}$ are the only nonzero leptonic couplings of the ALP ($c^A_{\ell_i \ell_i}=0$). The charged LFV processes governed by the above Lagrangian can have different properties depending on the assumed chiral structure. We consider the three cases 
\begin{itemize} 
\item V+A: the ALP couples to right-handed leptons, i.e. $c_{\ell_i \ell_j}\equiv c^{V}_{\ell_i \ell_j}=c^A_{\ell_i \ell_j}$\,,
\item V$-$A: the ALP couples to left-handed leptons, i.e. $c_{\ell_i \ell_j}\equiv c^{V}_{\ell_i \ell_j}=-c^A_{\ell_i \ell_j}$\,,
\item V/A: either $c_{\ell_i \ell_j}\equiv c^{V}_{\ell_i \ell_j}\neq 0, c^{A}_{\ell_i \ell_j}=0$ or $c_{\ell_i \ell_j}\equiv c^{A}_{\ell_i \ell_j}\neq 0, c^{V}_{\ell_i \ell_j}=0$\,,
\end{itemize} 
and study each case independently. To ensure the validity of the effective Lagrangian, the symmetry breaking scale $f_a$ is required to be much larger than the typical energy scale of the processes under consideration. If kinematically allowed, LFV ALPs can be produced through LFV decays of the SM leptons, e.g. $\tau^\pm\rightarrow e^\pm\,a$ and $\tau^\pm\rightarrow\mu^\pm\,a$. In this study, the ALP mass $m_a$ is assumed to be in the range 100 eV to 1 MeV. Such decays are always possible in this mass range. 

The produced ALP can decay into visible SM particles if relevant ALP couplings are assumed to be non-zero and if the decay is kinematically allowed. With the assumed ALP mass range $100\ \mathrm{eV}\leq m_a \leq 1\ \mathrm{MeV}$, the ALP decay is kinematically restricted and the ALP can only decay into a pair of photons. Assuming a non-zero ALP-photon coupling $c_{\gamma\gamma}$, the width of the ALP decay into a pair of photons is given by \cite{Bauer:2017ris}
\begin{equation} 
\Gamma(a \rightarrow \gamma\gamma) = \frac{m_a^3}{4\pi} \Bigl(\frac{c_{\gamma\gamma}}{f_a}\Bigr)^2.
\label{widtha2gaga}
\end{equation}
The cubic dependence of the decay width on the ALP mass implies that the di-photon decay mode is substantially suppressed for light ALPs. Light ALPs, if produced at a collider, are therefore likely to escape the detector before they decay. Using Eq. \ref{widtha2gaga}, the ALP decay length $L_a$ is obtained to be 
\begin{equation} 
L_a = \gamma\beta\tau = \frac{|\vec{p}_a|}{m_a}\frac{1}{\Gamma(a\rightarrow\gamma\gamma)} = \frac{4\pi}{m_a^4}\Bigl(\frac{ f_a}{c_{\gamma\gamma}}\Bigr)^2 |\vec{p}_a|\,, 
\label{estimate1}
\end{equation} 
where $\gamma$ is the usual boost factor, $\tau$ is the ALP proper lifetime, $\beta$ is the ALP speed and $\vec{p}_a$ is the three-momentum of the ALP. The region $c_{\gamma\gamma}/f_a>10^{-8}$ TeV$^{-1}$ of the ALP parameter space has been experimentally excluded \cite{Bauer:2017ris} for the ALP mass range considered in this work. As a result, using Eq. \ref{estimate1} and assuming an ALP mass of 1 MeV, one finds 
\begin{equation} 
L_a > \left(\frac{2.3\,|\vec{p}_a|}{\mathrm{GeV}}\right) 10^{19} \,\mathrm{m}\, .
\label{estimate2} 
\end{equation}
$|\vec{p}_a|$ depends on the experimental setup. In a collider experiment, it varies with the center-of-mass energy, applied triggers, etc. Assuming that $|\vec{p}_a|$ is of the order of magnitude of 10 GeV or larger (which is typical at high energy colliders), one finds $L_a > 2.3\times10^{20}\,\mathrm{m}$. This lower limit implies that the ALPs travel a distance many orders of magnitude larger than the typical size of a detector ($\sim 10\,\mathrm{m}$) before they decay. Such ALPs manifest themselves as missing energy. As seen in Eq. \ref{estimate1}, the decay length is inversely related to $m_a^4$. Consequently, an ALP lighter than 1 MeV has larger decay length and therefore more tendency to escape the detector. Although the majority of the ALPs are invisible at the detector, one should consider the effect of the small fraction of ALPs which decay before leaving the detector. The probability that the ALP decays inside the detector $P_a^{\mathrm{\,det}}$ is given by 
\begin{equation} 
 P_a^{\mathrm{\,det}}=1-e^{-L_{\mathrm{det}}/L_a},
\label{adecaylengthprob}
\end{equation} 
where $L_{\mathrm{det}}$ is the distance of the calorimeter from the collision point. We take into account the effect of decaying ALPs in an event-by-event manner in this work.

%***********************************************************
\section{ALP production through $\mu^-\mu^+\ra \tau^-\tau^+$, $\tau\ra e/\mu+a$} 
\label{sec:productionprocess}
\begin{figure}[t]
\centering
    \begin{subfigure}[b]{0.36\textwidth} 
    \centering
    \includegraphics[width=\textwidth]{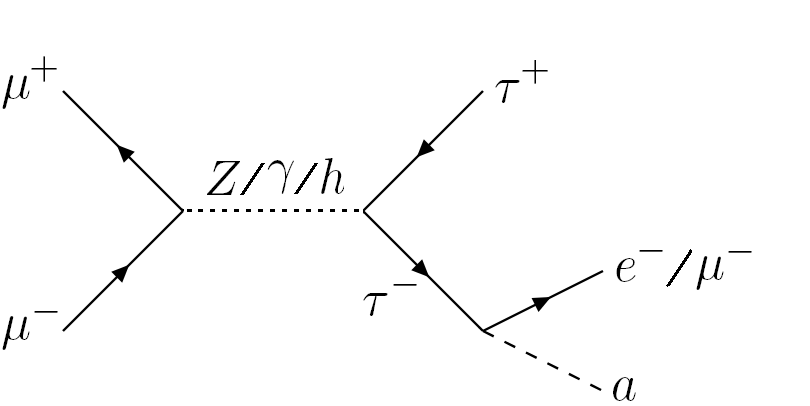}
    \end{subfigure} 
%\par\bigskip
    \begin{subfigure}[b]{0.36\textwidth} 
    \centering
    \includegraphics[width=\textwidth]{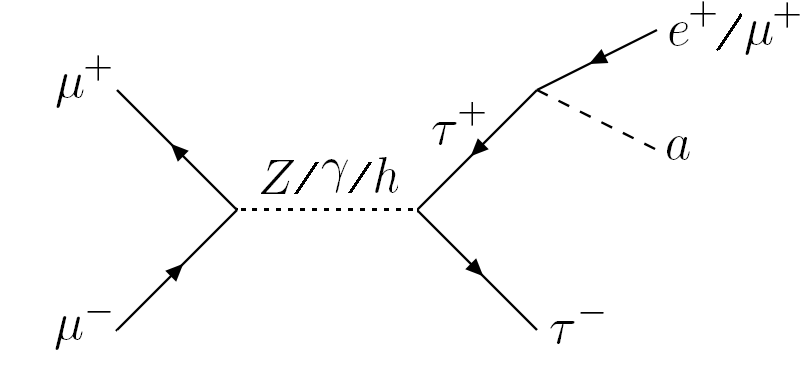}
    \end{subfigure}
  \caption{Feynman diagrams for the process $\mu^-\mu^+\ra \tau^-\tau^+$ followed by the LFV decay $\tau\ra e/\mu+a$.}
\label{feynmanDiagrams}
\end{figure}

LFV ALPs described by Eq. \ref{lagrangian} can be produced through the process $\mu^-\mu^+\ra \tau^-\tau^+$ with one of the tau leptons undergoing the decay $\tau\ra e/\mu+a$ and the other one undergoing a SM decay. Fig. \ref{feynmanDiagrams} shows Feynman diagrams contributing to this process. 
Among SM backgrounds relevant to this process, the $\tau^+\tau^-$ production followed by the leptonic decay of one of the tau leptons ($\tau\ra e/\mu+\nu \bar{\nu}$) is the most important one. In what follows in this section, we summarize main features of the assumed ALP production process and its dominant SM background, and then demonstrate that the kinematical variables of the final state lepton in the polarized tau decays $\tau\ra e/\mu+a$ and $\tau\ra e/\mu+\nu \bar{\nu}$ can be utilized to reduce the $\tau^+\tau^-$ background. 

%***********************************************************
\subsection{Tau pair production through $\mu^-\mu^+\ra \tau^-\tau^+$} 
\label{sec:mumutautau}
\begin{figure}[!t]
\centering
    \begin{subfigure}[b]{0.55\textwidth} 
    \centering
    \includegraphics[width=\textwidth]{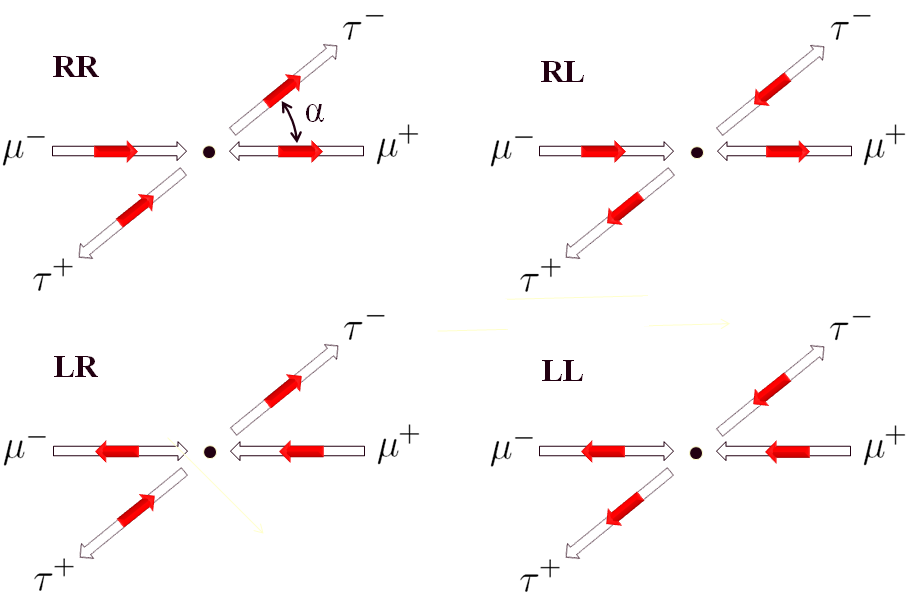}
    \end{subfigure} 
  \caption{Helicity combinations contributing to the process $\mu^-\mu^+\ra \tau^-\tau^+$ in the limit $m_\tau\ll E_\tau$. Hollow (red) arrows show the direction of the momentum (spin) of particles. RR, RL, LR and LL denote the helicity of the incoming $\mu^-$ and the outgoing $\tau^-$.}
\label{helicityCombinations}
\end{figure}

The process of muon-antimuon annihilation into a tau pair proceeds through a $Z/\gamma/h$ $s$-channel mediator as seen in Fig. \ref{feynmanDiagrams}. How much each of the mediators contributes to the total cross section varies with the center-of-mass energy and the polarizations of the colliding muon beams. Each of the initial and final state leptons in this process can either be in a left- or right-handed helicity state, leading to 16 distinct helicity combinations. In general, all of the helicity combinations contribute to the process. However, in the ultra-relativistic limit, where the energy of the particles is much larger than their mass ($m_\tau\ll E_\tau$), the chiral and helicity eigenstates become identical and therefore, only four helicity combinations have non-zero contributions (other combinations are substantially suppressed by a factor of $m_\tau/E_\tau$). This is a consequence of the fact that only certain combinations of chiral eigenstates contribute to the interaction (this statement holds always, not only in the ultra-relativistic limit). Allowed helicity combinations in the ultra-relativistic limit are provided in Fig. \ref{helicityCombinations}. As seen, the outgoing tau leptons have opposite helicities.

Parity conservation is violated in the production and decay of the $Z$ boson leading to asymmetries in the final state tau leptons. Asymmetry properties of the tau leptons can be understood using the asymmetry observables, i.e. forward-backward asymmetry $A_{\mathrm{FB}}$, polarization asymmetry $P_\tau$, forward (backward) polarization asymmetry $P_\tau^{\mathrm{F}}$ ($P_\tau^{\mathrm{B}}$ ) and forward-backward polarization asymmetry $A_{\mathrm{FB}}^{\mathrm{P}_\tau}$. Forward-backward asymmetry is defined as 
\begin{equation}
A_{\mathrm{FB}}=\frac{\sigma^{\mathrm{F}}-\sigma^{\mathrm{B}}}{\sigma_{\mathrm{total}}} \, ,
\label{eq:afb}
\end{equation}
where the total cross section ${\sigma_{\mathrm{total}}}={\sigma^{\mathrm{F}}+\sigma^{\mathrm{B}}}$, and $\sigma^{\mathrm{F}}$ and $\sigma^{\mathrm{B}}$ are given by
\begin{eqnarray}
\begin{aligned}
\sigma^{\mathrm{F}}=\sigma(h_{\tau^-}=+1,\cos\alpha>0)+\sigma(h_{\tau^-}=-1,\cos\alpha>0)\, , 
\\
\sigma^{\mathrm{B}}=\sigma(h_{\tau^-}=+1,\cos\alpha<0)+\sigma(h_{\tau^-}=-1,\cos\alpha<0)\, ,
\label{eq:sigmafb}
\end{aligned}
\end{eqnarray}
with $h_{\tau^-}$ being the helicity of the $\tau^-$ and $\alpha$ the angle between the momentum of the outgoing $\tau^-$ and the momentum of the incoming $\mu^-$ (see Fig. \ref{helicityCombinations}). The polarization asymmetry for the $\tau^-$ and $\tau^+$ is defined as 
\begin{eqnarray}
\begin{aligned}
P_{\tau^-}=\frac{(\sigma_{++}+\sigma_{+-})-(\sigma_{-+}+\sigma_{--})}{\sigma_{\mathrm{total}}}\, , 
\\
P_{\tau^+}=\frac{(\sigma_{++}+\sigma_{-+})-(\sigma_{+-}+\sigma_{--})}{\sigma_{\mathrm{total}}}\, , 
\label{eq:ptau+-}
\end{aligned}
\end{eqnarray}
where $\sigma_{++}$, $\sigma_{+-}$, $\sigma_{-+}$ and $\sigma_{--}$ are the cross sections corresponding to the four allowed helicity combinations of the tau pair denoted as
\begin{equation}
h_{\tau-}h_{\tau+}=++, +-, -+, -- \, .
\label{eq:taupairhelicityCombinations}
\end{equation}
In the ultra-relativistic limit, $\sigma_{++}$ and $\sigma_{--}$ are negligible and it follows from Eq. \ref{eq:ptau+-} that
\begin{equation}
P_{\tau-}=-P_{\tau+} \, .
\end{equation}
The $\tau^-$ polarization asymmetry will hereafter be referred to as polarization asymmetry $P_\tau\equiv P_{\tau^-}$. The forward and backward polarization asymmetries are defined as
\begin{eqnarray}
\begin{aligned}
P_\tau^{\mathrm{F}}=\frac{\sigma^{\mathrm{F}}(h_\tau=+1)-\sigma^{\mathrm{F}}(h_\tau=-1)}{\sigma^{\mathrm{F}}}\, , 
\\
P_\tau^{\mathrm{B}}=\frac{\sigma^{\mathrm{B}}(h_\tau=+1)-\sigma^{\mathrm{B}}(h_\tau=-1)}{\sigma^{\mathrm{B}}}\, ,
\label{eq:ptaufb}
\end{aligned}
\end{eqnarray}
and finally, the forward-backward polarization asymmetry is given by
\begin{equation}
A_{\mathrm{FB}}^{\mathrm{P}_\tau}=\frac{[\sigma^{\mathrm{F}}(h_\tau=+1)-\sigma^{\mathrm{F}}(h_\tau=-1)] - [\sigma^{\mathrm{B}}(h_\tau=+1)-\sigma^{\mathrm{B}}(h_\tau=-1)]}{\sigma_{\mathrm{total}}}\, .
\label{eq:aptaufb}
\end{equation}

\begin{figure}[t]
  \centering
  \includegraphics[width=1\textwidth]{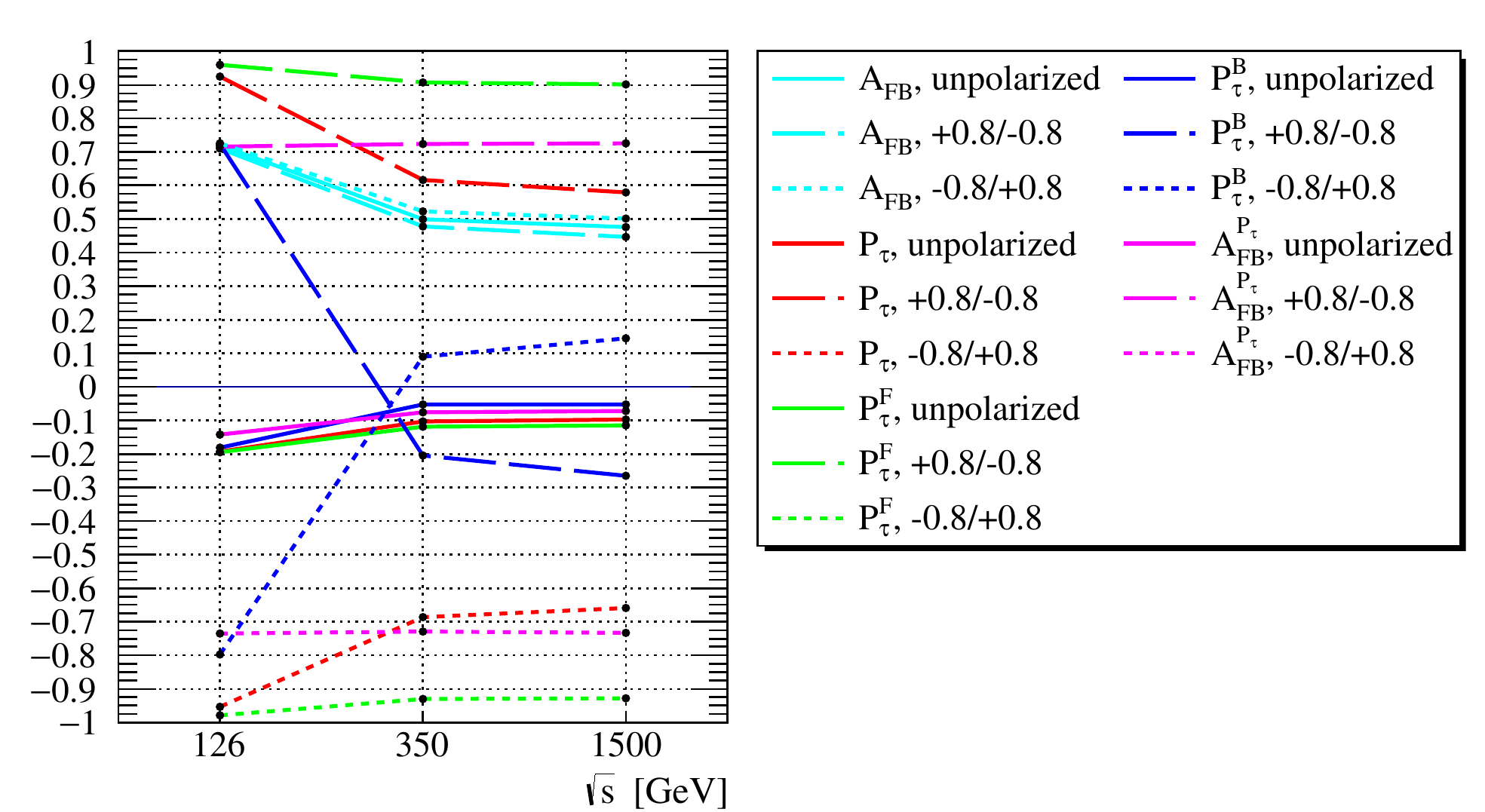}
  \caption{Asymmetry variables of the tau leptons produced in the process $\mu^-\mu^+\ra \tau^-\tau^+$ assuming the center-of-mass energies $\sqrt{s}=126, 350, 1500$ GeV. Solid lines show the variables assuming unpolarized muon beams, dashed lines show the variables assuming that the $\mu^-$ ($\mu^+$) beam is $+0.8$ ($-0.8$) polarized and dotted lines show the variables assuming the $\mu^-$ ($\mu^+$) beam to be $-0.8$ ($+0.8$) polarized. Different colors correspond to different asymmetry variables.}
\label{tauAsymmetries}
\end{figure}

Useful information can be extracted from the asymmetry variables introduced above. Simulating the $\mu^-\mu^+\ra \tau^-\tau^+$ process with the use of \texttt{MadGraph5\_aMC@NLO} \cite{Alwall:2011uj}, the results provided in Fig. \ref{tauAsymmetries} are obtained for the variables defined through Eqs. \ref{eq:afb}-\ref{eq:aptaufb}.
The obtained results correspond to the center-of-mass energies of 126, 350 and 1500 GeV. These center-of-mass energies have been proposed for a future muon collider as discussed in more details in Section \ref{sec:muoncollider}. Asymmetry variables have been computed for three different cases of the polarization of the initial muon beams, i.e. unpolarized beams, longitudinally polarized beams with $+0.8$ ($-0.8$) polarization for the $\mu^-$ ($\mu^+$) beam and longitudinally polarized beams with $-0.8$ ($+0.8$) polarization for the $\mu^-$ ($\mu^+$) beam (assuming that the beam polarization varies in the range $-1$ to $+1$). Positive and negative polarization signs denote the right- and left-handed helicity states, respectively. As seen, the forward-backward asymmetry $A_{\mathrm{FB}}$ changes very slightly with the polarization of the muon beams and varies in a range roughly between 0.45 and 0.7 for different center-of-mass energies. The majority of the $\tau^-$ ($\tau^+$) leptons are therefore emitted in the forward (backward) direction with respect to the momentum direction of the initial $\mu^-$ beam. This is an important feature, especially when the polarization asymmetry properties are considered. The polarization asymmetry variables $P_\tau$, $P_\tau^{\mathrm{F}}$, $P_\tau^{\mathrm{B}}$ and $A_{\mathrm{FB}}^{\mathrm{P}_\tau}$ are sensitive to the polarization of the initial muon beams. Let's concentrate on the results obtained for the unpolarized muon beams case and the case with $\mu^-$ ($\mu^+$) beam being $+0.8$ ($-0.8$) polarized. As seen, the polarization asymmetry $P_\tau$ is always negative in the unpolarized case, while it is always positive with values roughly between $+0.55$ and $+0.95$ in the polarized muon beams case. The $\tau^-$ ($\tau^+$) leptons are therefore $+0.55$ to $+0.95$ ($-0.55$ to $-0.95$) polarized on average in the polarized case. $P_\tau$ shows the average behavior of the polarization and doesn't depend on the momentum direction of the outgoing tau leptons. However, as the majority of the $\tau^-$ ($\tau^+$) leptons are forwardly (backwardly) emitted, we are more interested in the forward polarization asymmetry $P_\tau^{\mathrm{F}}$ which shows the behavior of the polarization of the $\tau^-$ ($\tau^+$) leptons emitted forwardly (backwardly). As seen, $P_\tau^{\mathrm{F}}$ is negative ($>-0.2$) in the unpolarized case, while it is positive with values roughly in the range $+0.90$ to $+0.98$ with polarized muon beams. This means that the $\tau^-$ ($\tau^+$) leptons emitted forwardly (backwardly), which constitute the majority of the produced $\tau^-$ ($\tau^+$) leptons, are highly polarized with a polarization roughly between $+0.90$ and $+0.98$ ($-0.90$ and $-0.98$) in the polarized muon beams case. This is an important feature of the $\mu^-\mu^+\ra \tau^-\tau^+$ process at the considered center-of-mass energies because it provides a way to produce highly polarized tau leptons. As it will be discussed in the rest of this section, kinematical variables of the tau decay (in both the $\tau\ra e/\mu+a$ and $\tau\ra e/\mu+\nu \bar{\nu}$ decay modes) depend on the tau polarization and can be utilized to improve the search for the new physics decay $\tau\ra e/\mu+a$. The mean polarization of the backwardly (forwardly) emitted $\tau^-$ ($\tau^+$) is given by the backward polarization asymmetry $P_\tau^{\mathrm{B}}$. As seen, $P_\tau^{\mathrm{B}}$ is always negative ($>-0.2$) for unpolarized muon beams. In the case of polarized beams, it is $\approx +0.7$ at the center-of-mass energy of 126 GeV and roughly in the range $-0.3$ to $-0.2$ at the center-of-mass energies of 350 and 1500 GeV. The backwardly (forwardly) emitted $\tau^-$ ($\tau^+$) leptons are not highly polarized at the center-of-mass energies of 350 and 1500 GeV and are therefore of no interest if one aims at utilizing the tau polarization to discriminate between the new physics and the SM decays of the tau lepton. This is, however, not the case at the center-of-mass energy of 126 GeV where the mean polarization of the backwardly (forwardly) emitted $\tau^-$ ($\tau^+$) is $\approx +0.7$ ($-0.7$). At this center-of-mass energy, the polarization asymmetry $P_\tau>+0.9$ and is much higher than the polarization asymmetry at other center-of-mass energies.

Dotted lines in Fig. \ref{tauAsymmetries} show that highly polarized tau leptons with a similar degree of polarization can also be produced assuming reverse polarizations for the initial muon beams, i.e. $-0.8$ ($+0.8$) polarization for $\mu^-$ ($\mu^+$) beam. As will be discussed, both cases considered for the muon beams polarization are expected to be equally useful in the search for LFV decays of the tau lepton. Comparing asymmetry variables obtained at different center-of-mass energies, it is seen that, going from $\sqrt{s}=350$ GeV to $\sqrt{s}=1500$ GeV, the asymmetry variables experience slight changes. This is the case for both the unpolarized and polarized muon beams cases. The asymmetry variables obtained at $\sqrt{s}=126$ GeV are, however, substantially different from the asymmetry variables at 350 and 1500 GeV center-of-mass energies.

%***********************************************************
\subsection{Lepton-flavor-violating $\tau\ra e/\mu+a$ decay} 
\label{sec:BSMtaudecay}
The products of the two-body decay $\tau\ra \ell\, a$ ($\ell=e,\mu$) are monoenergetic in the $\tau$ rest frame with 
\begin{equation}
p_{\ell}= \sqrt{ \left(\frac{m_\tau^2+m_{\ell}^2-m_a^2}{2 m_\tau}\right)^2-m_{\ell}^2}\, ,
\label{eq:mono}
\end{equation}
where $p_{\ell}$ represents the momentum of the final state lepton. If $m_{a}\ll m_\tau$, one finds $E_{\ell}\simeq m_\tau/2$, where $E_{\ell}$ denotes the energy of the final state lepton. As the decay products appear in a back-to-back configuration, the ALP has the same momentum as the lepton $\ell$. Neglecting the mass of the final state lepton, the corresponding rest frame differential decay width is given by \cite{Calibbi:2020jvd}
\begin{align}
\label{eq:decayBSM}
\frac{\text{d} \Gamma(\tau^\pm  \to \ell^\pm  \, a)}{\text{d}\cos\theta} = \frac{m_{\tau}^3}{128 \pi f_a^2 } \left( 1 - \frac{m_a^2}{m_{\tau}^2} \right)^2 \left({  \vert c_{\tau  \ell }^V\vert^2 + \vert c_{\tau  \ell }^A\vert^2 } \mp 2 \mathcal{P}_{\tau} \cos \theta \, {\text{Re}(c_{\tau  \ell }^V c_{\tau  \ell }^{A*})} \right) \, , 
\end{align}
where $\mathcal{P}_{\tau}$ (defined in the range 0-1) is the degree of polarization of tau leptons and $\theta$ is the angle between the momentum of the final state lepton and the polarization vector of the decaying tau in the tau rest frame. According to Eq. \ref{eq:decayBSM}, the three assumed chiral structures for the LFV coupling of the ALP (see Section \ref{sec:lagrangian}) lead to the distinct angular distributions
\begin{align}
\frac{\text{d} \Gamma(\tau^\pm  \to \ell^\pm  \, a)}{\text{d}\cos\theta} = \frac{m_{\tau}^3}{128 \pi f_a^2 } \left( 1 - \frac{m_a^2}{m_{\tau}^2} \right)^2 \times\left\{
                \begin{array}{ll}
\medskip
                  2c_{\tau  \ell }^2\left(1 \mp  \mathcal{P}_{\tau} \cos \theta \, \right) \,\,\,\,\,\,\, \mathrm{V\!+\!A}  \\ \medskip
                  2c_{\tau  \ell }^2\left(1 \pm  \mathcal{P}_{\tau} \cos \theta \, \right) \,\,\,\,\,\,\, \mathrm{V\!-\!A}  \\
                 c_{\tau  \ell }^2 \ \ \ \ \ \ \ \ \ \ \ \ \ \ \ \ \ \ \ \ \ \ \ \ \mathrm{V/A}  \\
                \end{array}
              \right. .
\label{eq:decaychiral}
\end{align}
Eq. \ref{eq:decaychiral} shows some qualitative features which we summarize in what follows. In the first two cases, the angular distribution of the final state lepton depends on the polarization of the decaying tau lepton. In the V+A case, the angular distribution of the final state lepton coming from the $\tau^\pm$ decay is proportional to $1 \mp  \mathcal{P}_{\tau} \cos \theta$. This implies that the positively charged $\ell^+$ final state lepton coming from the $\tau^+$ decay is more likely to be emitted in the backward direction relative to the tau lepton polarization. On the other hand, the negatively charged $\ell^-$ lepton is more likely to be emitted in the forward direction. This angular orientation is exactly reversed for the V$-$A case since the respective angular distributions in the V+A and V$-$A cases only differ in the sign of $\theta$ dependent term. To be more precise, in the V$-$A case, the $\ell^+$ lepton from the $\tau^+$ decay is more likely to be emitted in the forward direction, and the $\ell^-$ lepton is more likely to be emitted in the backward direction. In the third case, the two choices $c^{V}_{\tau \ell}\neq0 \ \mathrm{and} \ c^A_{\tau \ell}\neq0$ lead to the same differential decay width. Furthermore, the angular distribution of the final state lepton is isotropic and does not depend on the polarization of the decaying tau lepton. 

\begin{figure}[t]
  \centering
  \includegraphics[width=0.6\textwidth]{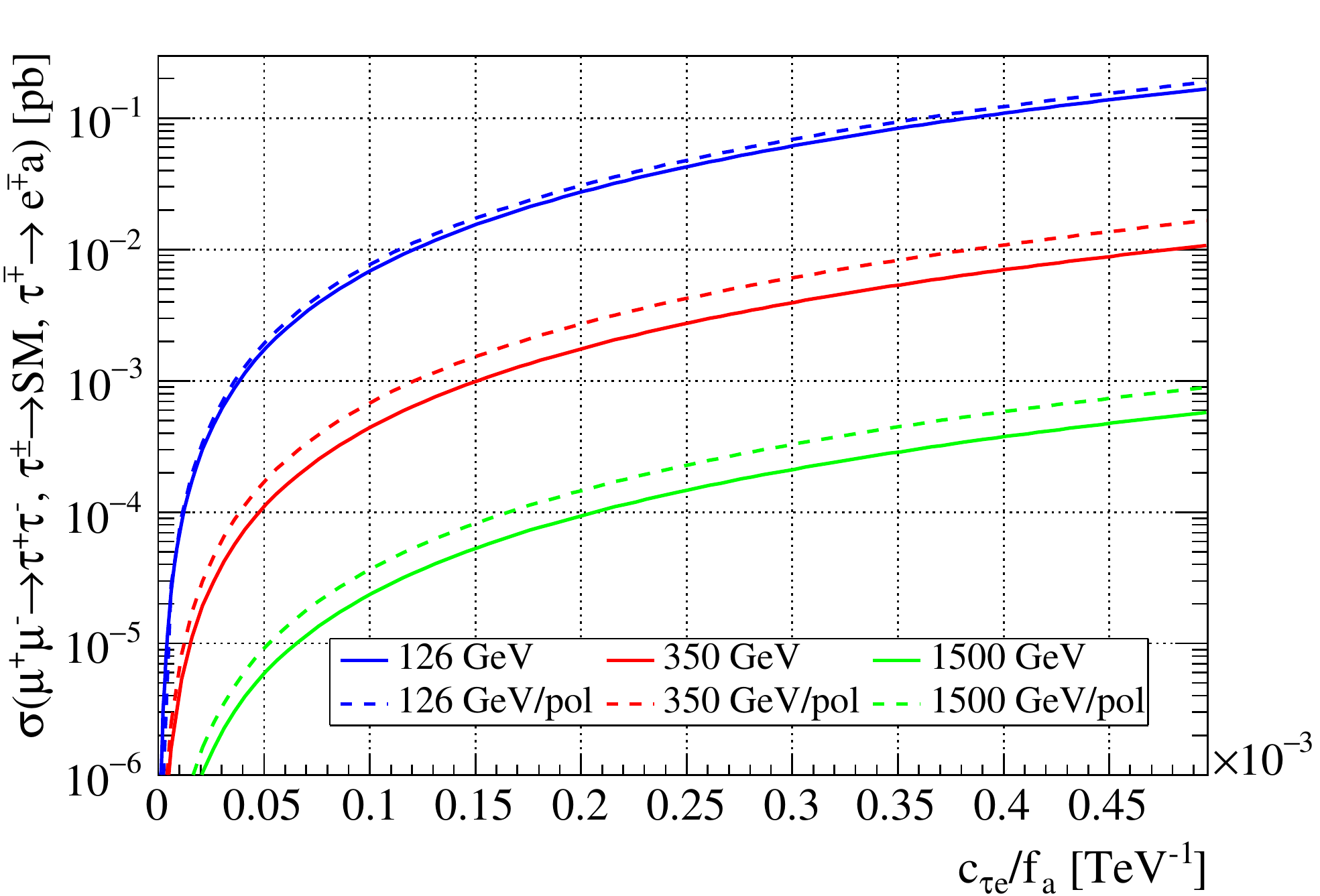}
  \caption{Cross section of the ALP production through the process $\mu^-\mu^+\ra \tau^-\tau^+$ followed by the decays $\tau^\pm\ra \mathrm{SM}$ and $\tau^\mp\ra e^\pm\,a$ as a function of the LFV coupling $c_{\tau  e}/f_a$. The ALP couples to right-handed leptons and the mass of ALP is assumed to be 1 MeV. Solid (dashed) lines show the cross section assuming unpolarized (polarized) muon beams at the center-of-mass energies 126 GeV (blue), 350 GeV (red) and 1500 GeV (green). In the case of polarized muon beams, the $\mu^-$ ($\mu^+$) beam is assumed to be $+0.8$ ($-0.8$) polarized. The plot for the cross section of the ALP production through the decay mode $\tau\ra \mu\,a$ is similar and is not displayed for brevity.}
\label{xsec}
\end{figure}

Fig. \ref{xsec} shows the cross section of the ALP production through the process $\mu^-\mu^+\ra \tau^-\tau^+$ with the subsequent decays $\tau^\pm\ra \mathrm{SM}$ and $\tau^\mp\ra e^\pm\,a$ against $c_{\tau  e}/f_a$. Cross sections have been obtained using \texttt{MadGraph5\_aMC@NLO} (see Section \ref{sec:eventgeneration} for further details). Different center-of-mass energies and the two cases of unpolarized and polarized muon beams are assumed. Moreover, it is assumed that the ALP couples to right-handed leptons.
As seen, the cross section decreases at higher center-of-mass energies which is expected because of the nature of $s$-channel processes. Furthermore, at the same center-of-mass energy, the cross section is higher when the muon beams are polarized which is also expected because of the spin-1 vector boson mediators involved in this production process. 

Only a fraction of tau leptons produced at a collider decays through the LFV decay mode $\tau\to\ell \, a$ in the detector. Integrating Eq. \ref{eq:decayBSM} with respect to $\theta$, the total decay width is obtained to be
\begin{align}
{ \Gamma(\tau  \to \ell  \, a)} = \frac{m_{\tau}^3}{64 \pi f_a^2 } \left( 1 - \frac{m_a^2}{m_{\tau}^2} \right)^2 \left({  \vert c_{\tau  \ell }^V\vert^2 + \vert c_{\tau  \ell }^A\vert^2 } \right) \, .
\label{eq:decayTotal}
\end{align}
Using Eq. \ref{eq:decayTotal}, the corresponding decay length of the tau lepton is given by 
\begin{equation} 
L_\tau = \gamma\beta\tau = \frac{|\vec{p}_\tau|}{m_\tau}\frac{1}{ \Gamma(\tau\to\ell \, a)} = \frac{ {64 \pi } }{\left( {m_{\tau}^2} - {m_a^2} \right)^2}\frac{f_a^2 }{ \left({  \vert c_{\tau  \ell }^V\vert^2 + \vert c_{\tau  \ell }^A\vert^2 } \right)} |\vec{p}_\tau|\,.
\label{taudecaylength}
\end{equation} 
The probability that the tau lepton decays through the LFV decay mode in the detector can then be obtained using $P_\tau^{\mathrm{\,det}}=1-e^{-L_{\mathrm{det}}/L_\tau}$. The decay probability depends on the mass and LFV couplings of the ALP and also on the tau momentum. In this study, we take into account the macroscopic decay length of the tau lepton event-by-event with the use of the decay probability $P_\tau^{\mathrm{\,det}}$. 

%***********************************************************
\subsection{Standard Model $\tau\ra e/\mu+\nu \bar{\nu}$ decay} 
\label{sec:SMtaudecay}
Tau leptons decay both hadronically and leptonically in the SM through an off-shell $W$ boson. In the $\tau$ rest frame, the final state $\ell^\pm$ lepton from the SM decays $\tau^+\to \ell^+\,\nu_\ell\,\bar{\nu}_\tau$ and $ \tau^- \to \ell^- \,\bar{\nu_\ell}\,\nu_\tau$ ($\ell=e,\mu$) obeys the Michel spectrum \cite{Tsai:1971vv}
\begin{equation}
\frac{\text{d}^2\Gamma\left(
     \begin{array}{ll}
     \tau^+\to \ell^+\,\nu_\ell\,\bar{\nu}_\tau  \\
     \tau^- \to \ell^- \,\bar{\nu_\ell}\,\nu_\tau
     \end{array}
\right)}{\text{d}p_\ell\, \text{d}\cos\theta} = \Gamma_{\tau} \frac{8\, p_\ell^2}{m_\tau^4} \left[3m_\tau-4E_\ell + \frac{3m_\ell^2}{m_\tau} - \frac{2m_\ell^2}{E_\ell} \pm \mathcal{P}_\tau \frac{p_\ell}{E_\ell}\left(4E_\ell - m_\tau - \frac{3m_\ell^2}{m_\tau} \right)\cos\theta\right]\, , 
\label{eq:decaySM}
\end{equation}
where $E_\ell$ and $p_\ell$ are the energy and the momentum of the final state lepton $\ell$, $\theta$ is the angle between the momentum of the final lepton $\ell$ and the polarization vector of the tau lepton, and $\Gamma_{\tau} = G_F^2 m_\tau^5/192\pi^3$ is the total leptonic decay width of the tau lepton when the mass of the final lepton $\ell$ is ignored. Neglecting the mass of the lepton $\ell$, Eq. \ref{eq:decaySM} can be simplified into
\begin{equation}
\frac{\text{d}^2\Gamma\left(
     \begin{array}{ll}
     \tau^+\to \ell^+\,\nu_\ell\,\bar{\nu}_\tau  \\
     \tau^- \to \ell^- \,\bar{\nu_\ell}\,\nu_\tau
     \end{array}
\right)}{\text{d}x_\ell\, \text{d}\cos\theta}\simeq \Gamma_{\tau}\big[\left(3-2x_\ell\right) \pm \mathcal{P}_\tau(2x_\ell-1)\cos\theta\big]x_\ell^2 \, , 
\label{eq:decaySMsimplified}
\end{equation}
where $x_\ell$ ($0\leq x_\ell \leq 1$) denotes the lepton $\ell$ energy fraction defined as $x_{\ell}=2E_{\ell}/m_\tau$. 

\begin{figure}[t]
  \centering
  \includegraphics[width=0.5\textwidth]{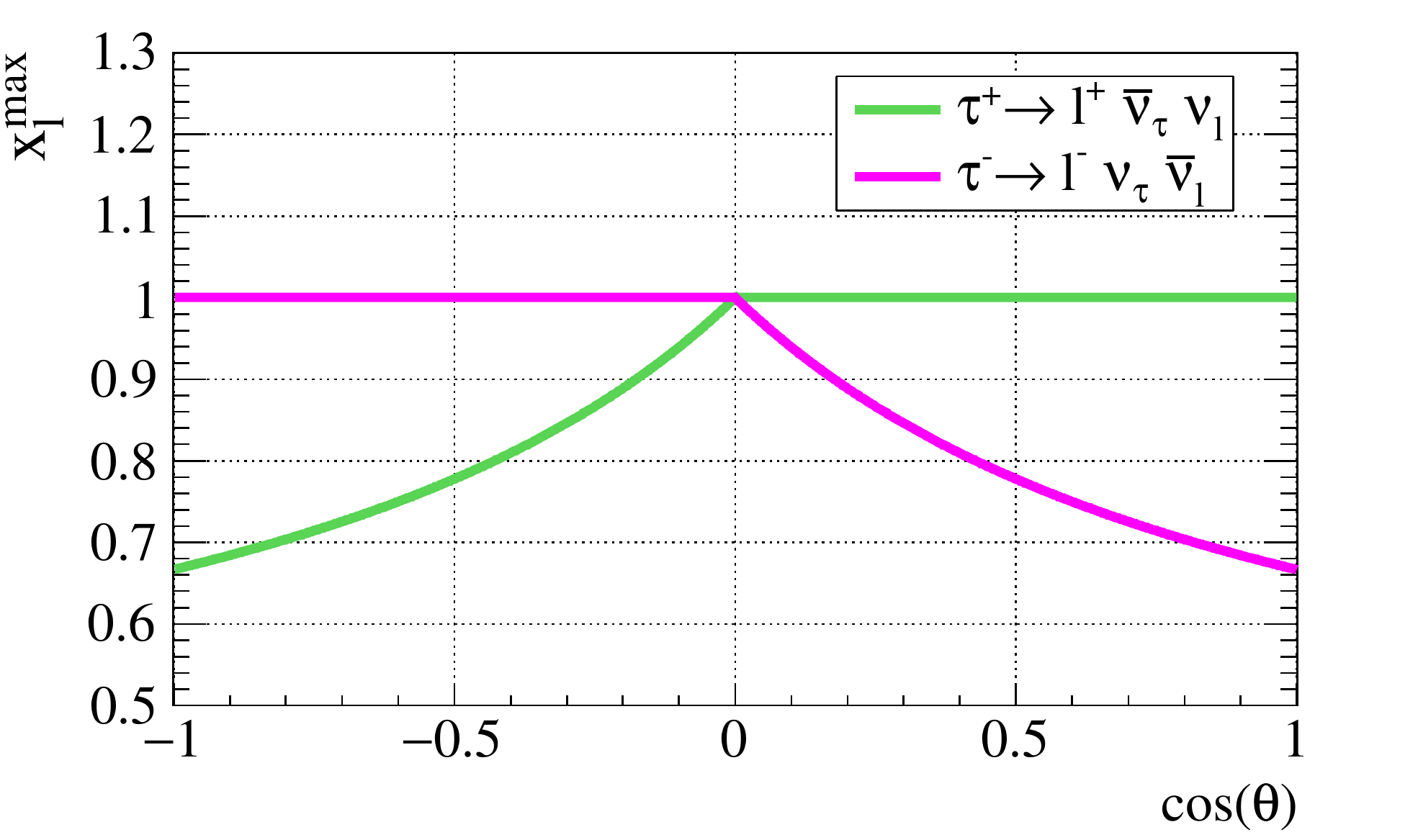}
  \caption{Position of the maximum of the Michel spectrum $x_\ell^{\mathrm{max}}$ as a function of $\cos\theta$, where $\theta$ is the angle between the momentum of the final state charged lepton and the polarization vector of the decaying tau assuming $\mathcal{P}_\tau=1$. The green (pink) line shows $x_\ell^{\mathrm{max}}$ for $\tau^+$ ($\tau^-$) decay.}
\label{xmax}
\end{figure}

Some important features can be understood from Eq. \ref{eq:decaySMsimplified}. For the unpolarized tau leptons ($\mathcal{P}_\tau=0$), angular dependence of the differential decay width vanishes leading to isotropic distribution of the final lepton $\ell$. In the case of polarized tau leptons, the final lepton $\ell^+$ ($\ell^-$) is more likely to be emitted in the forward (backward) direction relative to the polarization vector of the decaying tau. Using Eq. \ref{eq:decaySMsimplified}, one finds
\begin{equation}
x^{\text{max}}_{\ell^\pm}=\frac{3 \mp \mathcal{P}_\tau \cos{\theta}}{3(1 \mp  \mathcal{P}_\tau  \cos{\theta})}\ ,
\label{eq:xmax}
\end{equation}
where $x^{\text{max}}_{\ell^\pm}$ is the position of the maximum of the Michel spectrum in the $\tau^\pm$ decay. In this equation, $\cos\theta$ varies in the range $-1<\cos\theta<0$ ($0<\cos\theta<1$) for $\tau^+$ ($\tau^-$) decay. For the range $0<\cos\theta<1$ ($-1<\cos\theta<0$), $x^{\text{max}}_{\ell}=1$ holds for the decay of $\tau^+$ ($\tau^-$). Eq. \ref{eq:xmax} implies that for unpolarized tau leptons, the outgoing charged lepton peaks at $x_\ell=1$ regardless of the angle $\theta$, which coincides with the final state lepton in the decay $\tau^\pm  \to \ell^\pm a$ (see Section \ref{sec:BSMtaudecay}). In the case of polarized tau leptons, however, the angular dependence is preserved. Fig. \ref{xmax} shows the position of the maximum of the Michel spectrum against $\cos\theta$ for $\tau^-$ and $\tau^+$ decays.
As seen, in the $\tau^-$ ($\tau^+$) decay, $x^{\text{max}}_{\ell}$ decreases as $\cos\theta$ approaches 1 ($-1$) and becomes minimum when $\cos\theta$ is 1 ($-1$). In other words, final leptons $\ell^-$ ($\ell^+$) emitted in a region close to $\cos\theta= 1$ ($-1$) are less energetic than those emitted in other directions. Such leptons are mostly separated from the leptons resulting from the LFV $\tau^\pm  \to \ell^\pm  \, a$ decay which peak at $x_{\ell}\simeq 1$. This feature can be utilized to search for $\tau^\pm  \to \ell^\pm  \, a$ decays if polarized tau leptons with a high enough degree of polarization can be produced. Luckily, this is the case for the production process $\mu^-\mu^+\ra\tau^-\tau^+$ when the initial muon beams are polarized as discussed in detail in Section \ref{sec:mumutautau}. Highly polarized tau leptons can be produced by colliding polarized muon beams at a future muon collider. For events with the final state lepton $\ell^-$ ($\ell^+$) being emitted close to the direction (opposite direction) of the $\tau$ polarization vector, the background due to the SM tau decay can be reduced with the help of the energy spectrum of the final lepton. 

As discussed in Section \ref{sec:BSMtaudecay}, when the ALP couples to right-handed leptons, the final $\ell^-$ ($\ell^+$) lepton resulting from the LFV decay $\tau^\pm  \to \ell^\pm  \, a$ is more likely to be emitted in the forward (backward) direction. This angular behavior is exactly opposite to that of the final charged lepton in the SM tau decay leading to higher sensitivities for right-handed ALP couplings compared with the left-handed couplings. When the ALP couples to left-handed leptons, the angular distribution of the final lepton in the LFV tau decay is similar to that of the final charged lepton in the SM decay of the tau. As a consequence, the sensitivity is somewhat degraded. The sensitivity in the case of isotropic ALPs (V/A case) lies somewhere in between the left- and right-handed cases.

%***********************************************************
\subsection{Future muon collider}
\label{sec:muoncollider}
Synchrotron radiation from muon beams is suppressed by a factor of $10^9$ in comparison with electron beams of the same energy because of the large muon mass ($m_\mu / m_e \approx 207$). Muon beams can therefore be accelerated and brought into collision in a circular collider. In comparison with the hadron colliders where only a fraction of hadron energy is carried by the colliding partons, muon colliders are more efficient with a reasonable power consumption as all of the energy is carried by the colliding muons. As a consequence, the effective energy reach provided by a 14 TeV muon collider is similar to that of the 100 TeV FCC \cite{muoncollider2,Hinchliffe:2015qma}. Furthermore, muon colliders provide a much cleaner environment with less background compared with hadron colliders.

Until now, a great deal of time and effort has been dedicated to preparing a feasible design for a future muon collider \cite{muoncollider2,muoncollider1,muonColliders,Delahaye:2013jla,Ankenbrandt:1999cta,Gallardo:1996aa,Neuffer:1994bt,Skrinsky:1981ht}. Although no complete conceptual design has been reported yet, a finalized conceptual design report of a future muon collider is planned to be ready in the current decade. Muon Accelerator Program has developed a conceptual design for a potential muon collider \cite{muoncollider2,muoncollider1}. This muon collider will operate at different phases including the Higgs/top factory and multi-TeV phases. Tab. \ref{tab:muoncollider} presents the center-of-mass energy and the luminosity of different phases of this collider.
\begin{table}[!t]
\normalsize
    \begin{center}
         \begin{tabular}{cccccc}
& Higgs & Top-High Luminosity & & Multi-TeV &\\ \Xhline{1\arrayrulewidth}
Center-of-mass energy [GeV] & 126 & 350 & 1500 & 3000 & 6000 \\ 
Average luminosity [$10^{34} \mathrm{cm}^{-2} \mathrm{s}^{-1}$] & 0.008 & 0.6 & 1.25 & 4.4 & 12  \\
  \end{tabular}
\caption{The center-of-mass energy and the average luminosity of different phases of a future muon collider as proposed by the Muon Accelerator Program taken from \cite{muoncollider1}.}
\label{tab:muoncollider}
  \end{center}
\end{table}
In this work, considering the first three phases, i.e. the Higgs factory ($\sqrt{s}=126$ GeV), the top-high luminosity ($\sqrt{s}=350$ GeV) and the first Multi-TeV phase ($\sqrt{s}=1500$ GeV), we study the capability of this future muon collider to search for the LFV decay of the tau lepton.

%***********************************************************
\section{SM backgrounds and event generation}
\label{sec:eventgeneration}
There are different SM backgrounds relevant to the assumed ALP production process. The $\tau^+\tau^-$ production with one tau lepton experiencing the leptonic decay $\tau\ra e/\mu+\nu \bar{\nu}$ and the $W^-W^+$ production when one of the $W$ bosons decays through the $W\ra e/\mu+\nu$ decay mode have the dominant contribution to the total background. In this analysis, we consider the production of $\tau^+\tau^-$, $W^+W^-$, top pair $t\bar{t}$, $hZ$, $ZZ$, $Z\gamma$, dijet $q\bar{q}$ ($q=u,d,c,s,b$) and dilepton $e^-e^+/\mu^-\mu^+$ as the SM backgrounds. The generation of backgrounds is performed assuming all possible final states and decay modes.

The Lagrangian, Eq. \ref{lagrangian}, is implemented into \texttt{FeynRules} \cite{Alloul:2013bka} and the generated Universal FeynRules Output (UFO) model is passed to \texttt{MadGraph5\_aMC@NLO} to generate hard events. Parton showering, hadronization and decays of unstable particles are performed with the use of \texttt{Pythia 8.2.43} \cite{Sjostrand:2006za}. The detector effect is simulated using \texttt{Delphes 3.4.2}~\cite{deFavereau:2013fsa} and the delphes card for muon collider\footnote{\url{https://github.com/delphes/delphes/blob/master/cards/delphes_card_MuonColliderDet.tcl}}. \texttt{FastJet 3.3.2} \cite{Cacciari:2011ma} is used to reconstruct jets with the use of the inclusive Valencia jet algorithm \cite{valenciaalgorithm}. At future lepton colliders operating at the energy frontier, performance of classical lepton collider jet algorithms is degraded by non-negligible levels of background. The Valencia jet algorithm is a sequential recombination algorithm which combines good features of the lepton collider and hadron collider jet algorithms to achieve a greater performance in the presence of background. $R$ and $\beta$ parameters in this algorithm control the distance criterion and the robustness against background, respectively. These parameters can be optimized to achieve the best performance and background rejection. The optimum values found in this study are $R=1.2$ and $\beta=1$. Tau tagging is performed to identify jets likely to originate from a tau lepton. The tau tagging algorithm uses the distance criterion $\Delta R<0.5$, where $\Delta R = \sqrt{\Delta\eta^2+\Delta\phi^2}$, with $\eta$ and $\phi$ being the pseudorapidity and azimuth angle, respectively. $\tau$-jets are tagged with a $80\%$ efficiency for tau leptons with transverse momentum $p_T>10$ GeV. The mistag rate for light ($uds$) jets and electron is $2\%$ and $0.1\%$, respectively. Simulation of polarized tau decays is performed using the TauDecay package \cite{taudecayspackage} which is constructed in the \texttt{FeynRules} and \texttt{MadGraph5} frameworks.

Event generation is performed for the three center-of-mass energies $126, 350,$ and $1500$ GeV. For each center-of-mass energy, two cases are considered for the polarization of the initial muon beams. First, the beams are assumed to be unpolarized, and second, the beams are longitudinally polarized with $+0.8$ ($-0.8$) polarization for the $\mu^-$ ($\mu^+$) beam. For the center-of-mass energies of 126 and 1500 GeV, the ALP mass $m_a$ is assumed to be 1 MeV while for the center-of-mass energy of 350 GeV, different mass scenarios ranging from $100$ eV to 1 MeV are considered. It is assumed that only one LFV coupling of the ALP is non-zero at a time, i.e. $c_{\tau  e}\neq 0$, $c_{\tau  \mu}= 0$ or $c_{\tau  \mu}\neq 0$, $c_{\tau  e}= 0$. Event generation is performed for the two cases independently. In each case, the three chiral structures for the ALP LFV coupling (introduced in Section \ref{sec:lagrangian}) are assumed and event generation is independently performed. 

%***********************************************************
\section{Event selection}
\label{sec:eventselection}
The two cases of $c_{\tau  e}\neq 0$, $c_{\tau  \mu}= 0$ and $c_{\tau  \mu}\neq 0$, $c_{\tau  e}= 0$, hereafter referred to as $c_{\tau  e}\neq 0$ and $c_{\tau  \mu}\neq 0$ cases, respectively, are analyzed independently. In what follows, statements are true for both the cases unless stated otherwise. Reconstructed jets should satisfy the kinematical thresholds $p_T>30$ GeV and $\vert \eta \vert < 2.5$. $\tau$-tagged jets are required to only include exactly three charged hadrons, i.e. $\pi^+\pi^+\pi^-$ or $\pi^-\pi^-\pi^+$, and at most one photon. Restricting jet constituents to the three charged pion combination helps reconstruct the four-momentum of the decaying tau lepton more accurately. Furthermore, this restriction significantly suppresses jets stemming from sources other than tau leptons. Isolated electrons, muons and photons are identified with the use of isolation variable $I_{rel}$. Isolation variable for particle P is defined as $I_{rel}=\Sigma\, p_T^{\,i}/p_T^{\,\mathrm{P}}$, where the index $i$ runs over all particles (except for the particle P) with $p_T^{\,i}>p_T^{\,\mathrm{min}}$ and $\Delta R(i,\mathrm{P})<\Delta R^{\,\mathrm{max}}$. The particle P is then identified as an isolated particle if $I_{rel}<I_{rel}^{\,\mathrm{max}}$. The isolation parameters used here are $p_T^{\,\mathrm{min}}=0.5$ GeV, $\Delta R^{\,\mathrm{max}}=0.1$ and $I_{rel}^{\,\mathrm{max}}=0.2$. For isolated muons and electrons, the thresholds $p_T>10$ GeV and $\vert \eta \vert < 2.5$ are applied. For the $c_{\tau  e}\neq 0$ ($c_{\tau  \mu}\neq 0$) case, events should either have exactly one isolated muon (electron) or exactly one reconstructed jet and the only reconstructed jet should be also $\tau$-tagged (cut 1). Events with one $\tau$-tagged reconstructed jet cannot have any isolated muon (electron), and vice versa. That is, we select the decay modes $\tau\ra\nu+3\ \mathrm{charged\ pions}$ and $\tau\ra e/\mu+\nu \bar{\nu}$ among all possible SM decays of the tau lepton. Furthermore, events should have exactly one isolated electron (muon) in the $c_{\tau  e}\neq 0$ ($c_{\tau  \mu}\neq 0$) case (cut 2). The reconstructed objects should pass a relative electric charge sign criterion. In the $c_{\tau  e}\neq 0$ ($c_{\tau  \mu}\neq 0$) case, the charge of the isolated electron (muon) should be opposite to the charge of the isolated muon (electron) or the $\tau$-tagged jet (cut 3). The necessity of this sign criterion is rooted in the opposite charges of the tau leptons undergoing the LFV decay and the SM decay in the ALP production process. Isolated photons are reconstructed requiring $p_T>10$ GeV and $\vert \eta \vert < 2.5$. A plane perpendicular to the momentum direction of the tau leptons divides the space into two hemispheres, one on the side of the tau lepton undergoing the SM decay and one on the side of the tau lepton undergoing the LFV decay. The direction of the tau leptons is estimated using a procedure fully discussed in Section \ref{sec:analysis}. On the side of the tau lepton experiencing the SM decay, there can be at most one photon with the maximum allowed energy of $E_{max}^{\mathrm{\, SM}}$, and on the opposite side, there can be at most one photon with the maximum allowed energy of $E_{max}^{\mathrm{\, LFV}}$ (cut 4). The energies $E_{max}^{\mathrm{\, SM}}$ and $E_{max}^{\mathrm{\, LFV}}$ are optimized so as to achieve the strongest limits on the ALP LFV couplings. Optimum $E_{max}^{\mathrm{\, SM}}$ values found for the center-of-mass energies of 126, 350 and 1500 GeV are 45, 60 and 90 GeV, respectively. For $E_{max}^{\mathrm{\, LFV}}$, the optimum values 35, 50 and 70 GeV which respectively correspond to the center-of-mass energies of 126, 350 and 1500 GeV have been found. In counting the number of photons, the allowed photon inside the $\tau$-tagged jet is ignored. Imposing this restriction on the number of photons suppresses the $Z\gamma$ background significantly. Event selection efficiencies obtained for the $c_{\tau  e}\neq 0$ and $c_{\tau  \mu}\neq 0$ cases at the center-of-mass energy of 350 GeV (for example) are presented in Tab. \ref{seleff}. Efficiencies corresponding to the unpolarized and polarized muon beams cases are presented independently. As seen, the SM $\tau^+\tau^-$ and $W^+W^-$ production processes have higher event selection efficiencies than other backgrounds. The SM $\tau^+\tau^-$ production followed by the decay $\tau\ra e/\mu+\nu \bar{\nu}$ of one of the tau leptons resembles the signal signature very closely and thus the $\tau^+\tau^-$ production has the highest event selection efficiency.
\begin {table}[t]  
\centering
 \begin{tabular}{cccccccccccc}
 & & & signal & $\tau^+\tau^-$ & $W^+W^-$ & $t\bar{t}$ & $ZZ$ & $hZ$ & $Z\gamma$ & dijet & $ee/\mu\mu$ \\
\cline{1-12}
\parbox[t]{.05mm}{\multirow{5}{*}{\rotatebox[origin=c]{90}{Unpolarized}}} & \parbox[t]{3mm}{\multirow{5}{*}{\rotatebox[origin=c]{90}{$c_{\tau  e}\neq 0$}}} & 1 & 0.2067 & 0.1219 & 0.0518 & 0.0022 & 0.0065 & 0.0070 & 0.0009 & 0.0001 & 0.0731 \\
& & 2 & 0.8164 & 0.5255 & 0.3987 & 0.4679 & 0.0999 & 0.2244 & 0.0563 & 0.0654 & 0 \\
& & 3 & 0.9998 & 0.9998 & 0.9998 & 0.9039 & 0.8030 & 0.8860 & 0.9833 & 0.8011 & 0 \\
& & 4 & 0.9989 & 0.9980 & 0.9980 & 0.9991 & 0.9861 & 0.9966 & 0.0847 & 0.9975 & 0 \\
& & total & 0.1686 & 0.0639 & 0.0206 & 0.0009 & 0.0005 & 0.0014 & 4.0e-06 & 6.4e-06 & 0 \\
\cline{1-12}
\parbox[t]{.05mm}{\multirow{5}{*}{\rotatebox[origin=c]{90}{Polarized}}} & \parbox[t]{3mm}{\multirow{5}{*}{\rotatebox[origin=c]{90}{$c_{\tau  e}\neq 0$}}} & 1 & 0.2011 & 0.1131 & 0.0477 & 0.0021 & 0.0065 & 0.0068 & 0.0009 & 9.2e-5 & 0.0726 \\
& & 2 & 0.8818 & 0.5413 & 0.4518& 0.4656 & 0.1104 & 0.2334 & 0.0573 & 0.0672 & 0 \\
& & 3 & 0.9999 & 0.9997 & 0.9997 & 0.8933 & 0.7620 & 0.8943 & 0.9841 & 0.7501 & 0 \\
& & 4 & 0.9987 & 0.9982 & 0.9987 & 0.9991 & 0.9956 & 0.9978 & 0.0968 & 0.9974 & 0 \\
& & total & 0.1771 & 0.0611 & 0.0215 & 0.0009 & 0.0005 & 0.0014 & 4.8e-06 & 4.6e-06 & 0 \\
\cline{1-12}
\parbox[t]{.05mm}{\multirow{5}{*}{\rotatebox[origin=c]{90}{Unpolarized}}} & \parbox[t]{3mm}{\multirow{5}{*}{\rotatebox[origin=c]{90}{$c_{\tau  \mu}\neq 0$}}} & 1 & 0.1985 & 0.1178 & 0.0467 & 0.0021 & 0.0083 & 0.0066 & 0.0010 & 9.5e-5 & 0.0002 \\
& & 2 & 0.8901 & 0.5625 & 0.4441 & 0.4877 & 0.0771 & 0.2399 & 0.0499 & 0.0840 & 0 \\
& & 3 & 0.9998 & 0.9998 & 0.9998 & 0.9049 & 0.8271 & 0.8896 & 0.9844 & 0.8021 & 0 \\
& & 4 & 0.9987 & 0.9980 & 0.9983 & 0.9991 & 0.9895 & 0.9966 & 0.0794 & 0.9974 & 0 \\
& & total & 0.1764 & 0.0661 & 0.0207& 0.0009 & 0.0005 & 0.0014 & 4.0e-06 & 6.4e-06 & 0 \\
\cline{1-12}
\parbox[t]{.05mm}{\multirow{5}{*}{\rotatebox[origin=c]{90}{Polarized}}} & \parbox[t]{3mm}{\multirow{5}{*}{\rotatebox[origin=c]{90}{$c_{\tau  \mu}\neq 0$}}} & 1 & 0.1952 & 0.1095 & 0.0439 & 0.0021 & 0.0083 & 0.0066 & 0.0010 & 6.0e-5 & 0.0003 \\
& & 2 & 0.9401 & 0.5812 & 0.4945 & 0.4737 & 0.0884 & 0.2490 & 0.0548 & 0.1026 & 0 \\
& & 3 & 0.9998 & 0.9998 & 0.9997 & 0.8937 & 0.7719 & 0.8934 & 0.9828 & 0.7506 & 0 \\
& & 4 & 0.9988 & 0.9981 & 0.9987 & 0.9991 & 0.9957 & 0.9995 & 0.0866 & 0.9974 & 0 \\
& & total & 0.1833 & 0.0635 & 0.0217 & 0.0009 & 0.0006 & 0.0015 & 4.7e-06 & 4.6e-06 & 0 \\
\cline{1-12}
\label{efficiencies}
\end{tabular}
\caption {Event selection relative efficiencies corresponding to the cuts 1-4 (see text for the detailed description of the selection cuts) obtained for the ALP production process and different SM backgrounds. Results corresponding to both the cases of unpolarized and polarized muon beams and also both the $c_{\tau  e}\neq 0$ and $c_{\tau  \mu}\neq 0$ cases have been shown. It is assumed that the ALP mass is 1 MeV, the center-of-mass energy is 350 GeV and the ALP couples to right-handed leptons (V+A case).}
\label{seleff}
\end {table} 

%***********************************************************
\section{Analysis}
\label{sec:analysis}
Selected events are analyzed to compute the variables which discriminate the ALP production process from the SM backgrounds. We employ a multivariate technique utilizing the Boosted Decision Trees (BDT) algorithm~\cite{Hocker:2007ht}. A proper set of discriminating variables is given to the BDT as input and training is performed considering all background processes according to their respective weights. The BDT output is examined in terms of the discriminating power with the use of the receiver operating characteristic (ROC) curve. The BDT output distribution is used to constrain the ALP LFV couplings $c_{\tau  e}/f_a$ and $c_{\tau  \mu}/f_a$, independently. The two cases $c_{\tau  e}\neq 0$ and $c_{\tau  \mu}\neq 0$, and different cases of the ALP coupling chiral structure and also the unpolarized and polarized muon beams cases are analyzed independently and the upper limit on the LFV coupling is obtained in each case. 

\begin{figure}[t]
  \centering
  \includegraphics[width=0.41\textwidth]{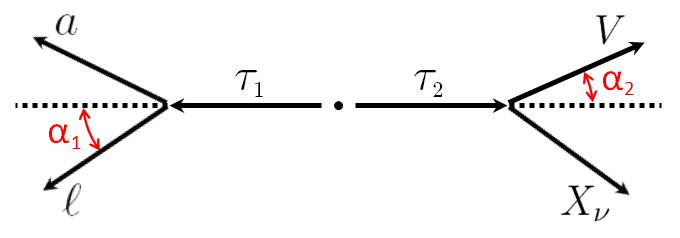}
  \caption{The angles $\alpha_1$ and $\alpha_2$ in the decays $\tau_1\ra \ell\,a$ and $\tau_2\ra V\,X_\nu$ of the tau leptons in the ALP production process in the laboratory frame. $V$ is either a system of hadrons (identified as a $\tau$-jet) or a lepton ($e,\mu$), $X_\nu$ is a system of neutrinos and $\ell=e,\mu$. $\tau_1$ ($\tau_2$) represents the tau lepton undergoing the LFV (SM) decay.}
\label{corrangles}
\end{figure}

Because of the invisible ALP and neutrinos in the final state of the tau decays, the momenta of the decaying tau leptons cannot be exactly determined and can only be estimated. How precise the momenta of the decaying taus are estimated is crucial to the sensitivity we achieve in the analysis. In what follows, we describe the procedure we use to estimate the momenta of the tau leptons. In the ALP production process, the two tau leptons decay through the LFV and SM modes $\tau_1\ra \ell\,a$ and $\tau_2\ra VX_\nu$, where $\tau_1$ ($\tau_2$) represents the tau lepton undergoing the LFV (SM) decay, $V$ is the visible part of the decay products which is either a system of hadrons (identified as a $\tau$-jet) or a lepton ($e,\mu$), $X_\nu$ is a system of neutrinos and $\ell=e,\mu$. The angle $\alpha_1$ is defined as the deviation angle of the momentum direction of the lepton $\ell$ from the direction of $\tau_1$ in the laboratory frame. Similarly, the angle $\alpha_2$ is defined as the deviation angle of the momentum direction of $V$ from the momentum direction of $\tau_2$. Fig. \ref{corrangles} depicts the decaying tau leptons in the ALP production process and the angles $\alpha_1$ and $\alpha_2$.
Using the conservation of momentum, one finds
\begin{equation}
\cos\alpha_1=\frac{2E_{\tau_1} E_\ell + m_a^2 - m_\tau^2 - m_\ell^2}{2\,p_\ell\sqrt{E_{\tau_1}^2-m_\tau^2}}\, , \ \ \ \ \cos\alpha_2=\frac{2E_{\tau_2} E_V - m_\tau^2 - m_V^2}{2\,p_V \sqrt{E_{\tau_2}^2-m_\tau^2}}\, ,
\label{eq:corrangles}
\end{equation}
where $E_x$, $p_x$ and $m_x$ are respectively the energy, momentum and mass of the object $x$ as measured in the laboratory frame. $E_{\tau_1}$ and $E_{\tau_2}$ in Eq. \ref{eq:corrangles} cannot be experimentally measured because of the invisible particles in the final state of tau decays. However, ignoring the radiative emissions, the conservation of energy implies that each of the tau leptons carries half of the center-of-mass energy of the experiment, i.e. $E_{\tau_1}=E_{\tau_2}=\sqrt{s}/2$. We assume that the unit vectors $\hat{v}_1$ and $\hat{v}_2$ define the momentum directions of the tau leptons $\tau_1$ and $\tau_2$, respectively. Using the defined deviation angles $\alpha_1$ and $\alpha_2$, we have the dot products $\hat{v}_1.\,\vec{p}_\ell=p_\ell\cos\alpha_1$ and $\hat{v}_2.\,\vec{p}_V=p_V\cos\alpha_2$ with $\vec{p}_x$ being the momentum vector of the object $x$ (see Fig. \ref{corrangles}). Furthermore, as the tau leptons are produced in a back-to-back configuration, we have $\hat{v}_1=-\hat{v}_2$. These relations form the system of equations 
\begin{eqnarray}
\begin{aligned}
& \hat{v}_1.\,\vec{p}_\ell=\frac{2E_{\tau_1} E_\ell + m_a^2 - m_\tau^2 - m_\ell^2}{2\,\sqrt{E_{\tau_1}^2-m_\tau^2}}\, , 
\\
&\hat{v}_1.\,\vec{p}_V=-\,\frac{2E_{\tau_2} E_V - m_\tau^2 - m_V^2}{2\,\sqrt{E_{\tau_2}^2-m_\tau^2}}\, ,
\\
&\vert\hat{v}_1\vert=1 \, .
\label{eq:}
\end{aligned}
\end{eqnarray}
This system of equations can be simultaneously solved and has two solutions for $\hat{v}_1$. We compute the average of the solutions and take it as $\hat{v}_1$. The momentum vectors $\vec{p}_{\tau_1}$ and $\vec{p}_{\tau_2}$ of the tau leptons can then be reconstructed using the obtained directions $\hat{v}_1$ and $\hat{v}_2$, the tau lepton mass and the energy of the tau leptons ($\sqrt{s}/2$).

\begin{figure*}[t]
  \centering  
    \begin{subfigure}[b]{0.48\textwidth} 
    \centering
    \includegraphics[width=\textwidth]{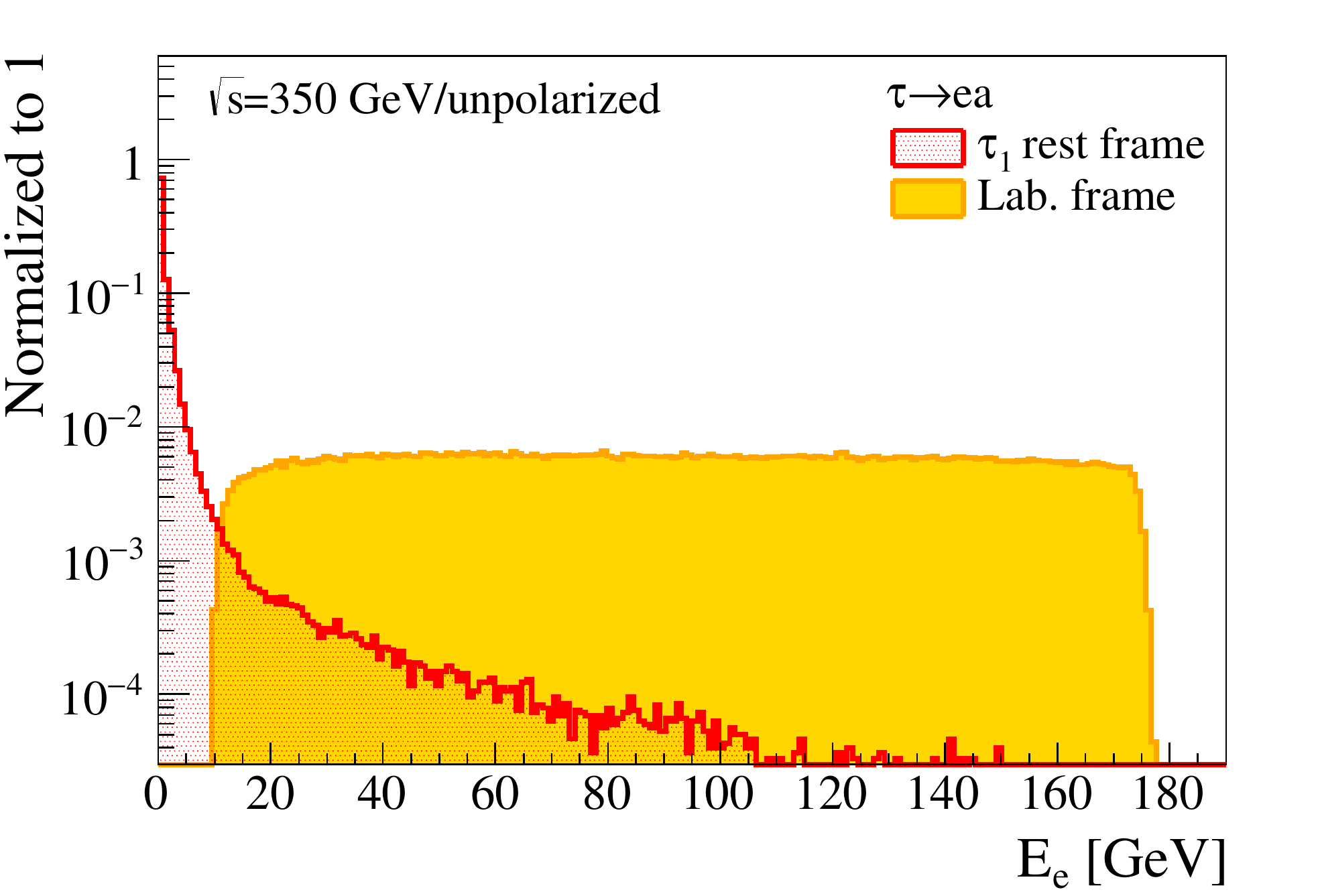}
    \caption{}
    \label{EeEeRFCORRcomparison0}
    \end{subfigure} \\
    \begin{subfigure}[b]{0.48\textwidth}
    \centering
    \includegraphics[width=\textwidth]{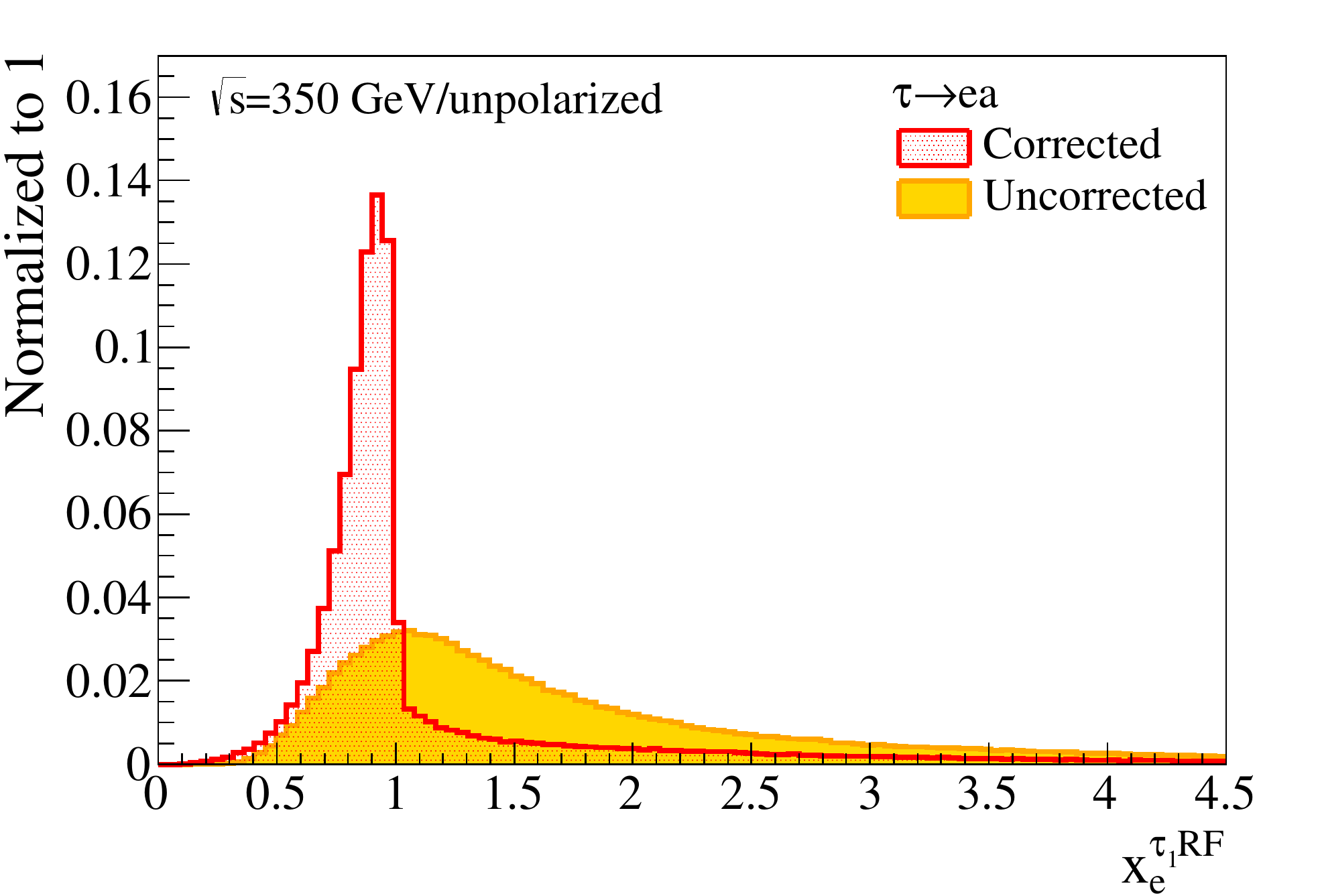}
    \caption{}
    \label{xeRFxeRFCORRcomparison0} 
    \end{subfigure} 
    \begin{subfigure}[b]{0.48\textwidth}
    \centering
    \includegraphics[width=\textwidth]{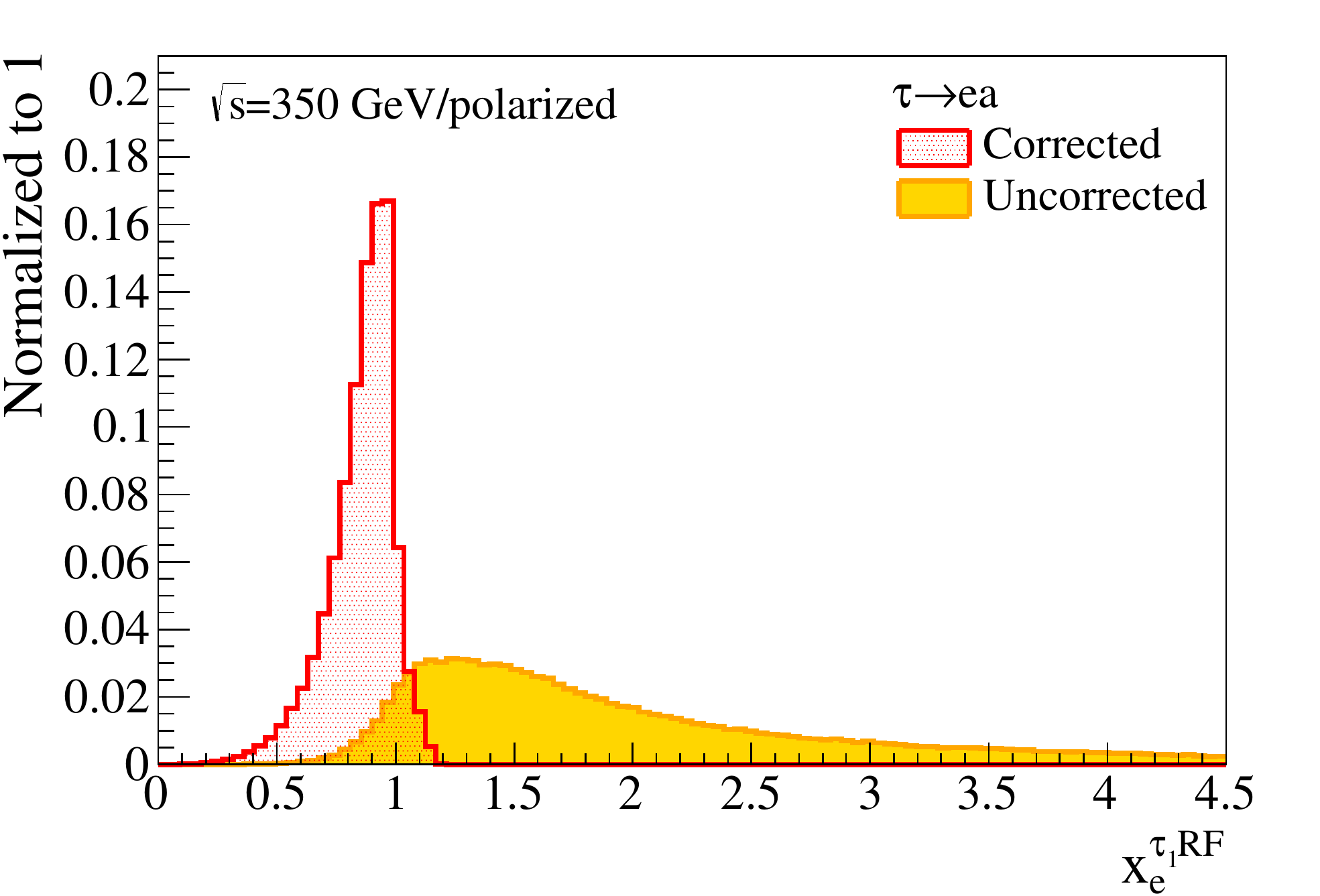}
    \caption{}
    \label{xeRFxeRFCORRcomparison80}
    \end{subfigure}
\caption{a) Energy spectrum of the isolated electron or positron in the laboratory frame and the $\tau_1$ rest frame for the unpolarized muon beams. Electron energy fraction in the $\tau_1$ rest frame $x_e^{\tau_1\mathrm{RF}}$ obtained with (corrected) and without (uncorrected) the use of the described tau momentum estimation procedure, for b) unpolarized and c) polarized muon beams. The distributions are obtained assuming the process $\mu^-\mu^+\ra \tau^-\tau^+$, $\tau^\pm\ra e^\pm\, a$, $\tau^\mp\ra \mathrm{SM}$ with $\sqrt{s}=350$ GeV and $m_a=1$ MeV. The ALP couples to right-handed leptons. }
  \label{fig:comparison}
\end{figure*}

The produced ALP is likely to escape the detector before its decay (see Section \ref{sec:lagrangian}). We reconstruct the four-momentum ($E_a,\vec{p}_a$) of the ALP using the relations
\begin{equation}
E_a = \sqrt{s} - E_{\tau_2} - E_\ell \overset{E_{\tau_2} = \frac{\sqrt{s}}{2}}{=} \frac{\sqrt{s}}{2} - E_\ell\, , \ \ \ \  \vec{p}_a = - (\vec{p}_{\tau_2} + \vec{p}_\ell)\, ,
\label{eq:epax}
\end{equation}
which are deduced from the energy-momentum conservation. The reconstructed ALP and $\tau$ leptons are used to compute some useful variables introduced in what follows. We found it effective to boost some variables into the $\tau_1$ rest frame because it provides a considerable discriminating power compared with the laboratory frame. In the two-body decay $\tau_1\ra \ell\,a$, the decay products are monoenergetic in the $\tau_1$ rest frame (see Section \ref{sec:BSMtaudecay}) while their energy in the laboratory frame varies over a wide range of values. Fig. \ref{EeEeRFCORRcomparison0} shows the energy of the isolated electron (or positron) $E_e$ in both the laboratory frame and the $\tau_1$ rest frame assuming unpolarized muon beams, $c_{\tau  e}\neq 0$ and $\sqrt{s}=350$ GeV. As seen, the energy of the electron, which is widely distributed in the laboratory frame, is narrowly distributed in the $\tau_1$ rest frame. The rest frame distribution separates a significant amount of background from the ALP production process as seen in Fig. \ref{i0}. This figure shows the distribution of the electron energy fraction in the $\tau_1$ rest frame $x_e^{\tau_1\mathrm{RF}}=2E_{e}^{\,\tau_1\mathrm{RF}}/m_\tau$, for signal and different background processes. Boosting into the $\tau_1$ rest frame is also fruitful for the variable $\Omega^{\tau_1\mathrm{RF}}(\ell,a)$ which is defined as the angle between the momentum vector of the lepton $\ell$ and the momentum vector of the ALP in the $\tau_1$ rest frame (see Fig. \ref{corrangles}). Fig. \ref{g0} shows the distributions obtained for this variable. The distribution peaks at $\pi$ radians for the ALP production process, which is expected since in the two-body decay $\tau_1\ra \ell\,a$, the decay products are emitted in a back-to-back configuration. 

The sensitivity that we achieve in this analysis strongly depends on how good the momenta of the decayed $\tau$ leptons are reconstructed. An easy and common approach is to take the momenta of the reconstructed $\tau$-jets as the momenta of the decayed $\tau$ leptons. In this analysis, however, we used a more sophisticated and accurate method to compute the $\tau$ leptons momenta (as described). To see how effective the employed method for estimating the momenta of the $\tau$ leptons is, we compare the distributions of the electron energy fraction in the $\tau_1$ rest frame $x_e^{\tau_1\mathrm{RF}}$ in the following two cases. First, $\tau$ momenta are reconstructed using the estimation procedure described in this section, and second, the momentum of $V$ (see Fig. \ref{corrangles}) is taken as the momentum of $\tau_2$ (ignoring the effect of $X_\nu$) and the momentum of $\tau_1$ is then reconstructed using the reconstructed $\tau_2$. Figs. \ref{xeRFxeRFCORRcomparison0} and \ref{xeRFxeRFCORRcomparison80} show the $x_e^{\tau_1\mathrm{RF}}$ distributions obtained in the two mentioned cases for unpolarized and polarized muon beams, respectively. 
As seen, $x_e^{\tau_1\mathrm{RF}}$ varies over a wide range in the uncorrected case (the latter case) while it is narrowly distributed with a peak near 1 (as expected) when a correction is applied using the described estimation procedure. The obtained narrow distribution helps improve the signal-background discrimination significantly. Distributions of other kinematical variables (used to discriminate the signal from background) are also sensitive to the approach used to compute the momenta of the decayed $\tau$ leptons and get improved in a similar way.

The list of the discriminating variables used in this analysis to feed the BDT is provided below. The variables are defined using the same notation as in Fig. \ref{corrangles}. Unless stated otherwise, the frame in which a variable is measured is the laboratory frame.
\begin{itemize}%[label=$\ast$]
    \item Missing transverse energy $\slashed{E}_T$.
    \item Transverse momentum and pseudorapidity of $V$, $p_{\,T}^V$ and $\eta^V$, where $V$ can be either a system of hadrons (identified as a $\tau$-jet) or an isolated lepton ($e,\mu$). 
    \item Transverse momentum of the isolated lepton $\ell$, $p_{\,T}^\ell$.
    \item Invariant mass of the isolated lepton $\ell$ and the ALP, $M_{\ell a}^{\mathrm{inv.}}$.
    \item Angle between the momentum vector of $V$ and the momentum vector of the isolated lepton $\ell$, $\Omega(V,\ell)$.
    \item Angle between the momentum vector of the isolated lepton $\ell$ and the momentum vector of the ALP in the $\tau_1$ rest frame, $\Omega^{\tau_1\mathrm{RF}}(\ell,a)$.
    \item Angle between the momentum vector of $\tau_1$ in the laboratory frame and the momentum vector of the isolated lepton $\ell$ in the $\tau_1$ rest frame, $\Omega^{\tau_1\mathrm{RF}}(\tau_1,\ell)$.
    \item Energy fraction of the isolated lepton $\ell$ in the $\tau_1$ rest frame, $x_\ell^{\tau_1\mathrm{RF}}=2E_{\ell}^{\tau_1\mathrm{RF}}/m_\tau$.
\end{itemize}
The distributions obtained for these variables in the unpolarized muon beams case assuming $c_{\tau  e}\neq 0$ and $\sqrt{s}=350$ GeV are provided in Fig. \ref{0}.
\begin{figure*}[!t]
  \centering  
    \begin{subfigure}[b]{0.48\textwidth} 
    \centering
    \includegraphics[width=\textwidth]{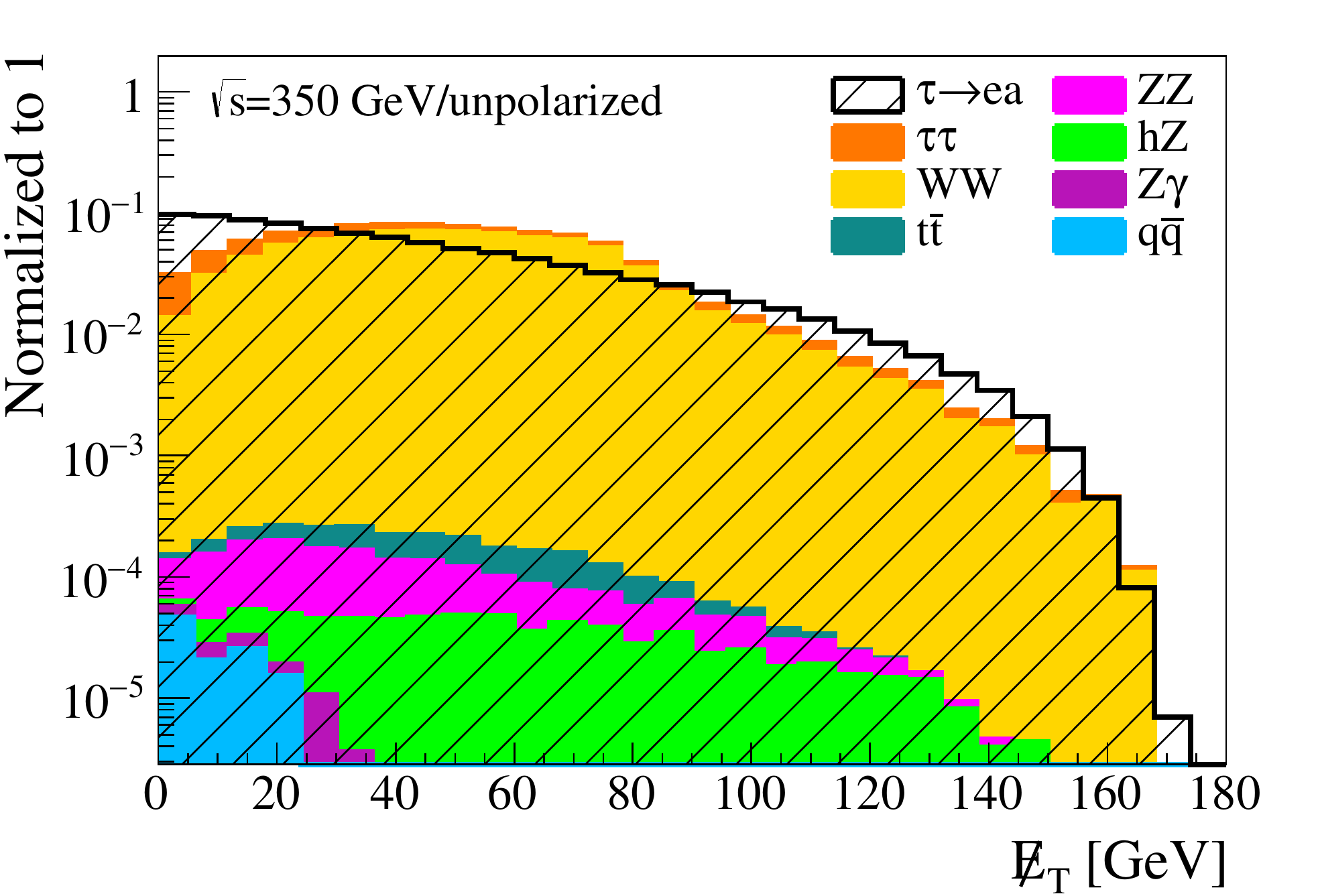}
    \caption{}
    \label{a0}
    \end{subfigure} 
    \begin{subfigure}[b]{0.48\textwidth}
    \centering
    \includegraphics[width=\textwidth]{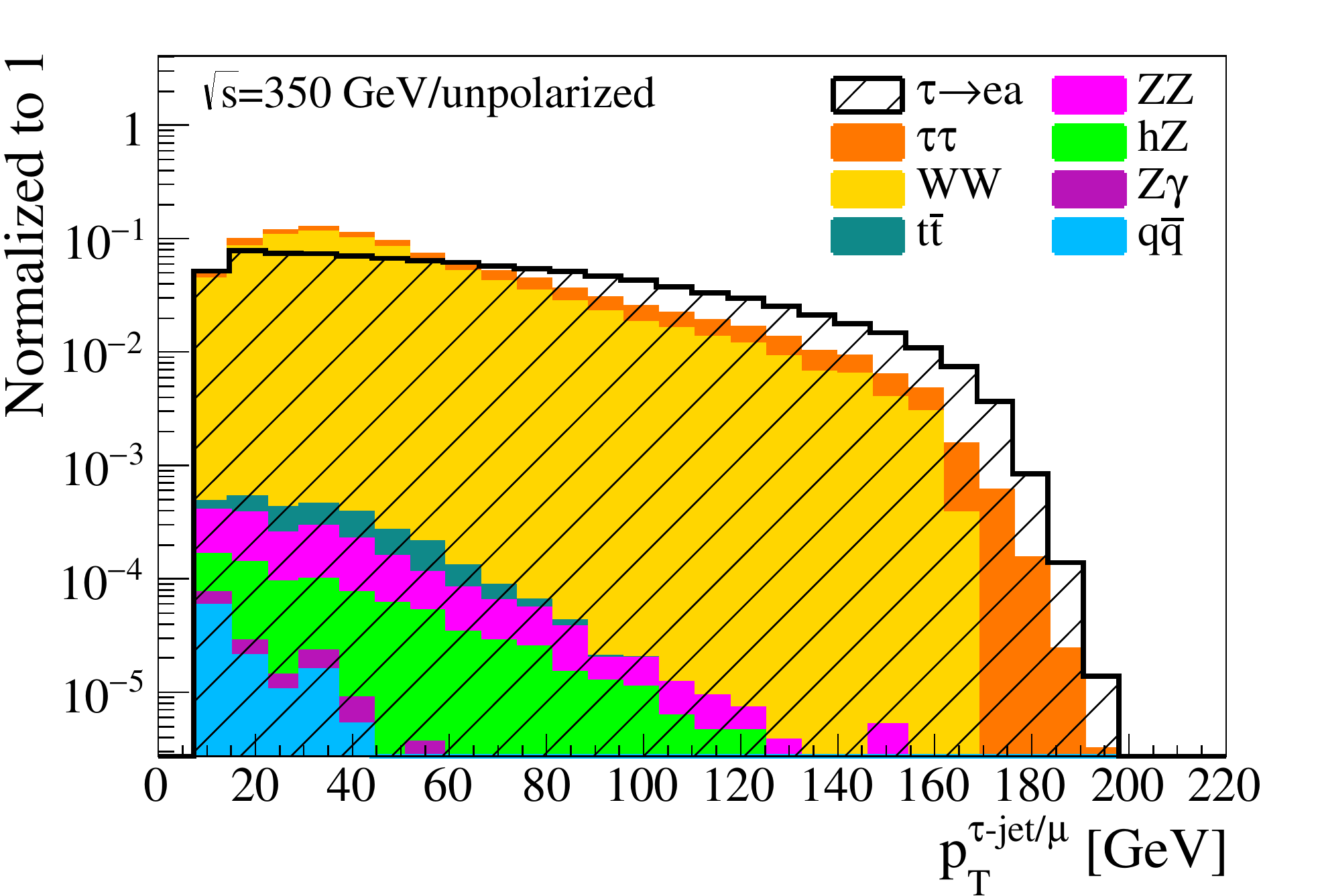}
    \caption{}
    \label{b0} 
    \end{subfigure} 
    \begin{subfigure}[b]{0.48\textwidth}
    \centering
    \includegraphics[width=\textwidth]{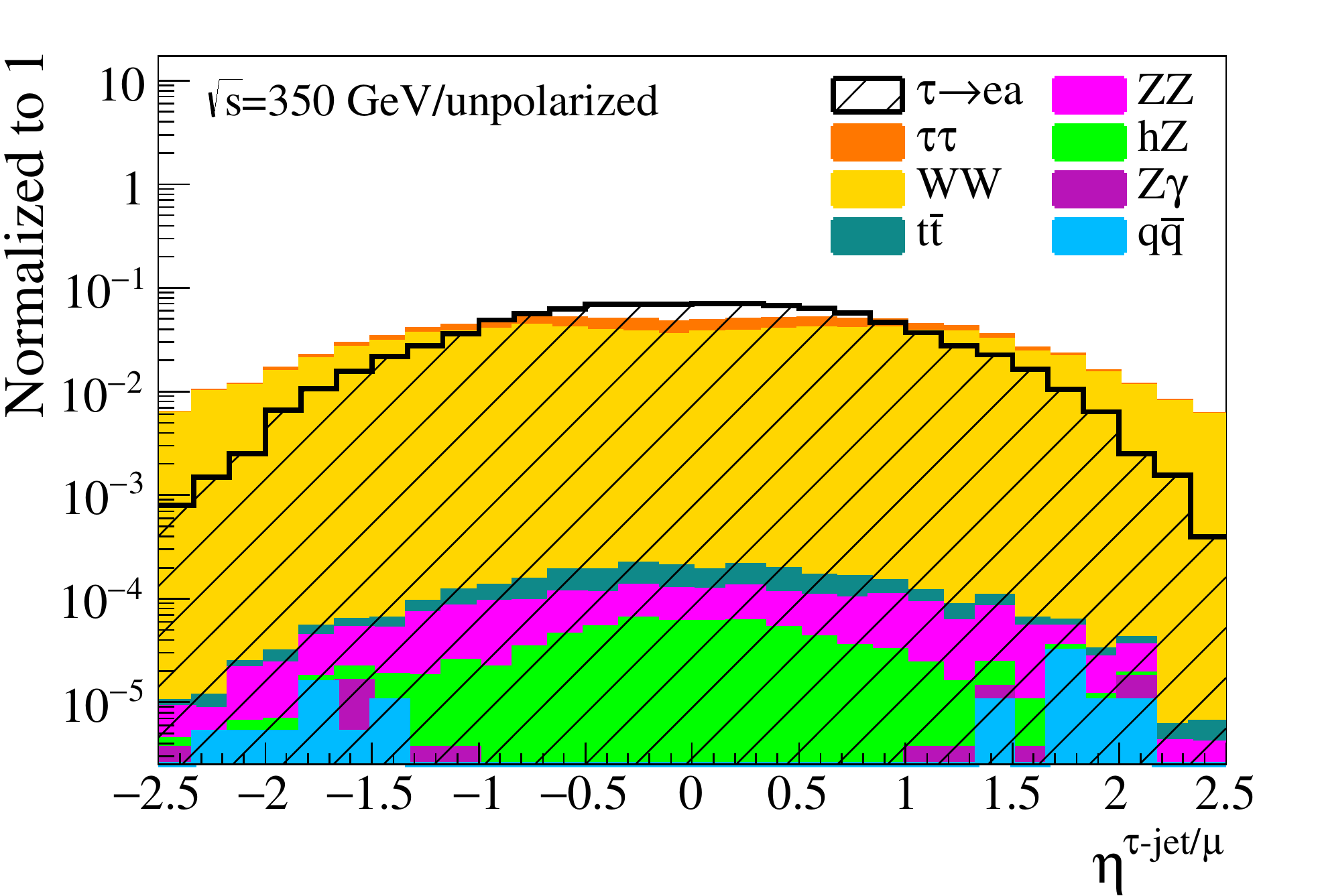}
    \caption{}
    \label{c0}
    \end{subfigure}
    \begin{subfigure}[b]{0.48\textwidth}
    \centering
    \includegraphics[width=\textwidth]{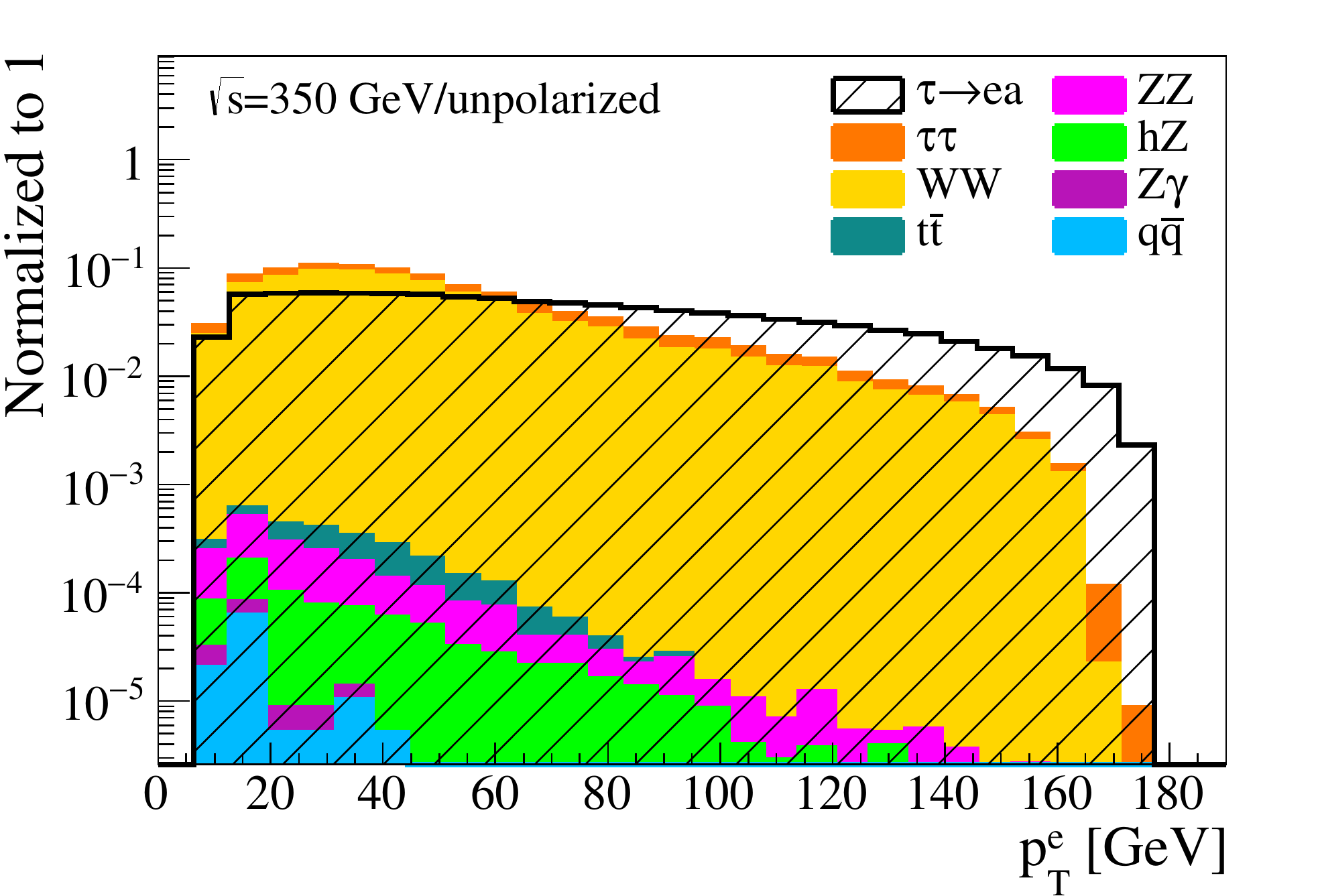}
    \caption{}
    \label{d0}
    \end{subfigure}
    \begin{subfigure}[b]{0.48\textwidth}
    \centering
    \includegraphics[width=\textwidth]{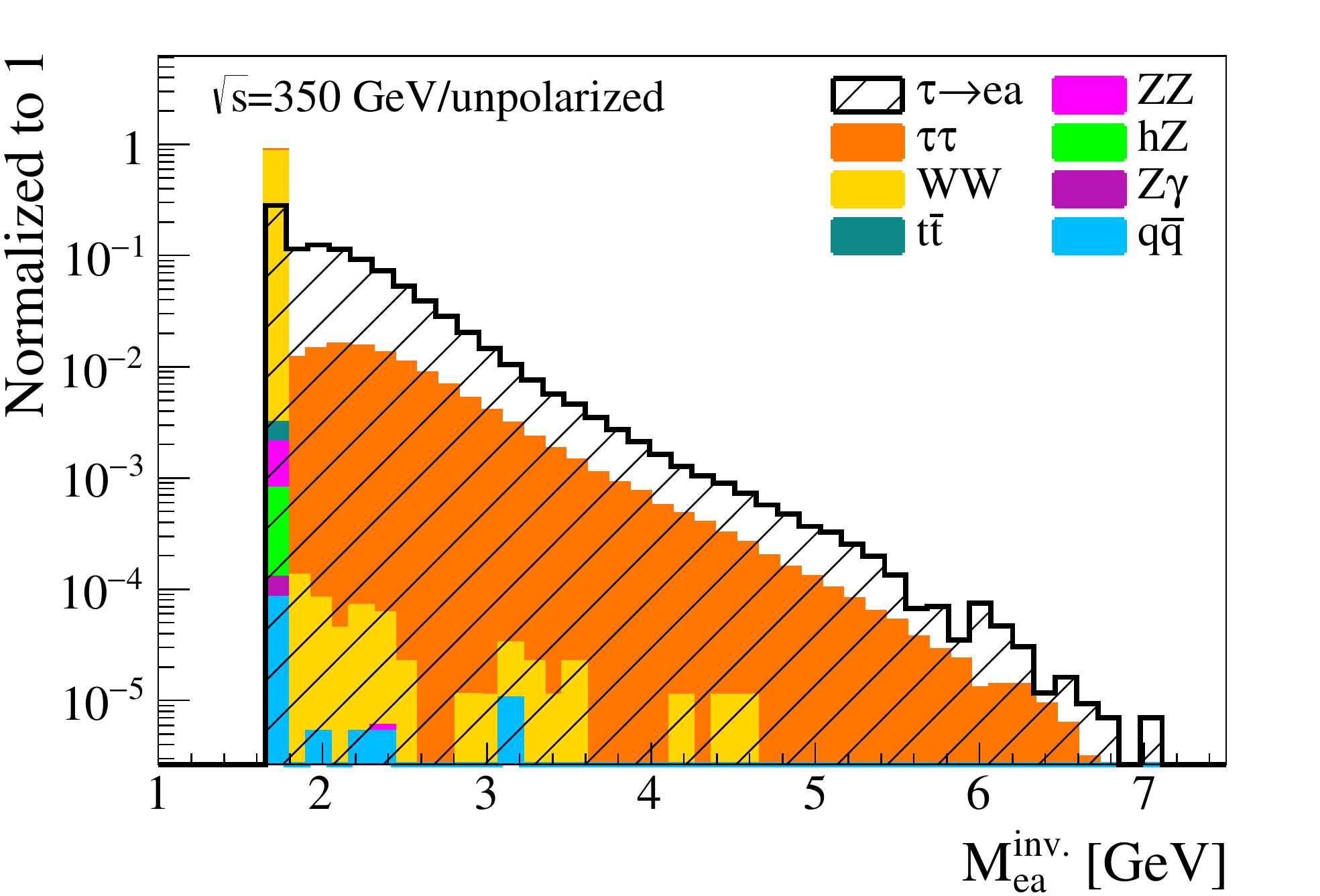}
    \caption{}
    \label{e0}
    \end{subfigure}
\begin{subfigure}[b]{0.48\textwidth}
    \centering
    \includegraphics[width=\textwidth]{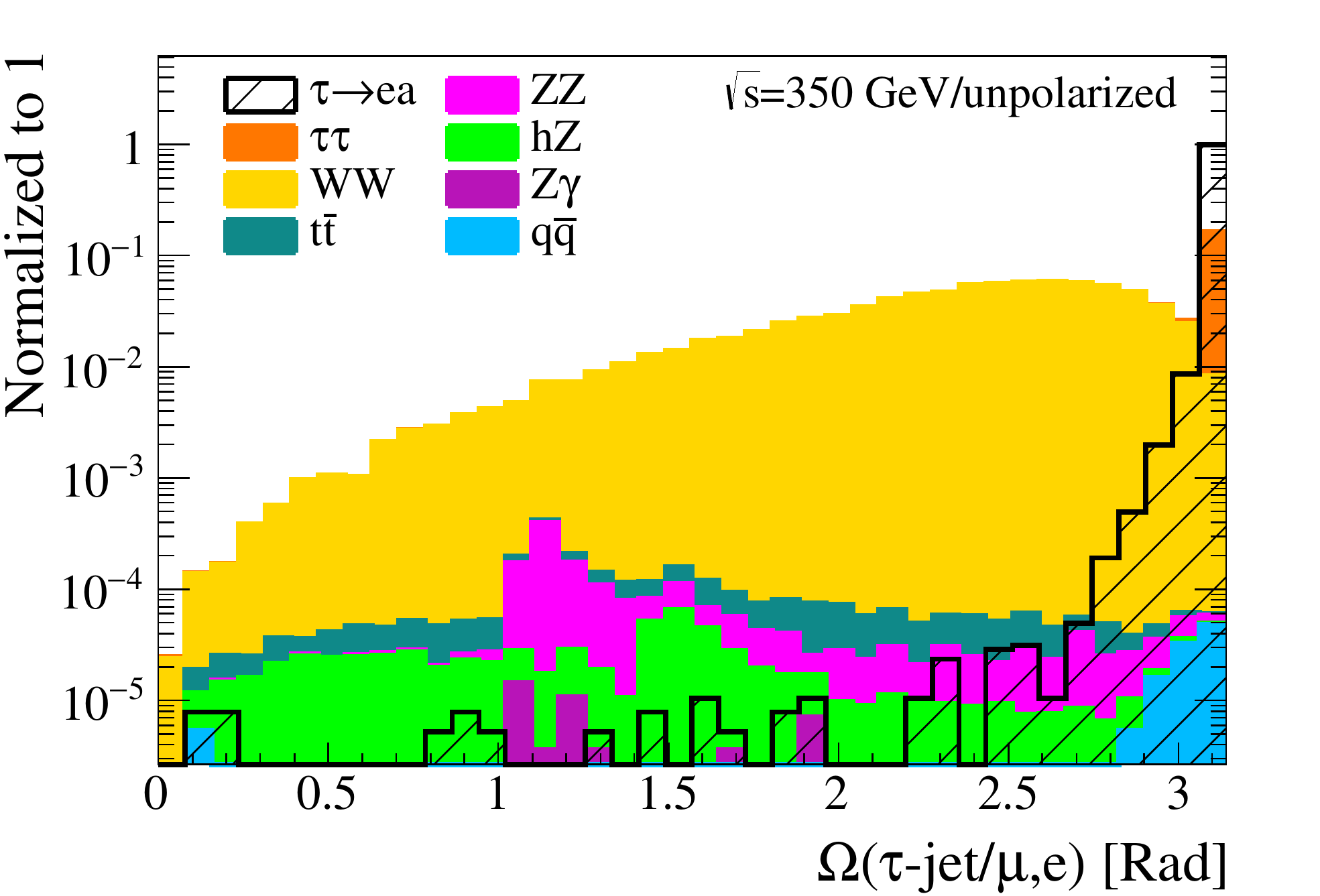}
    \caption{}
    \label{f0}
    \end{subfigure}
\caption{Distributions of the discriminating variables obtained for the unpolarized muon beams case assuming $c_{\tau  e}\neq 0$, $\sqrt{s}=350$ GeV and $m_a=1$ MeV. The ALP couples to right-handed leptons. All processes including the ALP production process ($\tau\ra e\, a$) and the SM backgrounds are shown.}
\label{0}
\end{figure*}
\begin{figure*}[!h]\ContinuedFloat
    \begin{subfigure}[b]{0.48\textwidth}
    \centering
    \includegraphics[width=\textwidth]{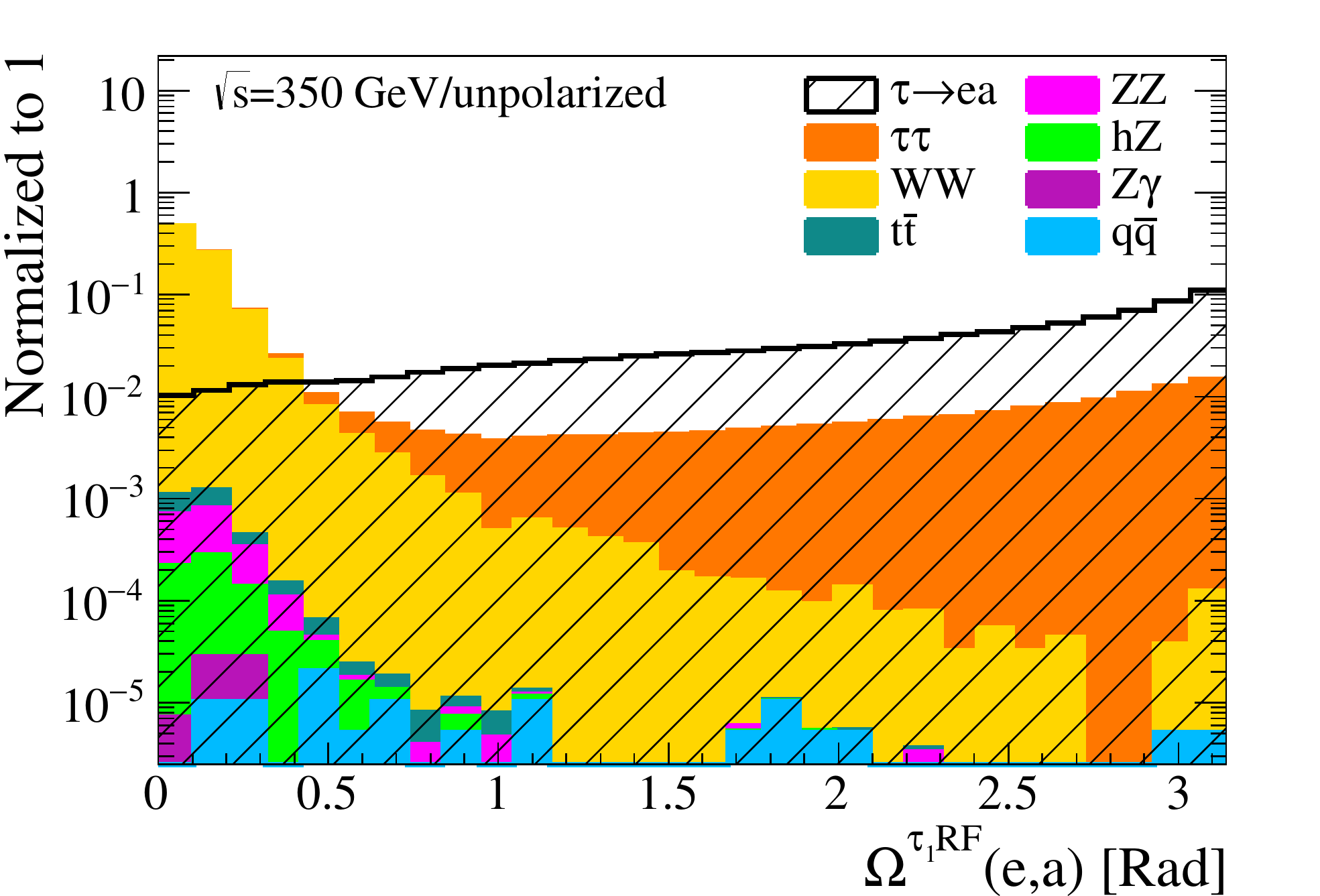}
    \caption{}
    \label{g0}
    \end{subfigure}
    \begin{subfigure}[b]{0.48\textwidth}
    \centering
    \includegraphics[width=\textwidth]{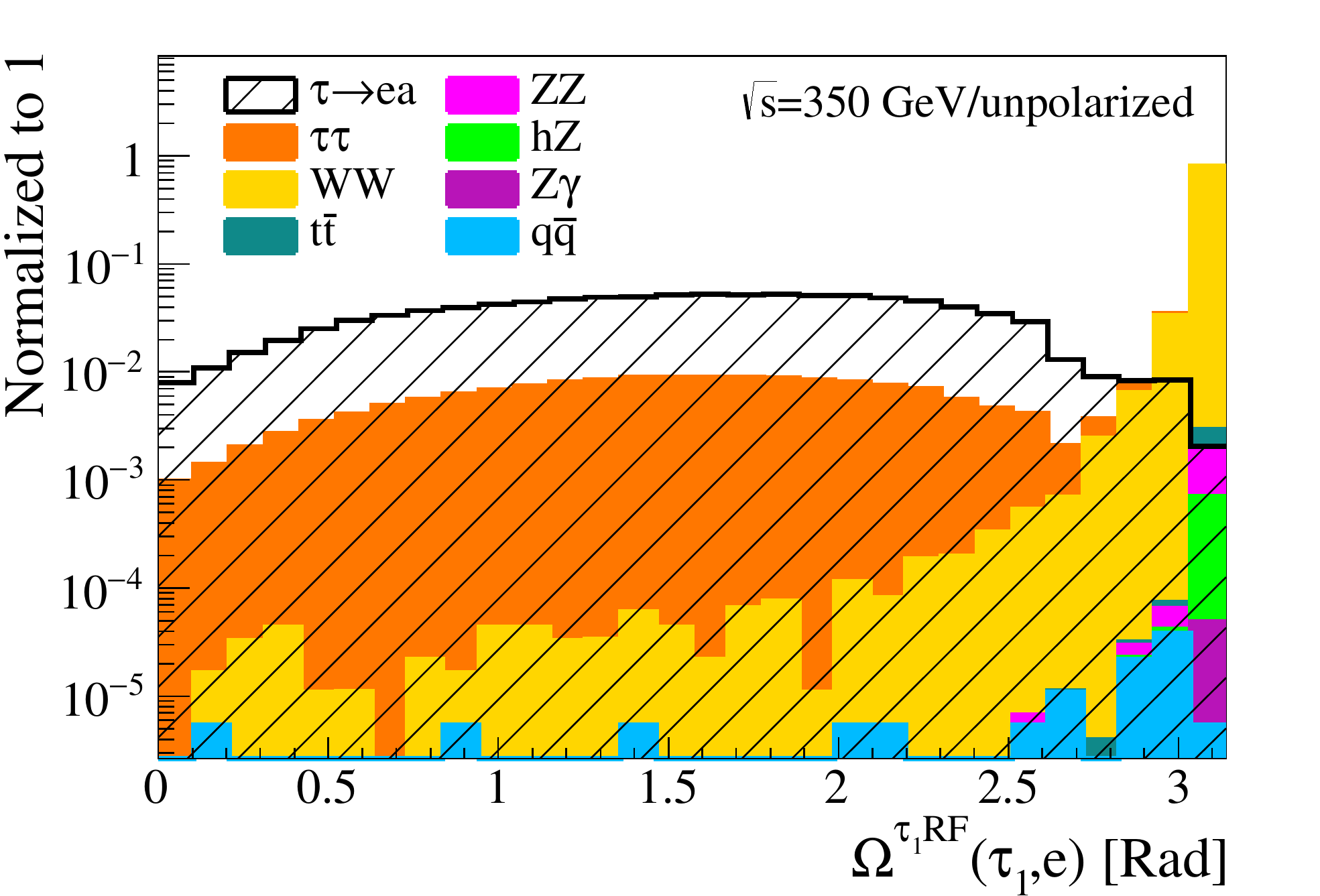}
    \caption{}
    \label{h0}
    \end{subfigure}
    \begin{subfigure}[b]{0.48\textwidth}
    \centering
    \includegraphics[width=\textwidth]{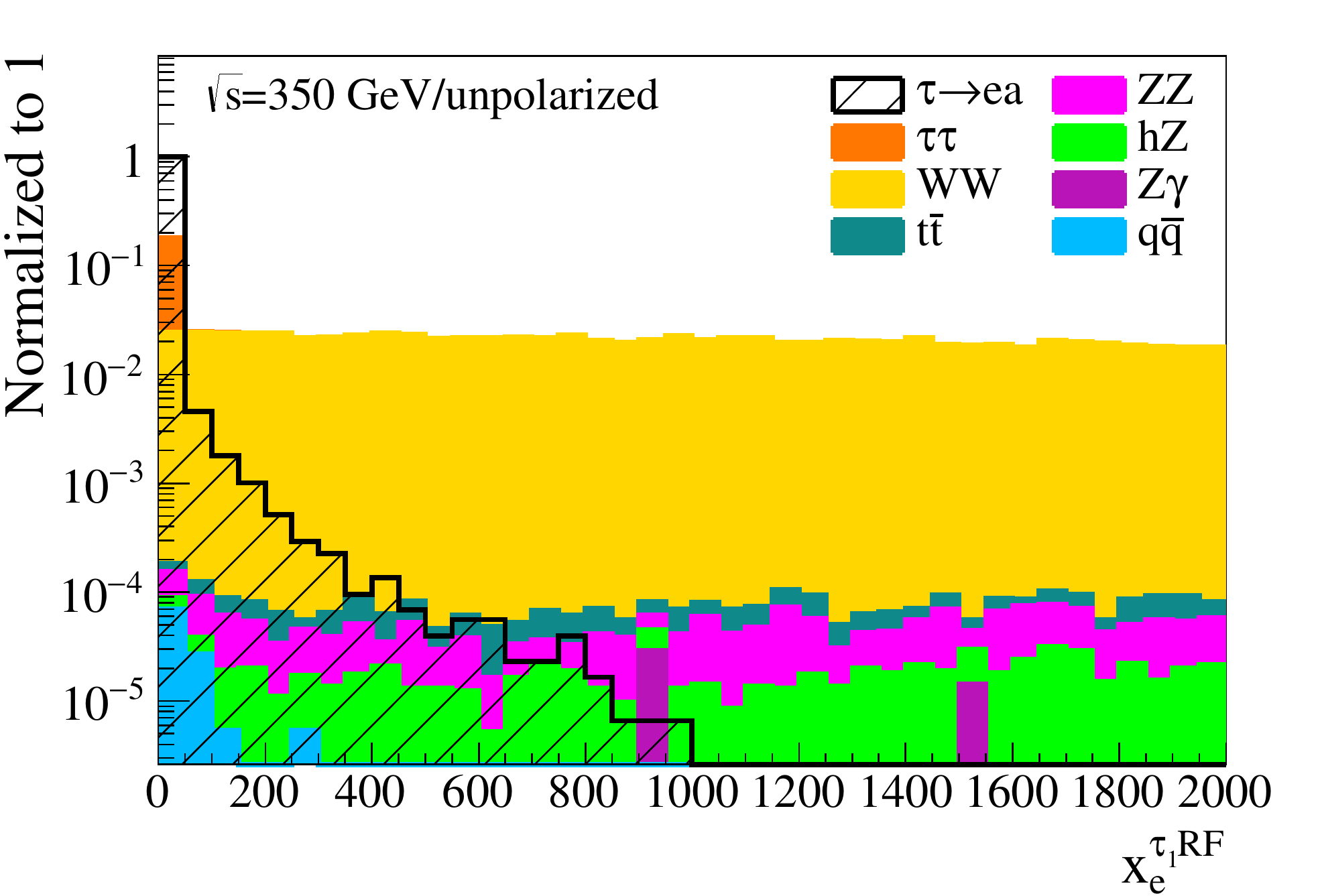} 
    \caption{}   
    \label{i0}
    \end{subfigure}
    \begin{subfigure}[b]{0.48\textwidth}
    \centering
    \includegraphics[width=\textwidth]{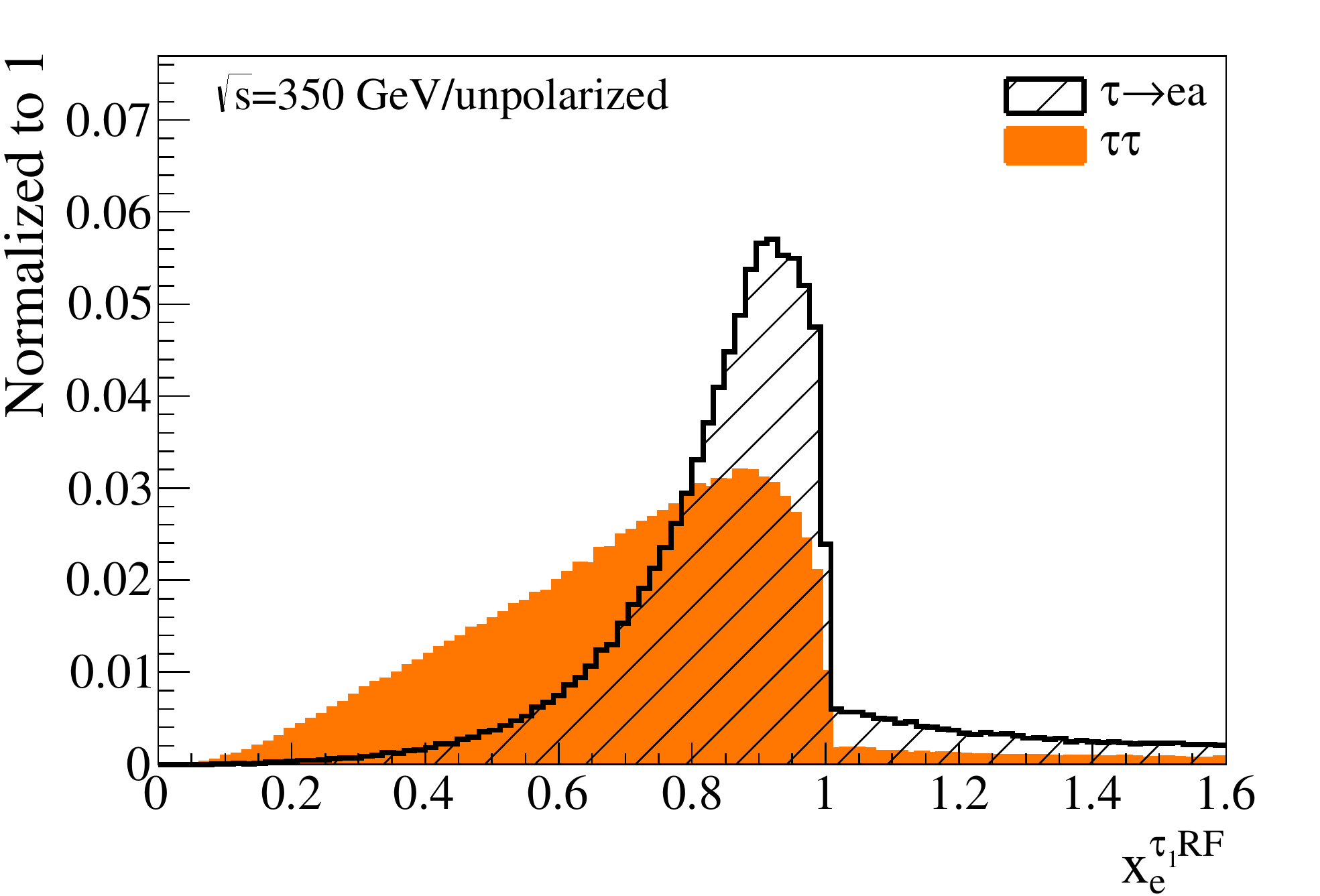}
    \caption{}
    \label{j0}
    \end{subfigure}
\caption{Distributions of the discriminating variables obtained for the unpolarized muon beams case assuming $c_{\tau  e}\neq 0$, $\sqrt{s}=350$ GeV and $m_a=1$ MeV. The ALP couples to right-handed leptons. All processes including the ALP production process ($\tau\ra e\, a$) and the SM backgrounds are shown.}
  \label{0}
\end{figure*}
In the presented plots, the distributions of SM backgrounds are normalized to $\sigma\times\epsilon\times L$, where $\sigma$ is the process cross section, $\epsilon$ is the event selection efficiency and $L$ is the integrated luminosity, and the total background is then normalized to unity. Furthermore, the signal distribution is normalized to unity. Distributions shown in the rest of the paper also use the same normalization convention. The obtained distributions indicate that the $\tau^+\tau^-$ production is the most severe SM background (which is expected as discussed in Section \ref{sec:productionprocess}). Fig. \ref{j0} shows the distribution of the energy fraction of the isolated lepton (electron or positron) in the $\tau_1$ rest frame $x_e^{\tau_1\mathrm{RF}}$ for the ALP production and the SM $\tau^+\tau^-$ production processes alone. As seen, the peaks of the two distributions coincide with each other. This is also the case for other introduced variables. Fig. \ref{cls0} shows the resulting BDT output for different processes.
\begin{figure}[!h]
  \centering
  \includegraphics[width=0.48\textwidth]{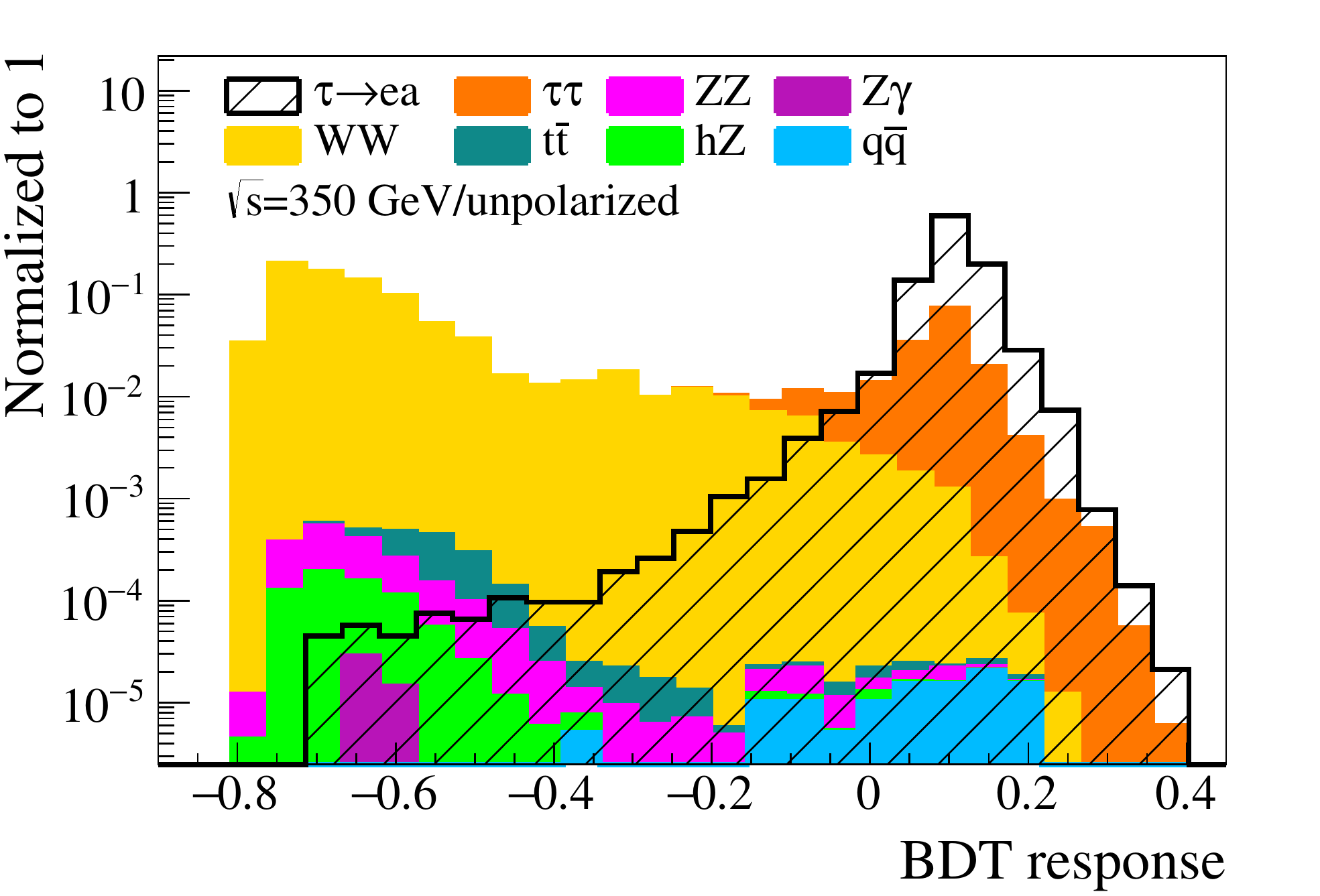}
  \caption{BDT response obtained for the ALP production process and the SM backgrounds. It is assumed that the muon beams are unpolarized, the ALP couples to right-handed leptons, $c_{\tau  e}\neq 0$, $\sqrt{s}=350$ GeV and $m_a=1$ MeV.}
\label{cls0}
\end{figure}
As seen, although a good discrimination is achieved for a significant amount of background, the new physics signal is overwhelmed by the SM $\tau^+\tau^-$ production in the case of unpolarized muon beams.

In case the initial muon beams are polarized, the signal-background discrimination can be improved using the kinematical variables of the decay of polarized tau leptons. In this case, tau leptons are highly polarized in a certain region of phase space. At the center-of-mass energies of 126, 350 and 1500 GeV, $\tau^-$ ($\tau^+$) leptons emitted in the forward (backward) region are highly polarized and thus help suppress the $\tau^+\tau^-$ background as discussed previously in Section \ref{sec:productionprocess}. In the $c_{\tau  e}\neq 0$ case, events with the isolated electron (positron) being forwardly (backwardly) directed are therefore of great interest. Similarly, in the $c_{\tau  \mu}\neq 0$ case, events  with forwardly (backwardly) directed isolated muon (antimuon) are important. As discussed in detail in Sections \ref{sec:BSMtaudecay} and \ref{sec:SMtaudecay}, in the leptonic SM decays of the $\tau^-$ ($\tau^+$), the final state leptons $\ell^-$ ($\ell^+$) directed close to the direction (opposite direction) of the polarization vector of the decaying tau are mostly separated from the leptons produced in the LFV decay $\tau^\pm  \to \ell^\pm  \, a$ which peak at $x_{\ell}\simeq 1$. The polarized decays with the final lepton being directed close to these directions can therefore improve the sensitivity. The forwardly (backwardly) directed $\tau^-$ ($\tau^+$) leptons are highly polarized with a mean polarization $>+0.9$ ($<-0.9$) assuming $+0.8$ ($-0.8$) polarization for the initial $\mu^-$ ($\mu^+$) beam. Consequently, the direction of the momentum of the decaying tau (for both the $\tau^-$ and $\tau^+$) is the desired direction around which the final state $\ell^\pm$ leptons from the SM and LFV tau decays are mostly separated. The signal-background discrimination for events with final state $\ell^\pm$ leptons being directed close to the momentum direction of the decaying tau is therefore expected to be improved in the polarized muon beams case compared with the unpolarized case. Fig. \ref{80} shows the distributions obtained for the discriminating variables in the polarized case assuming $c_{\tau  e}\neq 0$ and $\sqrt{s}=350$ GeV.
\begin{figure*}[!t]
  \centering  
    \begin{subfigure}[b]{0.48\textwidth} 
    \centering
    \includegraphics[width=\textwidth]{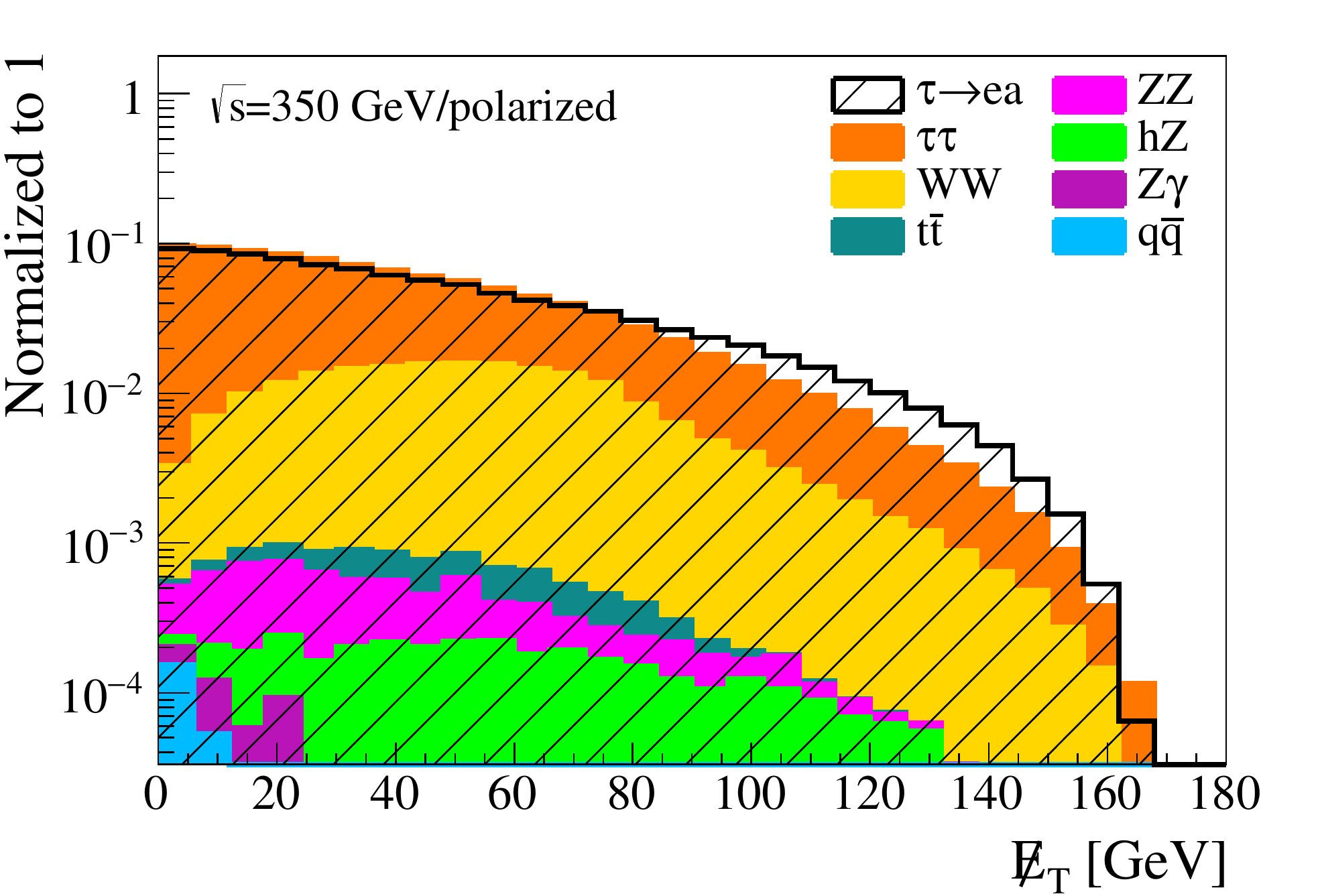}
    \caption{}
    \label{a80}
    \end{subfigure} 
    \begin{subfigure}[b]{0.48\textwidth}
    \centering
    \includegraphics[width=\textwidth]{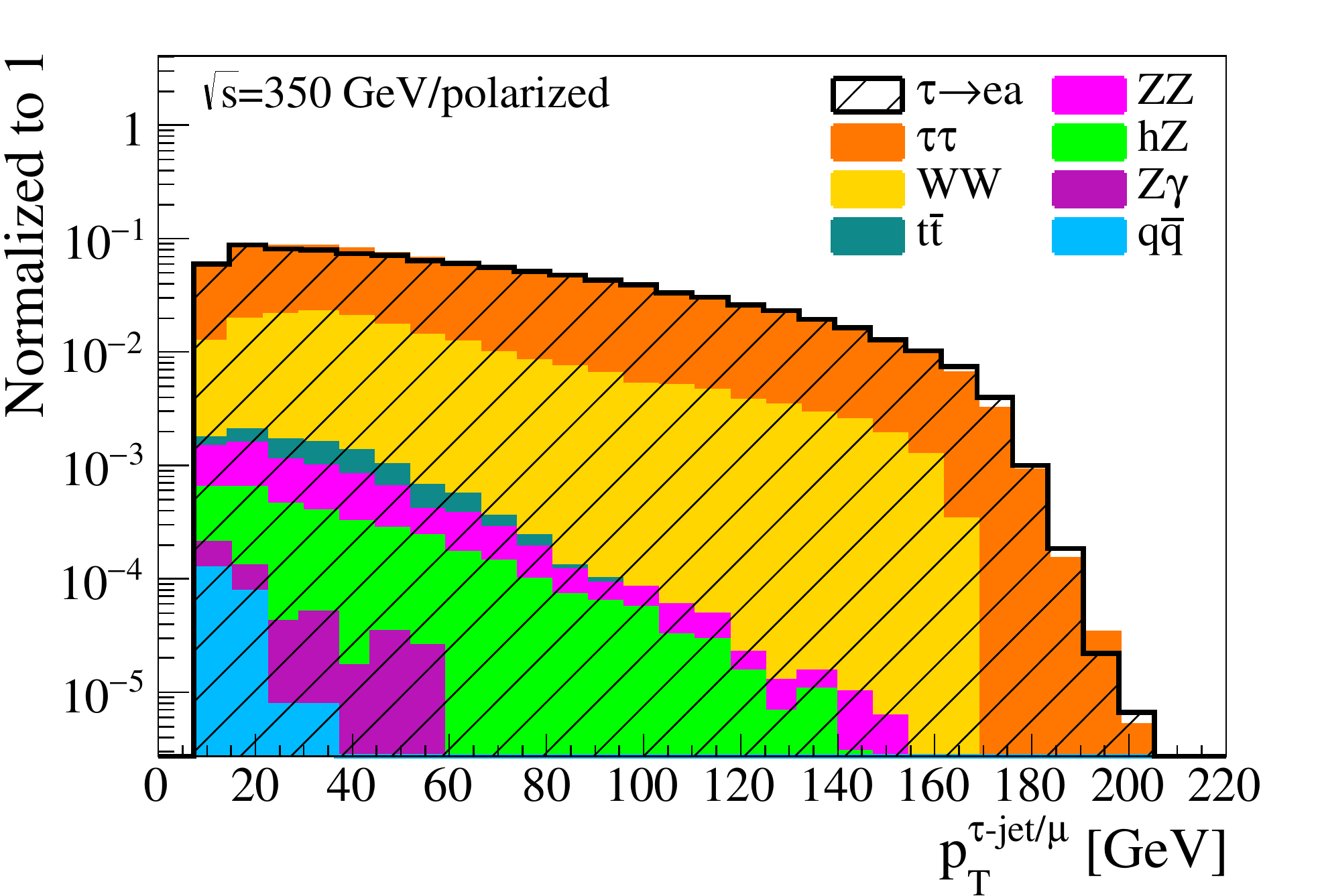}
    \caption{}
    \label{b80} 
    \end{subfigure} 
    \begin{subfigure}[b]{0.48\textwidth}
    \centering
    \includegraphics[width=\textwidth]{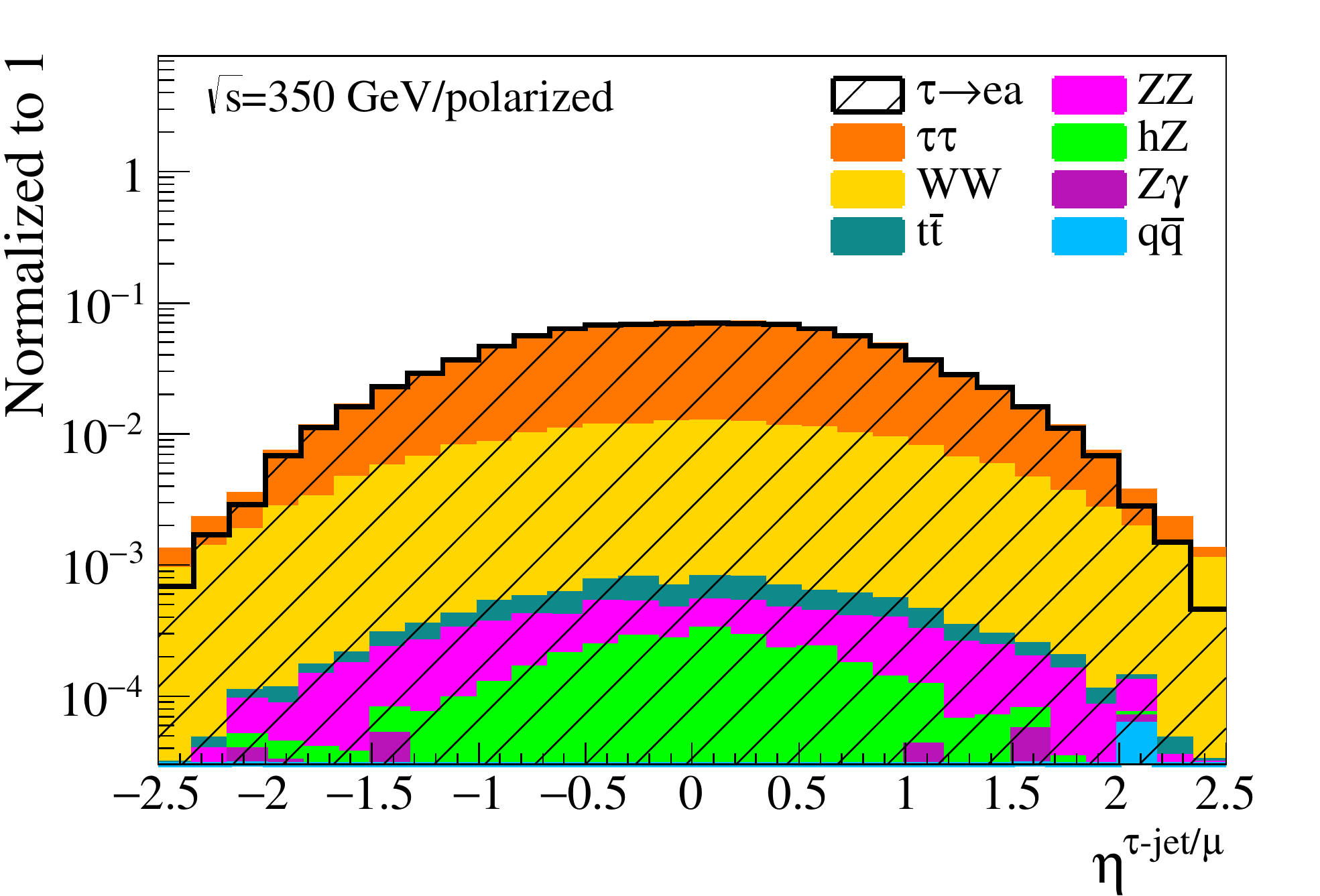}
    \caption{}
    \label{c80}
    \end{subfigure}
    \begin{subfigure}[b]{0.48\textwidth}
    \centering
    \includegraphics[width=\textwidth]{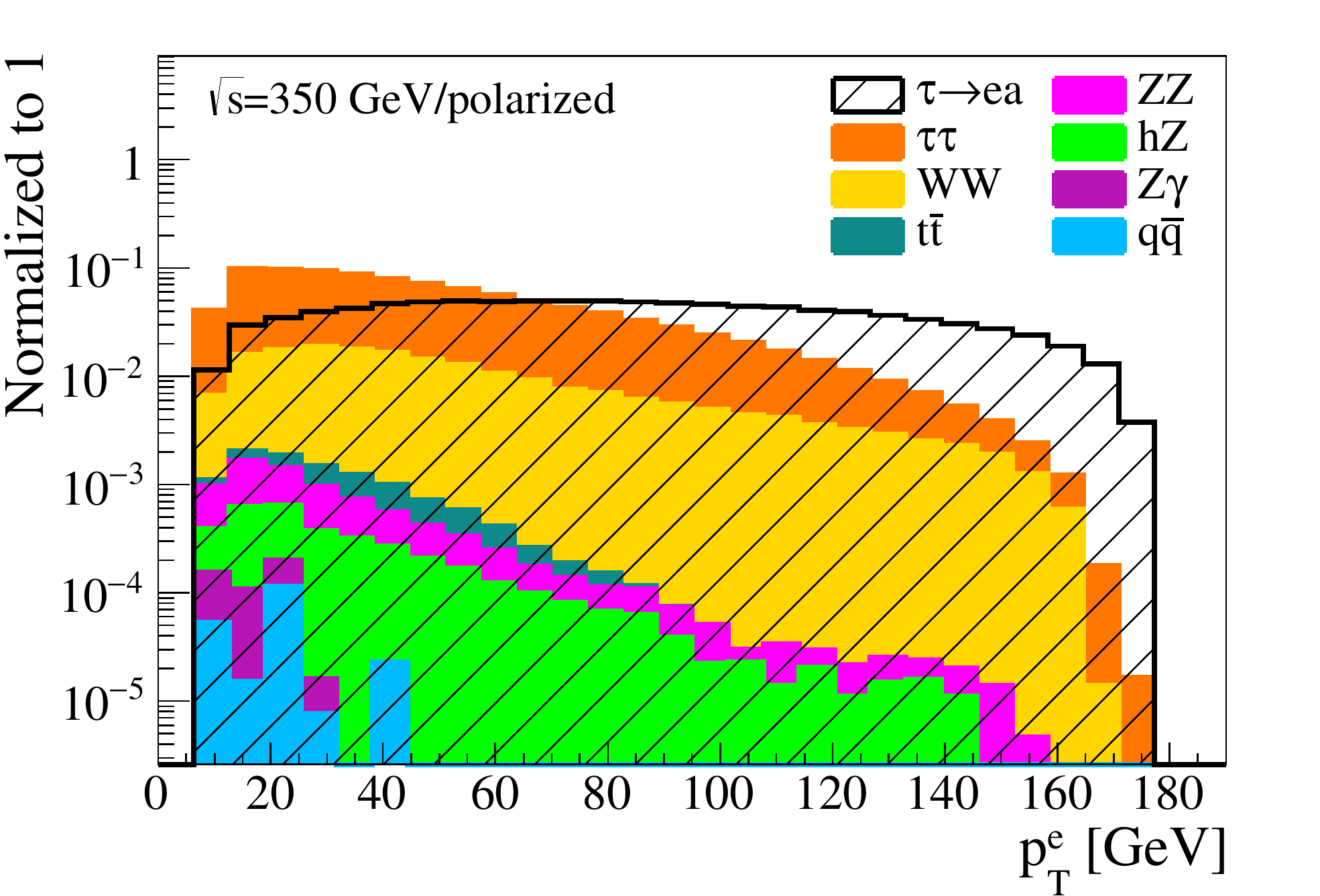}
    \caption{}
    \label{d80}
    \end{subfigure}
    \begin{subfigure}[b]{0.48\textwidth}
    \centering
    \includegraphics[width=\textwidth]{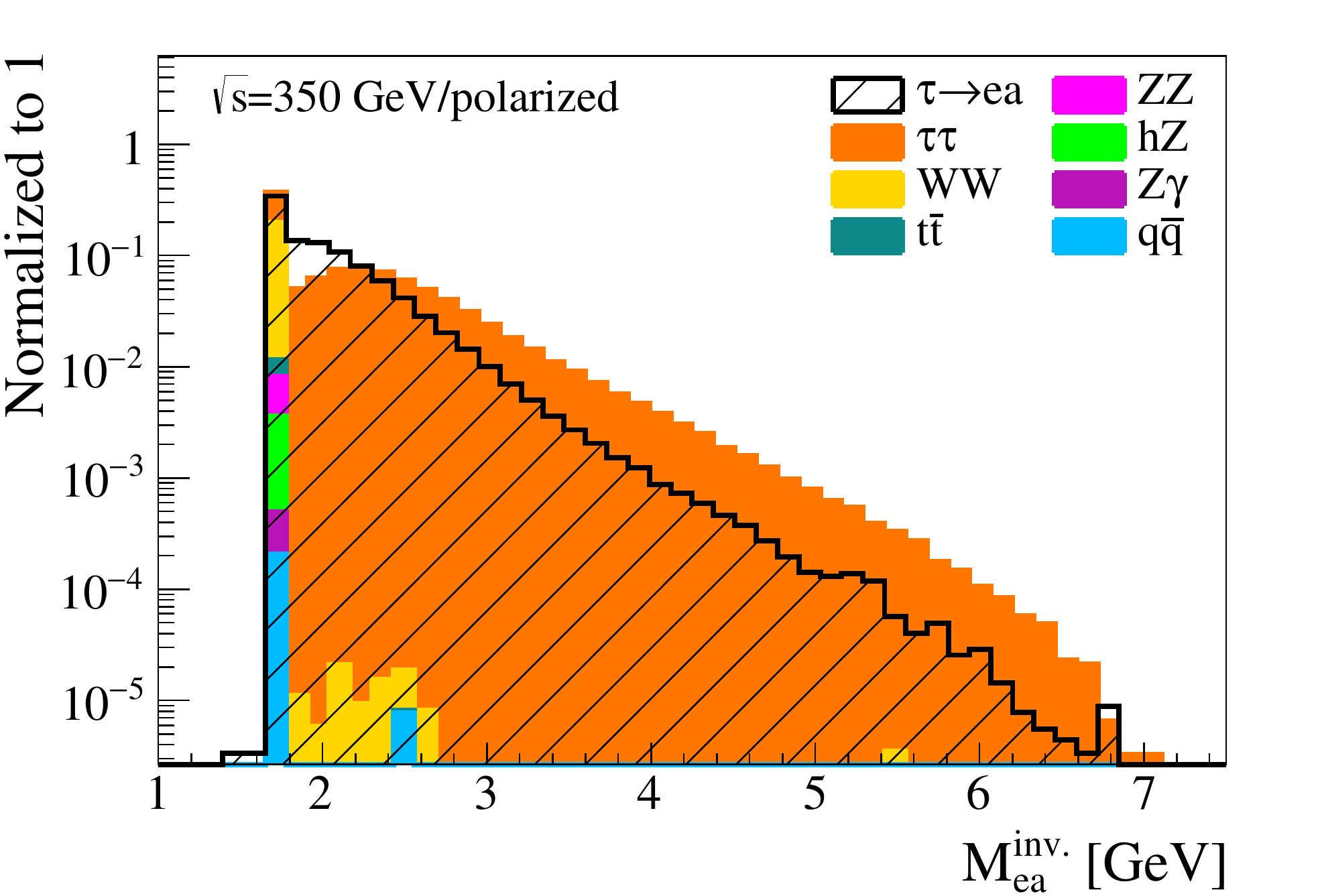}
    \caption{}
    \label{e80}
    \end{subfigure}
\begin{subfigure}[b]{0.48\textwidth}
    \centering
    \includegraphics[width=\textwidth]{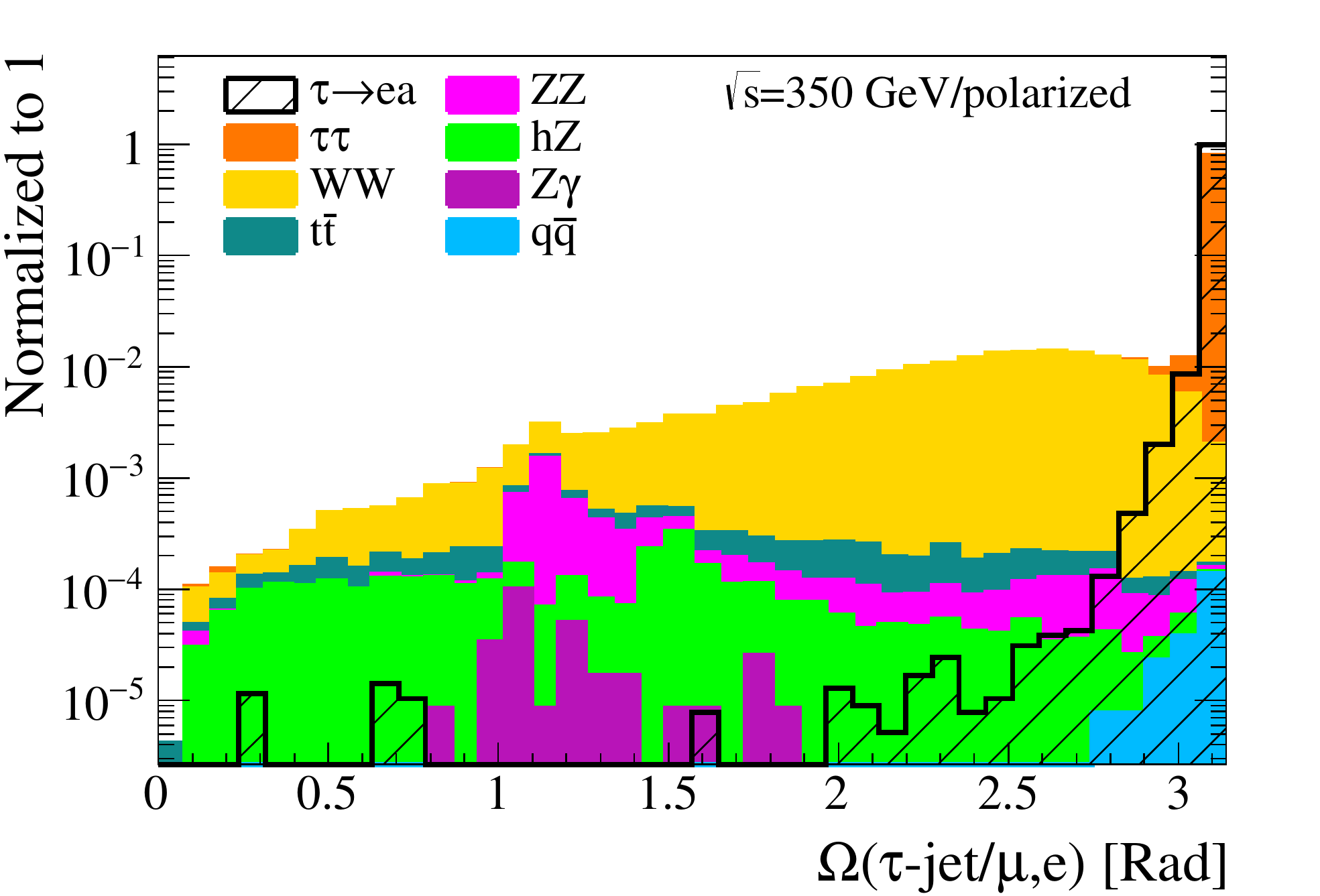}
    \caption{}
    \label{f80}
    \end{subfigure}
\caption{Distributions of the discriminating variables obtained for the ALP production process and the SM backgrounds. It is assumed that $c_{\tau  e}\neq 0$, $\sqrt{s}=350$ GeV, $m_a=1$ MeV, the muon beams are polarized and that the ALP couples to right-handed leptons. See text for further details.}
\label{80}
\end{figure*}
\begin{figure*}[!t]\ContinuedFloat 
    \begin{subfigure}[b]{0.48\textwidth}
    \centering
    \includegraphics[width=\textwidth]{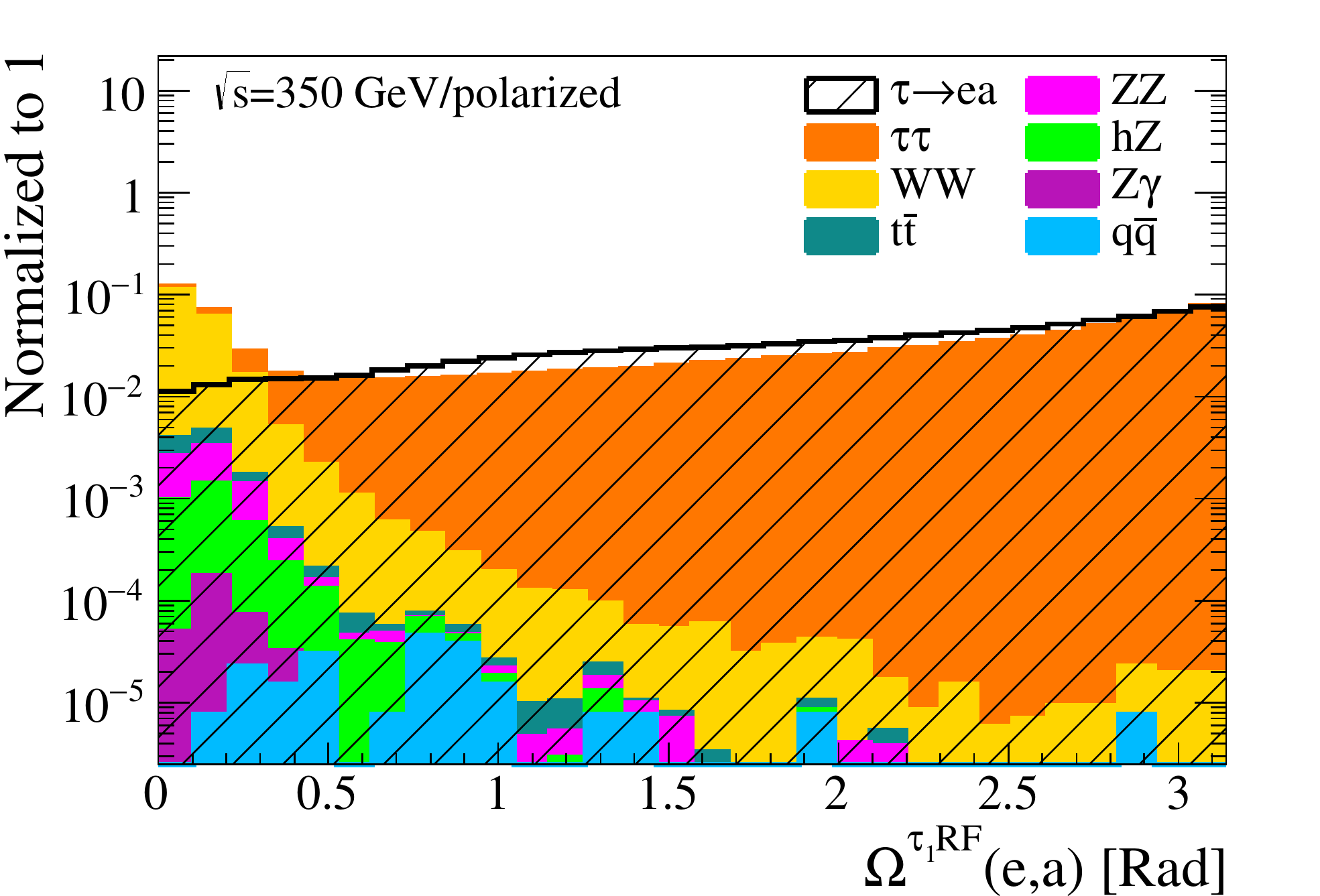}
    \caption{}
    \label{g80}
    \end{subfigure}
    \begin{subfigure}[b]{0.48\textwidth}
    \centering
    \includegraphics[width=\textwidth]{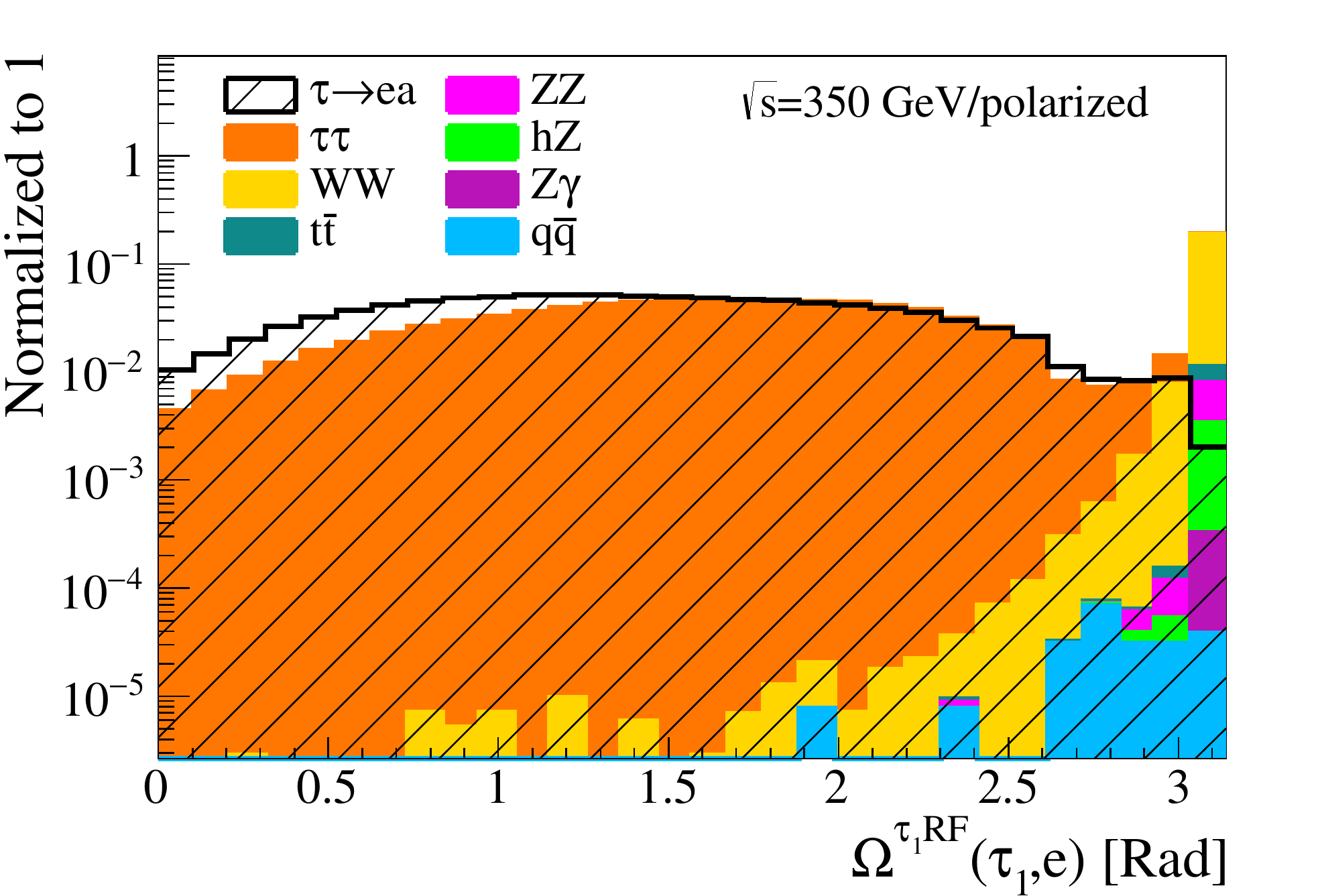}
    \caption{}
    \label{h80}
    \end{subfigure}
    \begin{subfigure}[b]{0.48\textwidth}
    \centering
    \includegraphics[width=\textwidth]{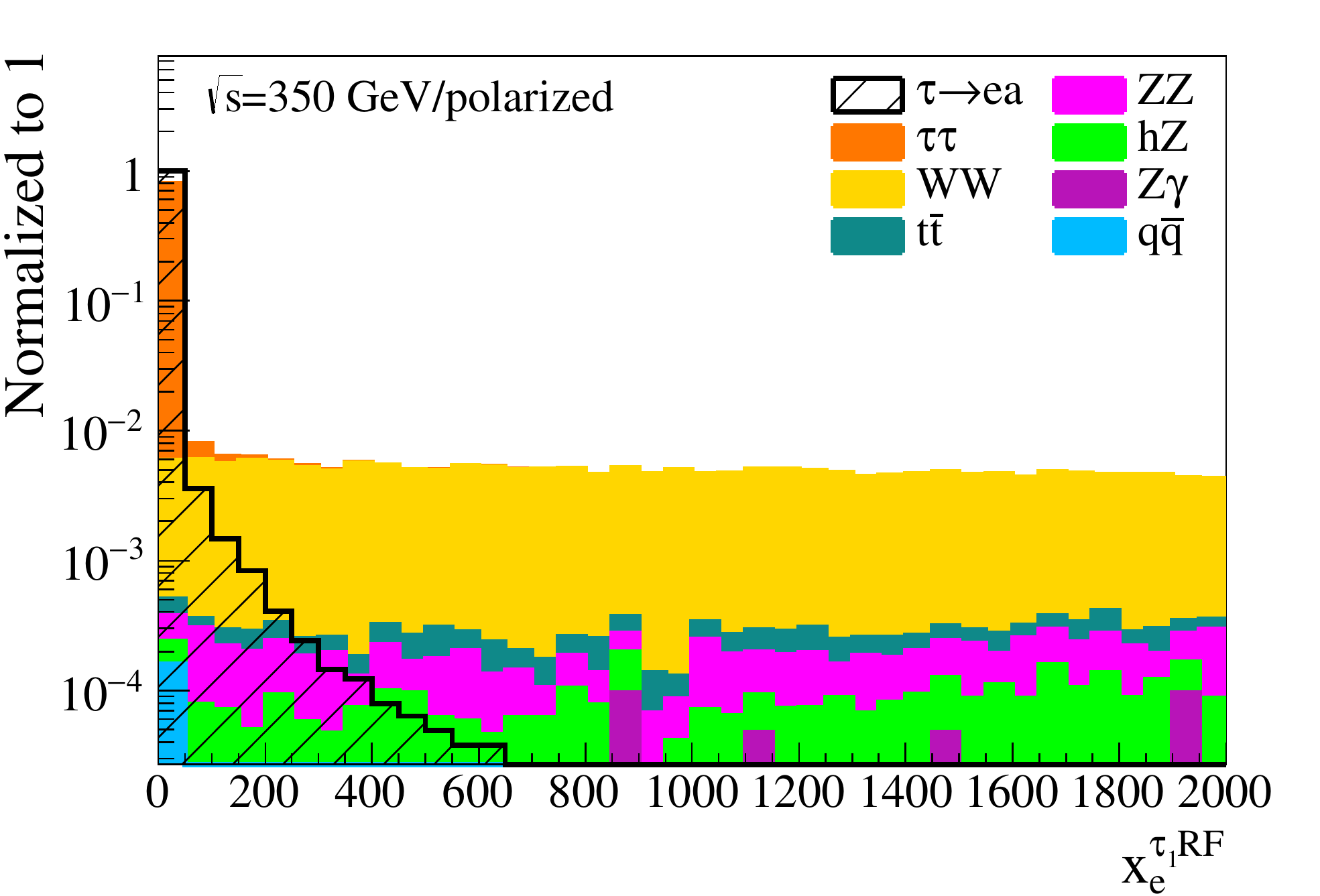} 
    \caption{}   
    \label{i80}
    \end{subfigure}
    \begin{subfigure}[b]{0.48\textwidth}
    \centering
    \includegraphics[width=\textwidth]{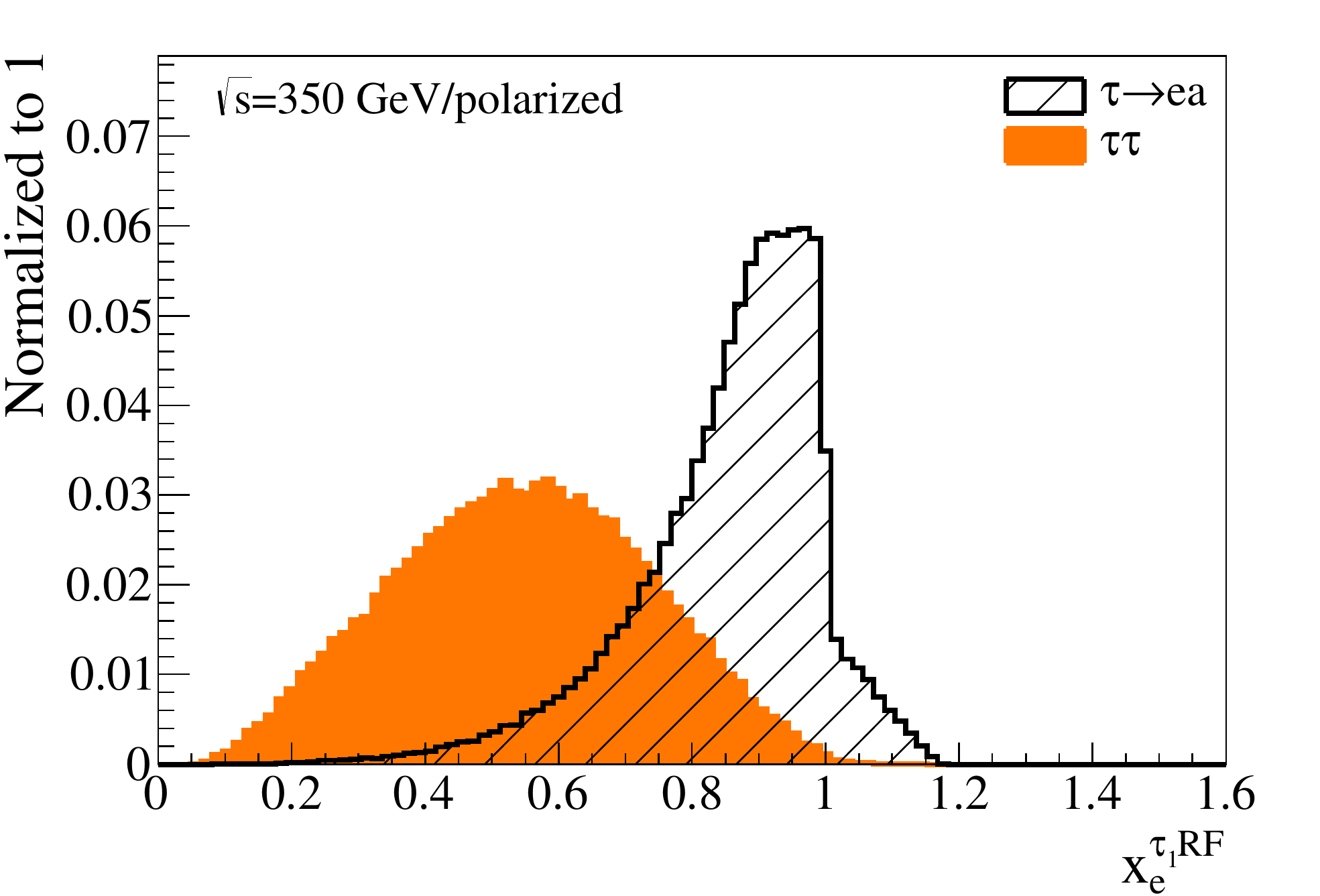} 
    \caption{}   
    \label{j80}
    \end{subfigure}
\caption{Distributions of the discriminating variables obtained for the ALP production process and the SM backgrounds. It is assumed that $c_{\tau  e}\neq 0$, $\sqrt{s}=350$ GeV, $m_a=1$ MeV, the muon beams are polarized and that the ALP couples to right-handed leptons. See text for further details.}
  \label{80}
\end{figure*}
Fig. \ref{j80} shows the distribution of the energy fraction of the isolated electron (or positron) in the $\tau_1$ rest frame in a certain region of phase space. The phase space considered to obtain the distributions shown in this figure is restricted by the following two conditions. First, only events with forwardly (backwardly) directed final electron (positron) are allowed, and second, the deviation of the electron or positron direction in the $\tau_1$ rest frame from the direction of $\tau_1$ in the laboratory frame should satisfy the condition $\Omega^{\tau_1\mathrm{RF}}(\tau_1,e)<\pi/4$ radians. The former condition restricts the phase space to a region with highly polarized tau leptons, and the latter condition keeps the deviation angle $\Omega^{\tau_1\mathrm{RF}}(\tau_1,e)$ in a limited range near zero. The $\pi/4$ upper limit on the deviation angle is just a choice and has not been deduced from any criterion. As seen in Fig. \ref{j80} and as expected, the peak of the energy fraction spectrum moves to the left for the $\tau^+\tau^-$ production process and doesn't coincide with the ALP production peak in the assumed region of phase space. This is in contrary to the unpolarized muon beams case where the peak of the $\tau^+\tau^-$ background coincides with the signal peak regardless of the region of phase space which is considered (compare Figs. \ref{j80} and \ref{j0}). The separated energy spectrums in the polarized muon beams case provide a significant discriminating power and help suppress the SM $\tau^+\tau^-$ background. The distributions in Figs. 10a-i are obtained by a full phase space analysis. The BDT output obtained using the distributions in Figs. 10a-i is shown in Fig. \ref{cls80}.
\begin{figure*}[!t]
  \centering  
    \begin{subfigure}[b]{0.48\textwidth} 
    \centering
    \includegraphics[width=\textwidth]{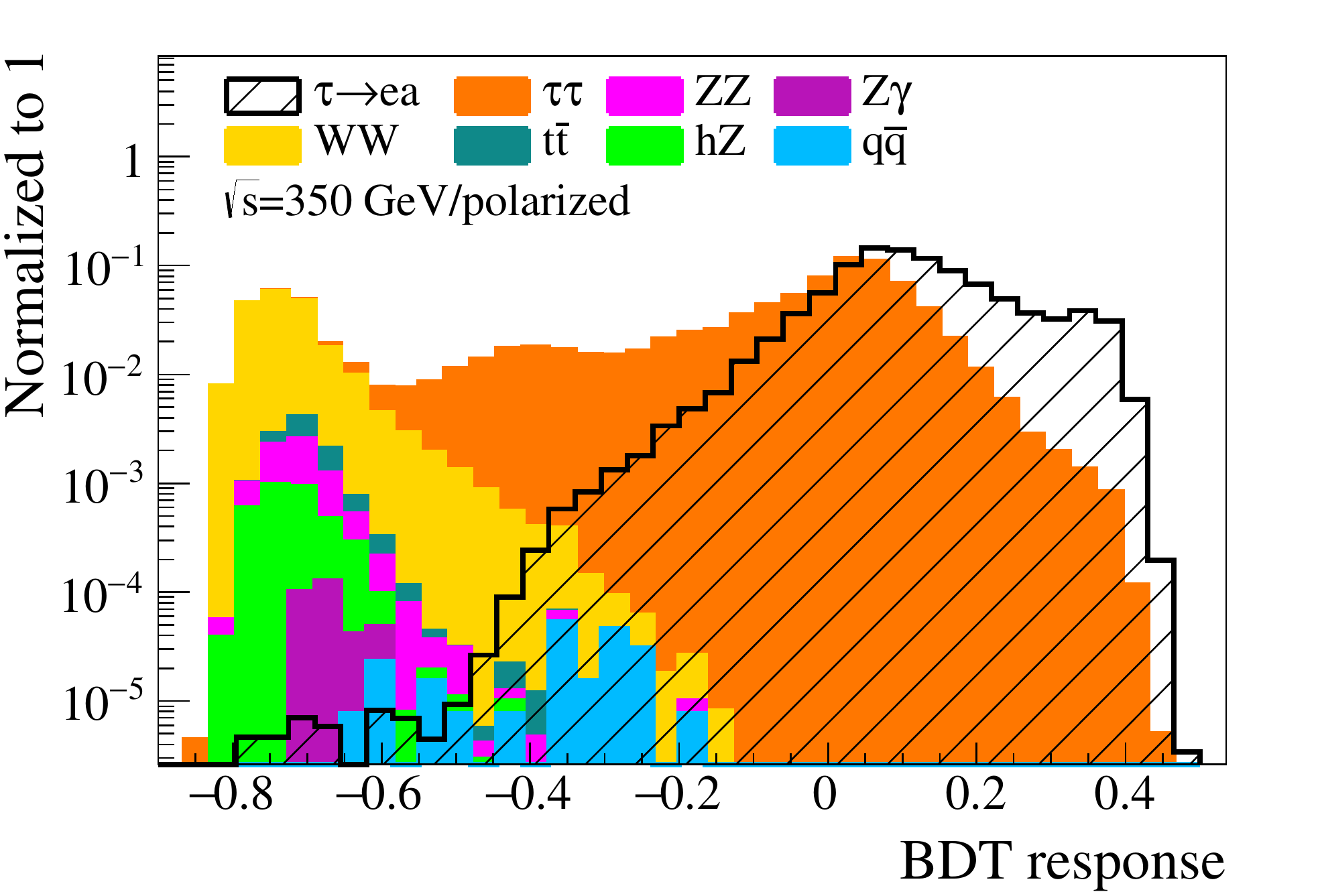}
    \caption{}
    \label{cls80}
    \end{subfigure} \\
    \begin{subfigure}[b]{0.48\textwidth}
    \centering
    \includegraphics[width=\textwidth]{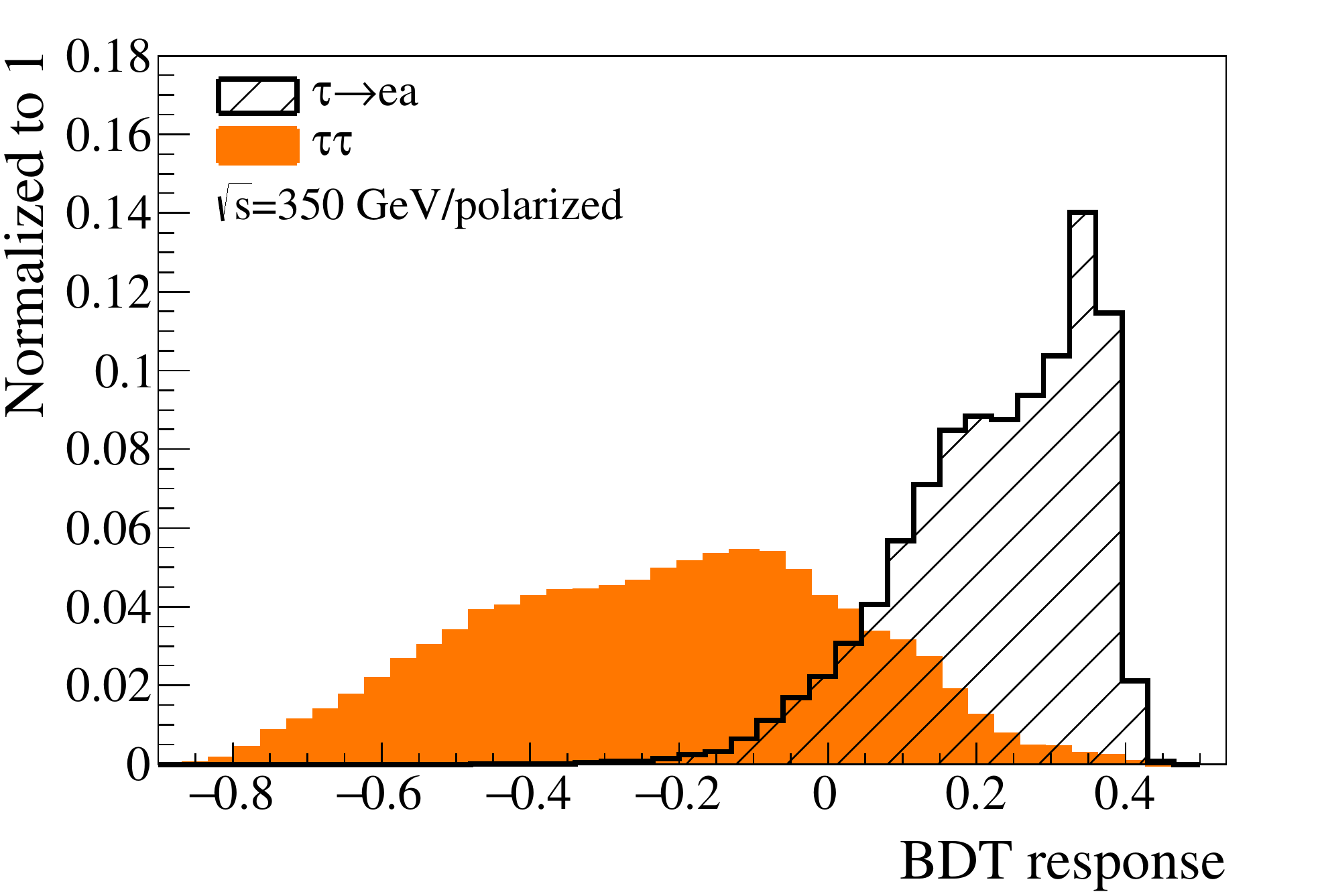}
    \caption{}
    \label{cls-in80} 
    \end{subfigure} 
    \begin{subfigure}[b]{0.48\textwidth}
    \centering
    \includegraphics[width=\textwidth]{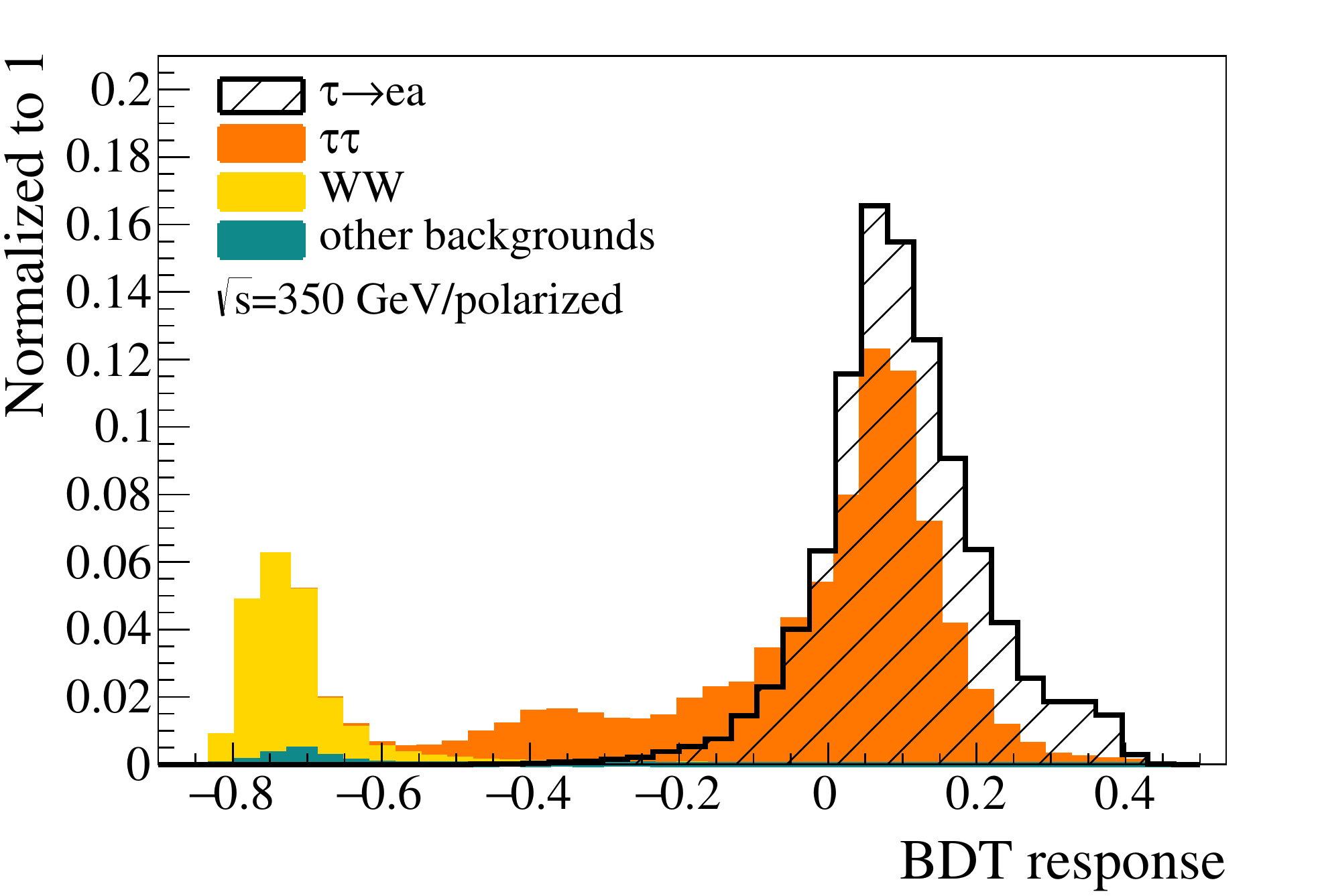}
    \caption{}
    \label{cls-out80}  
    \end{subfigure} 
    \caption{a) BDT response obtained for the ALP production process and the SM backgrounds assuming polarized muon beams. This result corresponds to $c_{\tau  e}\neq 0$, $\sqrt{s}=350$ GeV, $m_a=1$ MeV and the assumption that the ALP couples to right-handed leptons. Using the two conditions imposed to obtain the distributions of Fig. \ref{j80}, the distribution in a) is divided into two complementary distributions shown in b) and c). See text for further details.}
  \label{}
\end{figure*}
It is seen that in the polarized muon beams case, the discrimination between the ALP production process and the background is improved compared with the unpolarized muon beams case (compare Figs. \ref{cls0} and \ref{cls80}). This improvement mainly results from events belonging to a region of phase space where the $x_e^{\tau_1\mathrm{RF}}$ distributions of the signal and $\tau^+\tau^-$ production background are separated so that the $\tau^+\tau^-$ background can be suppressed. To see this more clearly, we divide the BDT response shown in Fig. \ref{cls80}, which is obtained by a full phase space analysis, into two parts corresponding to two complementary phase space regions. To do so, we use the aforementioned two conditions (on the angular orientation of the final state charged lepton and the deviation angle $\Omega^{\tau_1\mathrm{RF}}(\tau_1,e)$) assumed to obtain the distributions in Fig. \ref{j80}. The BDT response distributions for events satisfying the two above-mentioned conditions are shown in Fig. \ref{cls-in80}, and the BDT response distributions for events which don't satisfy at least one of these conditions are shown in Fig. \ref{cls-out80}. The distributions in Figs. \ref{cls-in80} and \ref{cls-out80} are complementary and their addition according to their respective weights forms the total distribution shown in Fig. \ref{cls80}. As seen, only the distributions of the signal and $\tau^+\tau^-$ production processes are shown in Fig. \ref{cls-in80}. This is because of the imposed two conditions which completely remove all the backgrounds other than the $\tau^+\tau^-$ production. As expected, the BDT response distributions in Fig. \ref{cls-in80}, which correspond to the phase space assumed to obtain the distributions in Fig. \ref{j80}, provide a much better signal-background discrimination compared with the BDT response distributions in Fig. \ref{cls-out80}. The BDT response shown in Fig. \ref{cls-out80} corresponds to a region of phase space where the $x_e^{\tau_1\mathrm{RF}}$ distributions of the signal and $\tau^+\tau^-$ background are less separated. As a result, the BDT response in this figure is similar to the BDT response obtained in the unpolarized muon beams case (shown in Fig. \ref{cls0}) in the sense that the peaks of the signal and $\tau^+\tau^-$ background distributions coincide. 

In case the polarization of the initial muon beams is reversed, i.e. $\mu^-$ ($\mu^+$) beam is $-0.8$ ($+0.8$) polarized, forwardly (backwardly) directed $\tau^-$ ($\tau^+$) leptons are highly polarized with a mean polarization $<-0.9$ ($>+0.9$) as discussed in Section \ref{sec:mumutautau}. The polarized decays due to such tau leptons can also help discriminate between the signal and the SM $\tau^+\tau^-$ background in a similar way to the case discussed above. The only difference is that as the polarizations of the produced tau leptons are reversed, the desired direction (around which the final state $\ell^\pm$ leptons from the SM and LFV tau decays are mostly separated) is the opposite direction of the momentum of the decaying tau (for both $\tau^-$ and $\tau^+$). It is therefore expected that imposing the condition $\Omega^{\tau_1\mathrm{RF}}(\tau_1,e)>3\pi/4$ radians in this case would yield a similar $x_e^{\tau_1\mathrm{RF}}$ distribution to that of Fig. \ref{j80} which has been obtained with reversely polarized muon beams and the condition $\Omega^{\tau_1\mathrm{RF}}(\tau_1,e)<\pi/4$ radians. Similar signal-background discriminations and sensitivities are therefore expected for the two cases with reverse muon beams polarizations. In this study, we only consider the case where $\mu^-$ and $\mu^+$ beams are respectively $+0.8$ and $-0.8$ polarized and provide the corresponding results.

%***********************************************************
\section{Prospects for constraints on the LFV couplings}
\label{sec:prospects}
We compute expected upper limits on the LFV couplings $c_{\tau e}/f_a$ and $c_{\tau \mu}/f_a$ at $95\%$ confidence level (CL) using the BDT response distributions. Tabs. \ref{tab:limitstaue} and \ref{tab:limitstaumu} respectively present expected limits on $c_{\tau e}/f_a$ and $c_{\tau \mu}/f_a$ obtained at the center-of-mass energies of 126, 350 and 1500 GeV for the unpolarized and polarized muon beams cases and also for different assumed chiral structures. Limits presented in the parentheses include an overall systematic uncertainty. To take potential systematic uncertainties into account, we consider an overall uncertainty of $10\%$ on the event selection efficiency of each analyzed process (including the signal and all background processes). At $\sqrt{s}=126$ and 1500 GeV, the presented limits correspond to the ALP mass of 1 MeV. At $\sqrt{s}=350$ GeV, different ALP mass scenarios ranging from 100 eV to 1 MeV have been considered. Limits presented for the center-of-mass energies 126, 350 and 1500 GeV correspond to the integrated luminosities 2.5, 189.2 and 394.2 fb$^{-1}$, respectively. The assumed integrated luminosities are total integrated luminosities collected in one year corresponding to the average luminosities provided in Tab. \ref{tab:muoncollider}.
\begin{table}[!t] %%%%%%%%%%%%%% limits
\begin{subtable}{1\textwidth}
\centering
\begin{tabular}{ccccccc}
 &  & {$m_a$ [MeV]}& V+A & V$-$A & V/A & \\
\cline{1-7}
\multirow{2}{*}{126 GeV} & {Unpolarized} & 1 & 9.02 (9.73) & 8.46 (9.12) & 12.24 (13.21) & \parbox[t]{.09mm}{\multirow{26}{*}{\rotatebox[origin=c]{270}{Expected 95$\%$ CL upper limit on $c_{\tau e}/f_a$  [$10^{-5}$ TeV$^{-1}$]}}} \\
\cline{2-6}
    & {Polarized} & 1 & 2.96 (3.18)  & 7.32 (7.90)  & 5.54 (5.97) & \\
\cline{1-6}
\multirow{22}{*}{350 GeV} & \multirow{11}{*}{Unpolarized} & 0.0001 & 6.63 (7.01) & 6.53 (6.90)  & 9.37 (9.90) &  \\
 &  & 0.1 & 6.68 (7.06)  & 6.57 (6.94)  & 9.37 (9.90)  & \\
 &  & 0.2 & 6.69 (7.07)  & 6.52 (6.88)  & 9.39 (9.92)  &  \\
 &  & 0.3 & 6.69 (7.08) & 6.55 (6.91)  & 9.38 (9.91)  & \\
 &  & 0.4 & 6.71 (7.10)  & 6.54 (6.91)  & 9.40 (9.93) &  \\
 &  & 0.5 & 6.71 (7.10)  & 6.55 (6.91)  & 9.36 (9.88) &  \\
 &  & 0.6 & 6.69 (7.08)  & 6.53 (6.90)  & 9.40 (9.93) &  \\
 &  & 0.7 & 6.71 (7.09)  & 6.53 (6.90)  & 9.35 (9.88) &  \\
 &  & 0.8 & 6.70 (7.09)  & 6.52 (6.89)  & 9.37 (9.91) &  \\
 &  & 0.9 & 6.73 (7.12)  & 6.51 (6.87)  & 9.39 (9.91) &  \\
 &  & 1 & 6.70 (7.08)  & 6.52 (6.89)  & 9.34 (9.87) &  \\
\cline{2-6}
& \multirow{11}{*}{Polarized} & 0.0001 & 2.45 (2.57)  & 4.59 (4.92)  & 4.11 (4.34) &  \\
 &  & 0.1 & 2.44 (2.56)  & 4.59 (4.92)  & 4.13 (4.35) &  \\
 &  & 0.2 & 2.40 (2.52)  & 4.57 (4.90)  & 4.16 (4.39) &  \\
 &  & 0.3 & 2.40 (2.52)  & 4.58 (4.91)  & 4.05 (4.27) &  \\
 &  & 0.4 & 2.41 (2.53)  & 4.63 (4.96)  & 4.08 (4.30) &  \\
 &  & 0.5 & 2.41 (2.53)  & 4.58 (4.90)  & 4.16 (4.39) &  \\
 &  & 0.6 & 2.39 (2.51)  & 4.56 (4.88)  & 4.14 (4.37) &  \\
 &  & 0.7 & 2.43 (2.55)  & 4.58 (4.90)  & 4.10 (4.32) &  \\
 &  & 0.8 & 2.40 (2.52)  & 4.56 (4.88)  & 4.16 (4.38) &  \\
 &  & 0.9 & 2.41 (2.53)  & 4.56 (4.88)  & 4.16 (4.38) &  \\
 &  & 1 & 2.36 (2.48)  & 4.58 (4.90)  & 4.14 (4.37) &  \\
\cline{1-6}
\multirow{2}{*}{1500 GeV} & {Unpolarized} & 1 & 12.48 (13.15)  & 12.24 (12.90)  & 17.41 (18.35) &  \\
\cline{2-6}
    & {Polarized} & 1 & 4.80 (5.04)  & 9.12 (9.75)  & 7.14 (7.51) &  \\
\cline{1-7}
\end{tabular}
\caption{}
\label{tab:limitstaue}
\end{subtable}
\caption{Expected $95\%$ CL upper limits on a) $c_{\tau e}/f_a$ and b) $c_{\tau \mu}/f_a$ obtained assuming different center-of-mass energies and ALP mass scenarios for the unpolarized and polarized muon beams cases. The three chiral structures assumed for the ALP coupling, i.e. V+A, V$-$A and V/A, have been analyzed independently and corresponding limits are shown. Limits at the center-of-mass energies of 126, 350 and 1500 GeV correspond to the integrated luminosities of 2.5, 189.2 and 394.2 fb$^{-1}$, respectively. Limits including an overall systematic uncertainty are presented in the parentheses.} 
\label{tab:limits}
\end{table}
\begin{table}[!t]
\ContinuedFloat
\begin{subtable}{1\textwidth}
\centering
\begin{tabular}{ccccccc}
  &  & {$m_a$ [MeV]}& V+A & V$-$A & V/A & \\
\cline{1-7}
\multirow{2}{*}{126 GeV} & {Unpolarized} & 1 & {8.87 (9.57) } & 8.50 (9.17)  & 12.26 (13.23) & \parbox[t]{.09mm}{\multirow{26}{*}{\rotatebox[origin=c]{270}{Expected 95$\%$ CL upper limit on $c_{\tau \mu}/f_a$  [$10^{-5}$ TeV$^{-1}$]}}} \\
\cline{2-6}
    & {Polarized} & 1 & 2.87 (3.08)  & 7.25 (7.83)  & 5.11 (5.51) &  \\
\cline{1-6}
\multirow{22}{*}{350 GeV} & \multirow{11}{*}{Unpolarized} & 0.0001 & 6.63 (7.01)  & 6.46 (6.82)  & 9.25 (9.77) &  \\
 &  & 0.1 & 6.62 (6.99)  & 6.43 (6.79)  & 9.25 (9.77) &  \\
 &  & 0.2 & 6.62 (6.99)  & 6.42 (6.78)  & 9.25 (9.77) &  \\
 &  & 0.3 & 6.60 (6.98)  & 6.44 (6.80)  & 9.26 (9.78) &  \\
 &  & 0.4 & 6.63 (7.01)  & 6.45 (6.81)  & 9.23 (9.74) &  \\
 &  & 0.5 & 6.62 (6.99)  & 6.42 (6.78) & 9.27 (9.80)  & \\
 &  & 0.6 & 6.63 (7.00)  & 6.43 (6.79)  & 9.27 (9.80)  & \\
 &  & 0.7 & 6.63 (7.01)  & 6.38 (6.74)  & 9.27 (9.80)  & \\
 &  & 0.8 & 6.66 (7.04)  & 6.43 (6.78)  & 9.25 (9.77)   &\\
 &  & 0.9 & 6.65 (7.03)  & 6.48 (6.84)  & 9.24 (9.76) &  \\
 &  & 1 & 6.62 (7.00)  & 6.43 (6.79)  & 9.19 (9.71)   &\\
\cline{2-6}
& \multirow{11}{*}{Polarized} & 0.0001 & 2.23 (2.34)  & 4.45 (4.75)  & 3.76 (3.96)  & \\
 &  & 0.1 & 2.15 (2.25)  & 4.46 (4.74)  & 3.77 (3.97)  & \\
 &  & 0.2 & 2.18 (2.29)  & 4.45 (4.75) & 3.79 (3.99)  &\\
 &  & 0.3 & 2.13 (2.23)  & 4.45 (4.75)  & 3.72 (3.92)  & \\
 &  & 0.4 & 2.18 (2.28)  & 4.49 (4.80)  & 3.76 (3.96)  & \\
 &  & 0.5 & 2.17 (2.28)  & 4.38 (4.68)  & 3.75 (3.95)  & \\
 &  & 0.6 & 2.14 (2.24)  & 4.47 (4.78)  & 3.76 (3.95)  & \\
 &  & 0.7 & 2.21 (2.31)  & 4.47 (4.78)  & 3.57 (3.75)  & \\
 &  & 0.8 & 2.15 (2.25)  & 4.44 (4.74)  & 3.75 (3.93)  & \\
 &  & 0.9 & 2.10 (2.20)  & 4.50 (4.81)  & 3.77 (3.98)  & \\
 &  & 1 & 2.17 (2.27)  & 4.38 (4.68)  & 3.74 (3.92)  & \\
\cline{1-6}
\multirow{2}{*}{1500 GeV} & {Unpolarized} & 1 & 12.29 (12.95)  & 12.13 (12.77)  & 17.29 (18.22)  & \\
\cline{2-6}
    & {Polarized} & 1 & 4.49 (4.71)  & 8.84 (9.44)  & 7.86 (8.27)  & \\
\cline{1-7}
\end{tabular}
\caption{}
\label{tab:limitstaumu}
\end{subtable}
\caption{Expected $95\%$ CL upper limits on a) $c_{\tau e}/f_a$ and b) $c_{\tau \mu}/f_a$ obtained assuming different center-of-mass energies and ALP mass scenarios for the unpolarized and polarized muon beams cases. The three chiral structures assumed for the ALP coupling, i.e. V+A, V$-$A and V/A, have been analyzed independently and corresponding limits are shown. Limits at the center-of-mass energies of 126, 350 and 1500 GeV correspond to the integrated luminosities of 2.5, 189.2 and 394.2 fb$^{-1}$, respectively. Limits including an overall systematic uncertainty are presented in the parentheses.} 
\end{table}

As seen in Tabs. \ref{tab:limitstaue} and \ref{tab:limitstaumu}, the expected limits on $c_{\tau e}/f_a$ and $c_{\tau \mu}/f_a$ obtained for the polarized muon beams case are significantly stronger than the limits obtained in the case of unpolarized muon beams. This is the case for all the three studied center-of-mass energies and also all the three assumed chiral structures. This improvement is due to the suppression of the main SM background with the help of the energy $x_\ell^{\tau_1\mathrm{RF}}$ ($\ell=e,\mu$) spectrums as discussed before. It is also seen that, in the polarized muon beams case, different chiral structures assumed for the ALP LFV coupling lead to different sensitivities. The best and worst limits respectively correspond to the V+A and V$-$A structures and the limit for the V/A case has a moderate value. As discussed in Section \ref{sec:SMtaudecay}, this difference in the limits obtained for different chiral structures is expected and stems from the difference in the angular orientations of the final charged lepton in the LFV and SM tau decays. In the case of unpolarized muon beams, the limits obtained for the V+A and V$-$A cases are similar and stronger than the limit obtained for the V/A case. The relative weakness of the limit for the V/A case results from the smaller cross section of the ALP production in this case. Integrating Eq. \ref{eq:decaychiral} with respect to $\theta$, it is seen that the total width of the decay $\tau  \to \ell  a$ in the V/A case is half of the total width in the V+A and V$-$A cases. The ALP production cross sections in these cases also follow the same ratio and thus the limit for the V/A case is degraded. The degradation of the limit for the V/A case, as a result of the smaller cross section, also occurs in the case of polarized muon beams. However, in this case, it is fully compensated by the discriminating power provided by the kinematical variable $\Omega^{\tau_1\mathrm{RF}}(\tau_1,\ell)$ and the limit for the V/A case becomes stronger than the limit for V$-$A case.

Results presented in Tabs. \ref{tab:limitstaue} and \ref{tab:limitstaumu} show that the expected limits on $c_{\tau e}/f_a$ and $c_{\tau \mu}/f_a$ obtained at the center-of-mass energy of 350 GeV (corresponding to the Top-High Luminosity operating stage of the assumed muon collider) are more stringent than the limits obtained at the center-of-mass energies 126 and 1500 GeV. The main reason lies in the combination of the center-of-mass energy and the integrated luminosity at this collider operating stage that enhances the signal statistics. It is also seen that the limits obtained for different ALP mass scenarios at $\sqrt{s}=350$ GeV are similar with nonsignificant discrepancies. This can be understood as a result of the smallness of the ALP mass in the considered range ($m_a\leq1$ MeV) when compared with the process center-of-mass energy and the mass of tau lepton ($\approx 1.777$ GeV). The cross section of the ALP production and also the distributions of the kinematical variables don't change significantly for different ALP masses in this mass range and thus the limit is not sensitive to the ALP mass.

Assuming unpolarized muon beams and an ALP mass of 1 MeV, the strongest expected limits (including systematic uncertainties) on $c_{\tau e}/f_a$ ($c_{\tau \mu}/f_a$) obtained in this work for the V+A, V$-$A and V/A chiral structures are 7.08$\times10^{-5}$ (7.00$\times10^{-5}$), 6.89$\times10^{-5}$ (6.79$\times10^{-5}$) and 9.87$\times10^{-5}$ (9.71$\times10^{-5}$) $\mathrm{TeV}^{-1}$, respectively (see Tabs. \ref{tab:limitstaue} and \ref{tab:limitstaumu}). The corresponding limits obtained in the case of polarized muon beams are $2.48\times10^{-5}$ ($2.27\times10^{-5}$), $4.90\times10^{-5}$ ($4.68\times10^{-5}$) and $4.37\times10^{-5}$ ($3.92\times10^{-5}$) $\mathrm{TeV}^{-1}$. Based on our recast of the present experimental limits obtained by the ARGUS collaboration \cite{ARGUSlimit}, the limits corresponding to the V+A/V$-$A (V/A) chiral structure for $\leq1$ MeV ALP masses are $c_{\tau e}/f_a < 3.3\times 10^{-4}$ ($c_{\tau e}/f_a < 4.7\times 10^{-4}$) and $c_{\tau \mu}/f_a < 4.3 \times 10^{-4}$ ($c_{\tau \mu}/f_a < 6.1 \times 10^{-4}$) TeV$^{-1}$. Comparing the expected limits obtained in this study with the present experimental limits shows that the limits obtained in both the unpolarized and polarized muon beams cases in this work are significantly stronger than the present experimental limits, and the present limits can be improved by roughly one order of magnitude with the help of the present analysis.

%***********************************************************
\section*{Summary and conclusions} 
\label{sec:conclusions}
Spontaneously broken global $U\mathrm{(1)}$ symmetries produce CP-odd scalars called Axion-Like Particles (ALPs). Such particles provide the possibility to address some of the long-lasting SM problems. ALPs can have lepton-flavor-violating (LFV) couplings to the SM charged leptons. Recently, there have been some proposals for an explanation of the anomalous magnetic dipole moments of muon and electron using LFV ALPs. ALPs with LFV couplings can also solve the strong CP problem if couple to gluons. Motivated by the numerous applications that ALPs have found in many areas, we have studied the capability of a future muon collider suggested by the Muon Accelerator Program (MAP) to search for LFV ALPs production. The production process assumed in this work is the muon-anti muon annihilation into a tau lepton pair followed by the LFV decay $\tau\ra\ell a$ ($\ell=e,\mu$) of one of the tau leptons. With the use of some suitable discriminating variables and a multivariate technique, we tried to discriminate between the signal and backgrounds. We have obtained expected $95\%$ CL upper limits (including systematic uncertainties) on the ALP LFV couplings $c_{\tau e}/f_a$ and $c_{\tau \mu}/f_a$ for light ALPs ($m_a\leq1$ MeV) assuming three different chiral structures for the ALP LFV coupling. The obtained limits have been computed assuming the center-of-mass energies of 126, 350 and 1500 GeV, which respectively correspond to the integrated luminosities of 2.5, 189.2 and 394.2 fb$^{-1}$. The obtained limits are significantly stronger than present experimental limits. However, it was seen that, when the colliding muon beams are unpolarized, the ALP production is overwhelmed by the SM $\tau^+\tau^-$ production background although the rest of backgrounds are significantly suppressed. Assuming $-$0.8/+0.8 (or +0.8/$-$0.8) polarized $\mu^-\mu^+$ beams, we have suggested a procedure which utilizes tau polarization-induced effects to improve the sensitivity of the search. Highly polarized tau leptons resulting in such effects can be produced in the collision of polarized muon beams. The employed procedure is based on the differences in properties of the polarized LFV and SM tau decays. In the LFV decay $\tau\ra\ell a$, the lepton energy fraction $x_{\ell}\equiv2E_{\ell}/m_\tau \simeq 1$ and is independent of the momentum direction of the final lepton. However, in the SM decay $\tau\ra \ell\nu \bar{\nu}$, the energy fraction distribution of the final charged lepton depends on the polarization direction of the decaying tau. This dependence leads to separated $x_{\ell}$ distributions for the LFV and SM tau decays in certain momentum directions of the final charged lepton. Utilizing this effect, the SM $\tau^+\tau^-$ background has been suppressed to some extent and the limits have been improved. As the angular distribution of the charged lepton momentum in the LFV and SM tau decays depend on the polarization direction of the decaying tau, the limits obtained for different chiral structures of the ALP coupling are different from each other. The best and worst limits respectively correspond to the V+A and V$-$A cases and the limit for the V/A case lies somewhere in between. The improvement achieved by utilizing the polarization-induced effects is significant and suggests that this procedure can also be employed in similar collider searches to enhance the sensitivity. In the unpolarized muon beams case and assuming the ALP mass to be 1 MeV, the strongest expected limits on $c_{\tau e}/f_a$ ($c_{\tau \mu}/f_a$) obtained in this study for the V+A, V$-$A and V/A chiral structures are 7.08$\times10^{-5}$ (7.00$\times10^{-5}$), 6.89$\times10^{-5}$ (6.79$\times10^{-5}$) and 9.87$\times10^{-5}$ (9.71$\times10^{-5}$) $\mathrm{TeV}^{-1}$, respectively. The corresponding limits achieved in the case of polarized muon beams are $2.48\times10^{-5}$ ($2.27\times10^{-5}$), $4.90\times10^{-5}$ ($4.68\times10^{-5}$) and $4.37\times10^{-5}$ ($3.92\times10^{-5}$) $\mathrm{TeV}^{-1}$. A comparison shows that the limits achieved in both the unpolarized and polarized muon beams cases in this study are significantly stronger than the current experimental limits and the present analysis can improve the experimental limits on the $c_{\tau e}/f_a$ and $c_{\tau \mu}/f_a$ couplings by roughly one order of magnitude. It can be concluded that the present analysis can serve as a tool for searching for LFV ALPs as it provides the possibility to probe an unprecedented region of the parameter space. 

%***********************************************************
\section*{Acknowledgments}
The authors would like to thank Mohsen Dayyani Kelisani for fruitful discussions on particle accelerators.

%***********************************************************
\RaggedRight % removes extra spaces between words in bibliography

\end{document}